\documentclass[a4paper,11pt]{article}
\pdfoutput=1
\usepackage{jheppub}
\usepackage{amsmath}
\usepackage{amssymb}
\usepackage{color}
\usepackage{graphicx}
\usepackage{hyperref}
\usepackage{xspace,slashed}
\usepackage[utf8]{inputenc}
\usepackage{subcaption}
\usepackage{multirow}
\usepackage{algorithm}
\usepackage{algorithmic}
\usepackage{stackengine}
\usepackage{enumitem}
\usepackage{etoolbox}

\usepackage{booktabs}

\makeatletter
\g@addto@macro\bfseries{\boldmath}
\makeatother
\newcommand\FM{F}
\newcommand{\cTlambda}{{\cal T}_\lambda}
\newcommand{\kl}{{kl}}
\newcommand\qtran{q_t}
\newcommand{\ourphi}{\varphi}
\newcommand{\zone}{z_1}
\newcommand{\ztwo}{z_2}
\newcommand{\zthree}{z_3}
\newcommand{\zedi}{z_i}
\newcommand{\bnotf}{b_{0,f}}
\newcommand{\etal}{\eta}
\newcommand{\Tmap}{R}
\newcommand{\lone}{l}
\newcommand{\exlone}{l}
\newcommand{\exloneperp}{l_\perp}
\newcommand{\exloneperpmod}{l_t}
\newcommand{\ltwo}{{\bar l}}
\newcommand{\ltone}{\tilde{l}}
\newcommand{\pt}{{\tilde{p}}}
\newcommand{\eesv}{EESV}
\newcommand{\as}{\alpha_s}
\newcommand{\tp} {{\tilde p}}
\newcommand{\mathd}{\mathrm{d}}
\newcommand\nf{n_{f}}
\newcommand\Ca{C_A}
\newcommand\Cf{C_F}
\newcommand{\V}{V}
\renewcommand{\v}{v}
\newcommand{\nm}{n_{\rm m}}
\newcommand{\nmperp}{n_{{\rm m},\perp}}
\newcommand{\nmt}{n_{{\rm m},t}}
\newcommand{\nmv}{\vec n_{\rm m}}
\newcommand{\philn}{\ourphi_{ln}}
\newcommand{\ot}{\overline t}

\DeclareMathOperator{\Gmpl}{G}
\DeclareMathOperator{\Empl}{E}

\allowdisplaybreaks

\newcommand{\OXaff}{Rudolf Peierls Centre for Theoretical Physics,
  Clarendon Laboratory, Parks Road, Oxford OX1 3PU, UK}
\newcommand{\WADaff}{Wadham College, Oxford OX1 3PN, UK}
\newcommand{\TTPaff}{Institute for Theoretical Particle Physics,
  KIT, 76128 Karlsruhe, Germany}
\newcommand{\MILaff}{INFN, Sezione di Milano-Bicocca, and Universit\`a di Milano-Bicocca, Piazza della Scienza 3, 20126 Milano, Italy}
\newcommand{\MPIaff}{ Max-Planck-Institut für Physik, 80805 M\"unchen, Germany}

\preprint{
  \begin {flushright}
    OUTP-22-04P, TTP22-022, P3H-22-036, MPP-2022-36 
  \end{flushright}
}
\title{
Linear power corrections to $e^+e^-$ shape variables in the three-jet region
}
\author[a,b]{Fabrizio Caola,}
\author[a]{Silvia Ferrario Ravasio,}
\author[c]{Giovanni Limatola,}
\author[d]{Kirill Melnikov,}
\author[c,e]{Paolo Nason,}
\author[d]{Melih Arslan Ozcelik}
\affiliation[a]{\OXaff}
\affiliation[b]{\WADaff}
\affiliation[c]{\MILaff}
\affiliation[d]{\TTPaff}
\affiliation[e]{\MPIaff}

\emailAdd{fabrizio.caola@physics.ox.ac.uk}
\emailAdd{silvia.ferrarioravasio@physics.ox.ac.uk}
\emailAdd{g.limatola@campus.unimib.it}
\emailAdd{kirill.melnikov@kit.edu}
\emailAdd{paolo.nason@mib.infn.it}
\emailAdd{melih.oezcelik@kit.edu}

\abstract{ We use an abelian model to study linear power corrections
  which arise from infrared renormalons and affect event shapes in
  $e^+e^-$ annihilation into hadrons. While previous studies explored
  power corrections in the two-jet region, in this paper we focus on
  the three-jet region, which is the most relevant one for the
  determination of the strong coupling constant.  We show that for a
  broad class of shape variables, linear power corrections can be
  written in a factorised form, that involves an
  analytically-calculable function, that characterises changes in the
  shape variable when a soft parton is emitted, and a constant
  universal factor.  This universal factor is proportional to the
  so-called Milan factor, introduced in earlier literature to describe
  linear power corrections in the two-jet region.  We find that the
  power corrections in the two-jet and in the three-jet regions are
  different, a result which is bound to have important consequences
  for the determination of the strong coupling constant from event
  shapes.  As a further illustration of the power of the approach
  developed in this paper, we provide explicit analytic expressions
  for the leading power corrections to the $C$-parameter and the
  thrust distributions in the $N$-jet region for arbitrary $N$, albeit
  in the abelian model.  }

\begin{document}

\maketitle 
%======================================================================
\section{Introduction}
% ======================================================================
The goal of this paper is to investigate non-perturbative
power-suppressed corrections to shape variable distributions in
$e^+e^-$ annihilation into hadrons in the three-jet region and
beyond (we will refer to shape variables in $e^+e^-$
annihilation into hadrons as \eesv{} from now on). This is
interesting for at least two reasons.  First, three-jet production in
$e^+e^-$ collisions is proportional to the strong coupling constant
$\as$, making it an important process for its determination.  Second,
even if currently other methods seem to be more promising for this
purpose~\cite{ParticleDataGroup:2020ssz}, studying three-jet
production in $e^+e^-$ collisions remains important from a theoretical
perspective. Indeed, it allows us to explore the interplay between
perturbative and non-perturbative effects in jet-production processes
in the simplest possible setting and, hopefully, understand how to
extend such studies to the more complex case of hadronic collisions.

Leading power-suppressed corrections  to shape variable distributions
are typically linear, i.e. they are of order $\Lambda_{\rm QCD}/Q$,
where $\Lambda_{\rm QCD}$ is a hadronic energy scale and $Q$ is the
hard scale of the process under consideration.  For typical $e^+e^-$ collider
energies these corrections are expected to be of the order of few
percent; therefore, they must be included in the calculation of
\eesv{}.
 Power corrections can be estimated either using Monte Carlo event
 generators~\cite{Bethke:2009ehn,Dissertori:2009ik,Kardos:2018kqj}, or
 with the help of analytic models~\cite{Akhoury:1995fb,Salam:2001bd,
   Hoang:2007vb,Abbate:2010xh,Gehrmann:2012sc,Mateu:2012nk,Hoang:2015hka,
   Gracia:2021nut}.  The Monte Carlo approach is certainly the most
 practical, but it is also very difficult to justify it theoretically
 in a convincing way.  Analytic models are conceptually more
 appealing, since they make contact with certain specific features
 that the full theory should have, such as infrared renormalons.
 Unfortunately, analytic calculations of linear power corrections are
 typically performed in the two-jet limit~\cite{Manohar:1994kq,
   Webber:1994cp, Dokshitzer:1995qm,Nason:1995np, Dasgupta:1996ki,
   Nason:1996pk, Beneke:1997sr, Dokshitzer:1997ew, Dokshitzer:1997iz,
   Dokshitzer:1998pt,Korchemsky:1999kt,Korchemsky:2000kp,
   Gardi:2001ny,Gardi:2003iv,Bauer:2003di,Lee:2006nr} and then
 extrapolated to the three-jet region. When applied to the
 determination of the strong coupling constant from event shape
 observables, this procedure leads to values of $\as$ that are several
 standard deviations lower than the world average, $\alpha_s(M_z)=
 0.1179(1)$~\cite{ParticleDataGroup:2020ssz}.  Significant ambiguities
 in the extrapolation of the power corrections from the two-jet to the
 three-jet region have been pointed out as a possible reason for this
 discrepancy~\cite{Luisoni:2020efy}.

The study presented in this paper is the continuation of  ref.~\cite{Caola:2021kzt} written by some of us recently.
The main result of that
reference can be summarised as follows: an observable that is 
inclusive with respect to soft QCD radiation does not exhibit linear
power corrections. Later we will review
the assumptions
required to reach this conclusion, but for now we emphasise that it
has important implications for the computation of \eesv{}.
Indeed, as shown in ref.~\cite{Caola:2021kzt}, this
result implies that no linear power corrections are
induced by the
recoil of hard partons caused by the emission of a soft
parton.\footnote{We note that in practical implementations of recoil
effects, as done e.g. in parton shower Monte Carlo programs, certain
loose requirements must be satisfied to ensure the validity of this
result. Several, but not all, commonly used recoil schemes do satisfy these
requirements, see ref.~\cite{Caola:2021kzt}. We will
comment more on this point in sec.~\ref{sec:cpar_pheno}.}  In
ref.~\cite{Caola:2021kzt} we have exploited this observation to
compute numerically the linear power corrections to both the
$C$-parameter and the thrust distributions in the three-jet region. 

In the present paper we build upon the results of
ref.~\cite{Caola:2021kzt} and further examine power corrections to
EESV. Our first goal is to obtain an analytic expression for the
linear power corrections to the $C$-parameter, in an abelian model,
for the generic three-jet final state. In ref.~\cite{Caola:2021kzt} we
performed an analytic computation of these corrections only for an
unrealistic scenario, where the shape variable was defined by first
clustering the $q\bar{q}$ pair that originated from the gluon
splitting.\footnote{This procedure only regards the analytic
  calculations. Numerical calculations were instead performed
  including the splitting.} In this paper we do not make this
simplifying assumption, derive analytic results accounting for the
$g^* \to q \bar q$ splitting and use the quark and anti-quark momenta
to compute the changes in the shape variables.  The availability of
analytical results is very helpful for obtaining robust
phenomenological predictions in an efficient way since, in contrast to
the numerical computations described in ref.~\cite{Caola:2021kzt},
they do not require a numerical extrapolation to small gluon
masses. Moreover, we will show that the analytic computation allows us
to uncover peculiar structures in our results, that may be interesting
to explore further in a field-theoretic context.

We use the approach described in ref.~\cite{Caola:2021kzt} to
construct a contribution to the $C$-parameter of the process $\gamma^*
\to q +\bar q + \gamma + (g^* \to q \bar q)$ which can lead to linear
power corrections.  We then compute the linear power correction to the
$C$-parameter by integrating over the energy of the virtual soft
gluon. Although this direct calculation is
quite complex, as it involves non-trivial elliptic integrals, the
final result is remarkably simple, suggesting that an alternative way
to compute it should exist. We then explain how to perform the
computation in such a way that the simplicity of the final result
becomes apparent.  We show that linear power corrections to the
$C$-parameter are given by a factorised expression, where one factor
depends on the properties of the shape variable and the kinematics of
soft partons, and the second factor is a universal constant that only
depends on the radiation dynamics.  At this point, it becomes
apparent that the factorisation property found for the $C$-parameter
is actually valid for a large class of shape variables, and that the constant
factor is the same for all of them.

We formulate  the conditions that an observable should satisfy
for this factorisation to happen, and demonstrate its power
by computing  linear
power corrections to the thrust distribution in the three-jet region,
in addition to the $C$-parameter.  We then extend these results 
to the general case where  $N$ hard jets are produced in $e^+e^-$ annihilation. 
While studies of the $C$-parameter and the thrust with  $N$-jet final states, for $N>3$,
have limited scope, they can still be useful for
phenomenology~\cite{Kluth:2006bw,OPAL:2005vad,Schieck:2006tc}. Furthermore,
we believe that the relative ease with which
power corrections to the $N$-jet
case can be computed is a
strong indicator of the efficacy of our approach.

We can also apply our procedure to the computation
  of shape variables in the two-jet region. Upon doing so, we obtain the same
  result as  in refs.~\cite{Dokshitzer:1997iz,Dokshitzer:1998pt} and, thus,
  find that the universal constant factor that we identify in the
context of the three-jet calculations is related to the so called ``Milan
factor'' of refs.~\cite{Dokshitzer:1997iz,Dokshitzer:1998pt}.
We note that, in the literature, the constant ``Milan factor'' is
often presented as a correction factor that should be applied to
calculations of \eesv{} performed with massive gluons but neglecting
the gluon splitting into $q \bar q$ pairs and the impact of such a
splitting on the observables.  In fact, this way of describing it,
although justified by its historical development, is slightly
misleading.  Computing power corrections to shape variables by
considering the emission of a massive gluon leads to wrong answers. In
fact, the deficiency of this approach can already be seen from the
fact that the results depend upon ambiguities in the definitions of
shape variables when there are massive partons in the final
state.\footnote{See for example the
discussion around eq.~(5.56) in
ref.~\cite{Beneke:1998ui}.\label{foot:amb}} These ambiguities do not
arise if the universal factor is applied to corrections to shape
variables caused by an emission of a massless soft parton in a
particular kinematic, as we will explain in this paper.\footnote{In
fact, the definition of Milan factor in the two-jet and symmetric
three-jet limit of ref.~\cite{Dokshitzer:1998pt} is fully consistent
with ours, so our novelty claim only regards factorisation in the
generic three-jet region. However, in our discussion, the Milan factor
emerges rigorously in the context of the large-$\nf$ framework, where
it is seen clearly that the final state with just a massive gluon
plays no role in the computation of power corrections.}

The remainder of this paper is organised as follows. In
sec.~\ref{sec:gen} we review the results obtained in
ref.~\cite{Caola:2021kzt} and set up the notation. In
sec.~\ref{sec:C3} we study linear power corrections to the
$C$-parameter in the three-jet region.  In sec.~\ref{sec:dirint} we
begin by computing these corrections using direct analytic integration
with a cut-off on the energy of the virtual gluon.  This calculation
involves elliptic integrals and it is highly non-trivial. In
principle, it can be performed using the formalism of elliptic
polylogarithms~\cite{Broedel:2017kkb,Broedel:2017siw,Broedel:2018iwv,Broedel:2018qkq,Broedel:2019hyg,Weinzierl:2022eaz}
and we show how to do this in appendix~\ref{sec:ellint}.  It turns out,
however, that the same calculation can be done without resorting to
this technology and we explain how to do this in the remaining part of
the section.

Although, as we mentioned earlier, the computation of
power-corrections to the $C$-parameter in the three-jet region is
quite demanding, its result is so simple that an explanation is called
for.  We provide such an explanation in sec.~\ref{sec:fact} where we
show how linear power corrections naturally factorise into a
process-dependent part and a universal factor that can be easily
computed.  We discuss subtleties related to this factorisation, since
it leads to divergent expressions whose regularisation needs to be
understood.

In sec.~\ref{sec:general} we show that the factorisation of power
corrections to the $C$-parameter found in sec.~\ref{sec:fact} applies
to a broader class of observables.  In sec.~\ref{sec:oncollinear}, we
formulate the conditions that the observables must satisfy for this
factorisation to happen, and in sec.~\ref{sec:Cexample} we show in
detail how they are satisfied in the case of the $C$-parameter.  In
sec.~\ref{sec:mil} we discuss the relation between our approach and
that of refs.~\cite{Dokshitzer:1997iz,Dokshitzer:1998pt} where the
Milan factor was originally introduced.

In sec.~\ref{sec:app}, we use our approach to compute linear power
corrections to thrust in the three-jet region (sec.~\ref{sec:thrust}),
and then to both the $C$-parameter and thrust for a generic $N$-jet
final state (sec.~\ref{sec:njet}).

In sec.~\ref{sec:numerics}, we perform preliminary phenomenological
studies of the power corrections in the three-jet region.
First, in sec.~\ref{sec:validation} we
validate the analytic results against the numerical ones of
ref.~\cite{Caola:2021kzt}. We then conjecture how
to generalise our results,
which are only valid for $q\bar q\gamma$ final states, 
to the phenomenologically interesting case of QCD jets. In
sec.~\ref{sec:cpar_pheno} we present phenomenological predictions
for the $C$-parameter and the thrust distributions in the three-jet
region, and compare results for the $C$-parameter with the findings of
ref.~\cite{Luisoni:2020efy}. We conclude in
sec.~\ref{sec:conc}.

Before continuing, we note that several parts of this paper present
complex calculations that rely on the use of sophisticated
analytic techniques. Although we believe that these techniques are
both interesting from a formal point of view and potentially useful for
calculations of other observables, their 
detailed presentation may obscure the main
results of the paper. For this reason, we would like to suggest an
alternative way to read this paper that avoids many
technicalities, yet makes our main result and message clear.

A reader not interested in technicalities should begin  with reading sec.~\ref{sec:gen} and
the beginning of sec.~\ref{sec:C3} up to  eq.~(\ref{eq:3.3}) and
then move   to eq.~(\ref{eq:tlIc}) where a remarkably simple result for power corrections
to the $C$-parameter in the three-jet region 
is shown.  One then  continues with  the reading of sec.~\ref{sec:fact} and, especially,
sec.~\ref{sec:F},   where
the phase space parametrisation that clarifies the origin of the
factorisation property is introduced. The factorisation formula derived in sec.~\ref{sec:F}
is divergent and ill-defined.   In section
\ref{sec:regF} a regularisation procedure is introduced that allows for
the full analytic calculation of the constant universal
factor. Alternatively, one can study  sec.~\ref{sec:AltF}, where a different regularisation 
procedure is presented which, besides showing that the universal factor
is a constant, also explains how to write it  as a finite integral that
can be easily  computed numerically.  Then, the  reader should  move
to sec.~\ref{sec:general} where 
sec.~\ref{sec:oncollinear}  illustrates how to compute
the observable-dependent coefficient by expressing it  as a finite
integral, which  is easily evaluated numerically.  In
the subsequent section~\ref{sec:Cexample} this  procedure is illustrated
in full detail using the $C$-parameter as an example. After that
the reader can move  to sec.~\ref{sec:numerics} where phenomenological
predictions in the three-jet region are discussed.

\section{Generalities and linear power corrections to shape variables}
\label{sec:gen}
Renormalons provide a robust framework for studying non-perturbative
corrections to QCD observables (see ref.~\cite{Beneke:1998ui} for a
review).  Linear power corrections $\mathcal O(\Lambda_{\rm QCD}/Q)$
arising from renormalons can be computed in a rigorous way in the so
called large-$\nf$ approximation, i.e. in the limit where the number
of light flavours is taken large and negative so that $\as |\nf|
\approx 1$.
Working within this framework, in ref.~\cite{Caola:2021kzt} we have
developed a formalism for studying linear power corrections to
$e^+e^-$ shape variables in the three-jet region.{\footnote{We emphasise that
  ``three-jet'' refers to an abelian model where a hard final state gluon is replaced by a photon.}
In this section, we
briefly recall the main results of ref.~\cite{Caola:2021kzt} and set
the stage for further analysis.

We are interested in linear power corrections to the
cumulant $\Sigma$ of a shape variable $\V$. We define the cumulant as
\begin{equation} \label{eq:CumulantDefinition}
  \Sigma (\v) = \sum_F\int \mathd \sigma_F \; \theta (\V(\Phi_F) - \v),
\end{equation}
where $F$ stands for  a particular 
final state, $\Phi_F$ denotes
the phase-space point of the state $F$ and $\V(\Phi_F)$ is the value of the shape
variable at the point $\Phi_F$.  We assume that the shape variable is
defined in such a way that it vanishes for a two-parton final
state.\footnote{For the $C$-parameter this feature follows from its
definition.  For the thrust $T$ we consider $\overline T = 1-T$ to ensure it.}
We note that we define the cumulant as the cross section for producing
a final state with the value of the shape variable $\V$ larger than a
 constant $\v$,
while it is common in the literature to define it as the cross section
for $\V<\v$.  Thanks to this  choice, the two-jet region does not
contribute to the cumulant and, since we are only interested in the
three-jet region, this simplifies the discussion.

According to refs.~\cite{FerrarioRavasio:2018ubr,Caola:2021kzt}, power
corrections to eq.~(\ref{eq:CumulantDefinition}) are obtained by
computing
\begin{align}
\label{eq:Tolambda}
\Sigma(v;\lambda) =& \int \mathd\Phi_{\rm b}
\left[\frac{\mathd\sigma^{\rm b}(\Phi_{\rm b})}{\mathd\Phi_{\rm
      b}}+\frac{\mathd\sigma^{\rm v}_\lambda(\Phi_{\rm
      b})}{\mathd\Phi_{\rm b}}\right] \theta(\V(\Phi_{\rm b})-v) +
\int
\mathd\Phi_{g^*}\frac{\mathd\sigma^{g^*}(\Phi_{g^*}^\lambda)}{\mathd\Phi_{g^*}}
\theta(V(\Phi_{g^*}^\lambda)-v)\nonumber \\ & -
\frac{\lambda^2}{\bnotf\as} \int \mathd\Phi_{q\bar{q}}
\frac{\mathd\sigma^{q\bar{q}}(\Phi_{q\bar{q}})}{\mathd\Phi_{q\bar{q}}}\delta(m^2_{q\bar{q}}-\lambda^2)\left[\theta(\V(\Phi_{q\bar{q}})-\v)
  -\theta(\V(\Phi_{g^*}^\lambda)-\v)
  \right],
\end{align}
where $\Phi_{\rm b}$ is the Born phase space (i.e. the
$q\bar{q}\gamma$ phase space in our case) and $\sigma^{\rm b}$ is the
corresponding cross section. In addition, $\sigma^{\rm v}_\lambda$
describes
corrections  to the cross section due to the exchange of a virtual  gluon with 
mass $\lambda$; $\Phi_{g^*}^\lambda$ is the phase space that describes
the emission of such a gluon,  and $\sigma^{g^*}_\lambda$ 
is the corresponding cross section. Finally, $\Phi_{q{\bar{q}}}$ is the phase
space that contains a $q\bar{q}$ pair with  
invariant mass $m_{q\bar q}$, and $\sigma^{q{\bar q}}$ is the
corresponding cross section. We have also
defined the beta function
coefficient in the large-$n_f$ limit 
\begin{equation}
  \bnotf = -\frac{n_f T_R}{3\pi}.
  \label{eq:bnotf}
\end{equation}
Eq.~\ref{eq:Tolambda} is the main ingredient for the computation of renormalon contribution to the cross
section, as illustrated in more detail in appendix~\ref{app:renstruct}.

We will refer to the momenta of the primary quark,
antiquark and photon as 
$p_1$, $p_2$, $p_3$, respectively, and to the momenta of 
quarks which  arise  from the gluon splitting   as  $\lone$, $\ltwo$.
Introducing the notation
\begin{align}
  [\mathd p]&=\frac{\mathd^3 p}{2p^0(2\pi)^3}\,,
\end{align}
we define the  phase space factors as follows 
\begin{align}
  \mathd\Phi_{\rm b}&=[\mathd p_1][\mathd p_2][\mathd p_3] (2\pi)^4\delta^{(4)}(p_1+p_2+p_3-q), \\
  \mathd\Phi_{g^*}^\lambda&=[\mathd p_1][\mathd p_2][\mathd p_3][\mathd k] (2\pi)^4\delta^{(4)}(p_1+p_2+p_3+k-q), \\
  \mathd\Phi_{q{\bar q}}&=[\mathd p_1][\mathd p_2][\mathd p_3][\mathd \lone] [\mathd \ltwo]
                   (2\pi)^4\delta^{(4)}(p_1+p_2+p_3+\lone+\ltwo-q)\,,
\end{align}
where $q$ is the momentum of the $e^+e^-$ system, and 
all momenta are light-like  except for $k$, that has $k^2=\lambda^2$. The normalisation
factor on the third line of eq.~(\ref{eq:Tolambda}) is such that
\begin{equation}\label{eq:inclusivesplit}
  - \frac{\lambda^2}{\bnotf\as} \int [\mathd \lone] [\mathd \ltwo]
  \frac{\mathd\sigma^{q\bar{q}}(\Phi_{q\bar{q}})}{\mathd\Phi_{q\bar{q}}}
  (2\pi)^4\delta^{(4)}(\lone+\ltwo-k) =2\pi
  \frac{\mathd\sigma^{g^*}(\Phi_{g^*}^\lambda)}{\mathd\Phi_{g^*}}.
\end{equation}
Terms linear in $\lambda$ that are implicitly present in
eq.~(\ref{eq:Tolambda}) are associated with renormalons and lead to
linear power corrections ${\cal O}(\Lambda_{\rm QCD}/Q)$.  It was
shown in ref.~\cite{Caola:2021kzt} that there are no ${\cal
  O}(\lambda)$ terms in the expansion of the virtual corrections in
powers of $\lambda$.  Furthermore, from eq.~(\ref{eq:inclusivesplit})
it follows that the contribution proportional to
$\theta(\V(\Phi_{g^*}^\lambda)-v)$ cancels in eq.~(\ref{eq:Tolambda})
among the second and third terms.  Thus, corrections linear in
$\lambda$ can only arise from the term proportional to
$\theta(V(\Phi_{q{\bar q}}^\lambda)-v)$.
 As explained in ref.~\cite{Caola:2021kzt},
for observables that satisfy certain criteria to be described later, 
these power corrections  can be found  by studying the emission of
a \emph{soft} $q\bar q$ pair. 

It is convenient to describe soft emissions by introducing a mapping
between the final state momenta with no extra emission, denoted as
$\pt_1$, $\pt_2$ and $\pt_3$, and $\lone$, $\ltwo$, to the
four-momenta $\{p_1,p_2,p_3,\lone$, $\ltwo\}$ of the full
five-particle final state,
\begin{equation}
  \pt_1,\, \pt_2,\, \pt_3,\, \lone,\, \ltwo \leftrightarrow p_1,\,
  p_2,\, p_3,\, \lone,\, \ltwo.
  \label{eq:mapp}
\end{equation}
Momentum conservation yields the constraints
\begin{equation}
  q= \pt_1+\pt_2+ \pt_3=p_1+ p_2+ p_3+ \lone+ \ltwo.
\end{equation}
In the following, we will refer to the $\pt_i$ momenta as the
\emph{underlying Born momenta}.  The final state momenta 
$p_i$ differ from the underlying Born momenta by recoil
effects, that are of the order of the momenta $\lone$, $\ltwo$. We
consider mappings of the form~\cite{Caola:2021kzt}
\begin{equation}
  p_i=p_i(\{\pt\},k),\quad i=1,\ldots, 3\,,
\end{equation}
where $\{\pt\}$ denotes
the set $\{\pt_1,\pt_2,\pt_3\}$,
and  $k=\lone+\ltwo$. We will also use the notation $\{p\}$ to
denote $\{p_1,p_2,p_3\}$.

We  consider mappings that are collinear-safe with respect to the directions of the radiating partons. This means
  that in the collinear limit
$k\propto \pt_1$ we must have $\pt_1=k+p_1$, $\pt_{2,3}=p_{2,3}$, and
analogous relations for $k\propto \pt_2$. Furthermore, we require that
for small $k$ the mapping is linear in $k$, i.e.\footnote{We stress
that these requirements do not restrict the generality of our
results. Indeed, it is always possible~\cite{Caola:2021kzt} to
explicitly construct recoil schemes with the above properties. The main
reason for working with this class of mappings is that they simplify
the investigation of linear power corrections~\cite{Caola:2021kzt}.}
\begin{equation}\label{eq:linearmap}
  p_i^\mu=\pt_i^\mu+\Tmap^\mu_{i,\nu}(\{\pt\}) k^\nu+{\cal
    O}\left(k_0^2\right).
\end{equation}

To expose linear power corrections, we follow
ref.~\cite{Caola:2021kzt} and introduce an operator $\cTlambda$ that
extracts the ${\cal O}(\lambda)$ terms from a given
expression. Then, writing
\begin{equation}
|M(\{p\},\lone,\ltwo)|^2 \equiv - \frac{\lambda^2}{
  2\pi\bnotf\as}
\frac{\mathd\sigma^{q\bar{q}}(\Phi_{q\bar{q}})}{\mathd\Phi_{q\bar{q}}},
\end{equation}
we obtain 
\begin{equation}
  \cTlambda[ \Sigma (\v;\lambda)] = \cTlambda \left[\int \mathd
    \Phi_{q{\bar q}} \,2\pi\,\delta(m_{q{\bar q}}^2-\lambda^2)
    |M(\{p\},\lone,\ltwo)|^2 \theta \left(\V(\{p\},\lone,\ltwo)-
    \v\right)\right].
\end{equation} 
Expanding the $\theta$ function around
$\V(\{p\},\lone,\ltwo)=\V(\{\pt\})$, we find
\begin{equation}
  \begin{split}
    &\cTlambda[ \Sigma (v;\lambda)] = \cTlambda \left[\int\mathd
      \Phi_{q{\bar q}} \,2\pi\, \delta(m_{q{\bar q}}^2-\lambda^2)
      \left|M(\{p\},\lone,\ltwo)\right|^2 \theta \left(\V(\{\pt\})- \v\right)
      \right]+
    \\
    &\cTlambda \bigg[\int \mathd \Phi_{q{\bar q}}
      \,2\pi\,\delta(m_{q{\bar q}}^2-\lambda^2)
      \left|M(\{p\},\lone,\ltwo)\right|^2 \delta\left(\V(\{\pt\}) - \v\right)
      \left[\V(\{p\},\lone,\ltwo)-\V(\{\pt\})\right]\bigg]\,.
    \label{eq:masterEquationM1}
  \end{split}
\end{equation}
As shown in ref.~\cite{Caola:2021kzt},
if the mapping eq.~\eqref{eq:mapp} satisfies eq.~\eqref{eq:linearmap},
terms involving an inclusive integration at fixed underlying Born
momenta  do not produce
${\cal O}( \lambda)$ contributions.
  This implies
  that no terms linear in $\lambda$ can arise in the first term
  on the right-hand side  of eq.~(\ref{eq:masterEquationM1}), and
  thus the $\cTlambda$ operator projects this term to zero.
  The second term on the right-hand side  of eq.~(\ref{eq:masterEquationM1}),
is not inclusive and, therefore, may induce linear power corrections. 
  We observe that this term involves  a factor  
$\V(\{p\},\lone,\ltwo)-\V(\{\pt\})$, that vanishes in the collinear
and soft limits for any IR-safe observable $\V$. Under these circumstances, a
linear term in $\lambda$ can only arise from the infrared-singular
part of $|M|^2$.  We are thus allowed to use the leading soft approximation 
and replace
\begin{equation}
  |M(\{p\},\lone,\ltwo)|^2 \rightarrow \mathcal{N} \frac{\mathd
    \sigma^{\rm b }(\tilde\Phi_{\rm b})}{\mathd \tilde\Phi_{\rm b}}
  \frac{J^\mu J^\nu}{\lambda^2} {\rm Tr}\left[\hat \lone \gamma^\mu
    \hat \ltwo \gamma^\nu\right],
 \label{eq:sigsoft}
\end{equation}
where $\hat p \equiv \gamma^\alpha\, p_\alpha$, 
$\mathd \tilde\Phi_{\rm b}$ is the underlying Born phase space
\begin{equation}
  \mathd\tilde\Phi_{\rm b} = [\mathd\pt_1][\mathd\pt_2][\mathd\pt_3]
  (2\pi)^4\delta^{(4)}(\pt_1+\pt_2+\pt_3-q),
\end{equation}
$J^{\mu}$ is the eikonal current for the emission of a soft
gluon with momentum $k$ 
\begin{equation}
J^\mu=\frac{\pt_1^\mu}{(\pt_1 k)} - \frac{\pt_2^\mu}{(\pt_2 k)},
\label{eq:eikcurr}
\end{equation}
and the trace arises from the inclusion of the gluon decay into
a quark-antiquark  pair. The normalisation factor ${\cal N}$
introduced in eq.~(\ref{eq:sigsoft}) evaluates to 
\begin{equation}
  \mathcal{N} = \left[-\frac{1}{2\pi\bnotf \alpha_s}\right]\times
  g_s^2 C_F \times g_s^2 T_Rn_f = 24 \pi^2 \alpha_s C_F.
\label{eq:norm}
\end{equation}
We use the phase-space factorisation in the soft limit and arrive
at~\cite{Caola:2021kzt} 
\begin{equation}
  \begin{split}
    \cTlambda[\Sigma(v;\lambda)] &= \int\mathd\tilde\Phi_{\rm
      b}\frac{\mathd \sigma^{\rm b}(\tilde\Phi_{\rm b})}{\mathd
      \tilde\Phi_{\rm b}} \delta\left(\V(\{\pt\}-v\right)\times
    \cTlambda\Bigg[ \mathcal N \int [\mathd k] \frac{J^\mu
        J^\nu}{\lambda^2} \theta\left(\omega_{\rm
        max}-\frac{(kq)}{\sqrt{q^2}}\right) \\
      &\times \int
      [\mathd \lone][\mathd\ltwo] (2\pi)^4 \delta^{(4)}(k-\lone-\ltwo) {\rm
        Tr}\left[\hat \lone \gamma^\mu \hat \ltwo
        \gamma^\nu\right]\left[
        \V(\{p\},\lone,\ltwo)-\V(\{\pt\})\right]\Bigg]\,.
    \label{eq:masterEquation0}
  \end{split}
\end{equation}
Note that  we 
introduced a cut-off
$\omega_{\rm max}$ on the energy of the intermediate gluon  in the
rest frame of $q$ to regulate the UV divergence of the eikonal
integral and make eq.~\eqref{eq:masterEquation0} well defined. Linear
power corrections do not depend on this
regulator~\cite{Caola:2021kzt}.

To further simplify eq.~\eqref{eq:masterEquation0}, we note that
it is natural to expect that the change of the shape variable due
to the emission of the two soft partons and the change due to recoil
effects separate as follows
\begin{equation}
  V(\{p\},\lone,\ltwo)-\V(\{\pt\}) = [\V(\{\pt\},\lone,\ltwo) -
    \V(\{\pt\})] + [\V(\{p\}) - \V(\{\pt\})] +
  \mathcal O(k_0^2).\label{eq:recoilsep}
\end{equation}
We will elaborate more on this point in
sec.~\ref{sec:oncollinear}. For now, we
just
note that, as explained in ref.~\cite{Caola:2021kzt}, 
both the $C$-parameter and the thrust satisfy this
condition~\cite{Caola:2021kzt}. If the separation as in
eq.~\eqref{eq:recoilsep} is possible, one can expand the second term
on the r.h.s. in $k$ using eq.~\eqref{eq:linearmap}.  In
ref.~\cite{Caola:2021kzt} it was shown that no linear terms in
$\lambda$ arise  from an unrestricted integral in the radiation
variables at fixed underlying Born kinematics even if we multiply the cross
section by an expression linear in $k$.  Thus, if we insert
eq.~(\ref{eq:recoilsep}) into eq.~(\ref{eq:masterEquation0}) the
second term does not lead to any linear power correction, and
eq.~(\ref{eq:masterEquation0}) can be further simplified by replacing 
\begin{equation}
  \V(\{p\},\lone,\ltwo)-\V(\{\pt\}) \to \V(\{\pt\},\lone,\ltwo) -
  \V(\{\pt\}).\label{eq:SimplifiedSfactor}
\end{equation}
One then obtains
\begin{equation}
  \cTlambda[\Sigma(v;\lambda)] =
  \int\mathd\sigma^{\rm b}(\tilde\Phi_{\rm b})
  \delta(\V(\{\tp\})-\v)
  \times\bigg[\mathcal N
    \cTlambda\left[I_{\V}(\{\pt\},\lambda)\right]\bigg],
  \label{eq:mastereqsimp}
\end{equation}
where we have introduced
\begin{equation}
  \begin{split}
    I_{\V}(\{\pt\},\lambda) =&
    \int [\mathd k] \frac{J^\mu
      J^\nu}{\lambda^2} \theta\left(\omega_{\rm
      max}-\frac{(kq)}{\sqrt{q^2}}\right) 
    \int
        [\mathd \lone][\mathd\ltwo] (2\pi)^4 \delta^{(4)}(k-\lone-\ltwo)
        \\
        &\times{\rm
          Tr}\left[\hat \lone \gamma^\mu \hat \ltwo
          \gamma^\nu\right]\left[
          \V(\{\pt\},\lone,\ltwo)-\V(\{\pt\})\right].
        \label{eq:Iv}
  \end{split}
\end{equation}
Eqs.~(\ref{eq:mastereqsimp}, \ref{eq:Iv}) provide the starting point for
the analytic investigations described in the next section.

\section{Power corrections to the $C$-parameter in the three-jet region}
\label{sec:C3}
In this section, we obtain an analytic result for linear power
corrections to the $C$-parameter distribution in the three-jet region.
We  consider the process
\begin{equation}
  \gamma^*(q)\to q(p_1)+\bar q(p_2) + \gamma(p_3),
\end{equation}
  and assume that all final-state particles are resolved. For a process with
  $N$ massless final-state particles with momenta $p_{1,..,N}$, the
  $C$-parameter is given by 
\begin{equation}
  C = 3 - 3 \sum \limits_{i>j}^{N} \frac{(p_i p_j)^2}{(p_i q) (p_j q)},
  \label{eq:cparm}
\end{equation}
where $q=\sum_{i=1}^N p_i$ is the momentum of the decaying virtual
photon. We  need to apply this formula to the case $N=5$, with
$p_4=\lone$ and $p_5=\ltwo$. From eq.~\eqref{eq:cparm}, it is easy
to see that the $C$-parameter satisfies the condition shown in
eq.~\eqref{eq:recoilsep}. To compute linear power corrections to the
cumulant, we can then use eqs.~(\ref{eq:mastereqsimp}, \ref{eq:Iv}). 

As mentioned in the introduction, we use two different approaches
to perform the analytic computation. First, in sec.~\ref{sec:dirint} we
directly integrate eq.~\eqref{eq:mastereqsimp}. Conceptually, this way
of obtaining linear power corrections is straightforward as it follows
directly from the results of ref.~\cite{Caola:2021kzt}. This result then provides a
solid benchmark for the following investigations.

The direct
analytic  integration of eq.~\eqref{eq:mastereqsimp} is quite
interesting from a technical point of view.  As 
mentioned in the introduction,
it involves elliptic structures and it is highly non trivial, yet
it yields a remarkably simple result.  It turns out that the complexity
of the calculation is a direct consequence of the way in which the
    integral over virtual-gluon energy  
is regulated, cf. eq.~(\ref{eq:mastereqsimp}). Indeed, while the
explicit $\omega_{\rm max}$ regulator in eq.~\eqref{eq:mastereqsimp}
arises naturally in the formalism of ref.~\cite{Caola:2021kzt}, it is not
optimal. We investigate this issue in sec.~\ref{sec:fact}, where we
show that it is possible to set up the calculation in a fully
factorised way, which both dramatically simplifies it  and
allows us to generalise our results to a wide class of observables. We
study such generalisation in sec.~\ref{sec:general}, where we also
comment on the relation between results obtained in this paper and calculations in the
two-jet and in the symmetric three-jet limits
performed within the 
Milan-factor approach of
refs.~\cite{Dokshitzer:1997iz,Dokshitzer:1998pt}.

\subsection{Direct integration with an explicit energy cut-off}
\label{sec:dirint}
We wish to analytically integrate
$I_{\V}(\{\pt\},\lambda)$ defined  in eq.~\eqref{eq:Iv}, for $\V=C$. For ease
of notation, from now on we will replace $\pt_i\to p_i$ in all expressions,
since the $p_i$ momenta, $i=1,2,3$,   do not appear in the calculation any longer.

We begin with the integration over the phase space of the emitted
quarks, keeping $k$ and $q$ fixed. It is convenient to perform this
integration in the rest frame of the decaying gluon with momentum $k$.
After that, we integrate over the direction of the vector $\vec{k}$ in
the rest frame of $q$, keeping $p_{1,2,3}$ fixed, and leaving the
integration over the gluon's energy to be done at the end.  Since the
angular integrations are straightforward,\footnote{All such
integrations can be performed in four dimensions as the gluon mass
protects from both soft and collinear divergences.}  we do not discuss
them further and just quote the result. We  write
\begin{equation}
  I_C(p_1,p_2,p_3,\lambda) = -\frac{3\lambda}{4 \pi^3 q}
  \sum_{i=1}^5 I_i(x,y,\lambda),
  \label{eq:3.3}
\end{equation}
where
\begin{equation}
I_i(x,y,\lambda) =  \int
\limits_{0}^{\beta_{\rm max}} {\rm d} \beta \; G_i(\beta,x,y).
\label{eq:Iidef}
\end{equation}
In eq.~\eqref{eq:3.3}, $q = \sqrt{q^2}$, $\beta$ is the velocity of the
massive gluon in the $q$ rest frame, $\beta_{\rm max} =
\sqrt{1-\lambda^2/\omega_{\rm max}^2}$ and the two variables $x$ and
$y$ parameterise the three-jet kinematics. They are
defined in terms of the scalar products  $q p_i$, $i=1,2,3$, as follows
\begin{equation}
q p_i  = \frac{q^2}{2} (1-\zedi),
\label{eq:zdef}
\end{equation}
so that $\sum \limits_{i=1}^{3} \zedi=1$. We then parameterise
$z_{1,2,3}$ as $\zone = xy$, $\ztwo = x(1-y)$, $\zthree = 1-x$.

The
explicit expressions for the functions $G_i$ are rather lengthy. We
report them in appendix~\ref{sec:funG5}.
A glance at these functions  shows that they are quite 
complex  and difficult to integrate. This happens for two reasons.
First, when taken separately, the $G_i$ functions exhibit very strong
    singularities at $\beta = 0$ and moderately strong singularities
    at $\beta = 1$.  The $\beta = 0$ singularities are unphysical, and
    cancel  in the sum. The $\beta=1$ singularities are physical
    and reflect the fact that the integral in eq.~(\ref{eq:Iv}) diverges
    at large values of the gluon energy, necessitating  the
    cut-off $\omega_{\rm max}$ or $\beta_{\rm max}$.
Second,  the integrand exhibits an \emph{elliptic}
    structure. Indeed, a glance at $G_{3,5}$ shows the 
    appearance of square roots of a degree-four polynomial,
    $\sqrt{(1-\beta^2)(1-c_{12}^2 \beta^2)}$, where $c_{12}^2 =
    1-s_{12}^2 = \cos^2 (\theta_{12}/2)$ and $\theta_{12}$ is the relative
    angle between three-momenta $\vec p_1 $ and $\vec p_2$ in the $q$
    rest frame.  It is well known that integration of square roots of
    degree-four polynomials leads to elliptic integrals.

In principle, methods exist that allow one to integrate over $\beta$
systematically.  To this end one defines a class of elliptic
polylogarithms with kernels which  close under integration by parts.
Integration then becomes an algebraic problem and can be performed in
a (relatively) straightforward way. We report such a calculation in
appendix~\ref{sec:ellint}. Unfortunately, the result of such
integration is very complicated. It involves both generalised and
elliptic polylogarithms, and its simplification is non-trivial.\footnote{We
  explain how to do it in appendix~\ref{sec:ellint}.
}

It turns out, however, that one can integrate over $\beta$ in a
different way, bypassing entirely the need for elliptic
polylogarithms. We  illustrate this point by using the function $G_5$ as
an example.  This allows us to expose all the key features of the
method, while keeping the discussion relatively short.

The explicit expression for $G_5$ is presented
in appendix~\ref{sec:funG5}, but we display it here one more time
for convenience,
\begin{align}
 G_5 & = \frac{\sqrt{1-\beta^2} \ln
   \left(\frac{1+\beta}{1-\beta}\right) \ln
   \left(\frac{\sqrt{1-\beta^2 c_{12}^2}+\beta s_{12}}{\sqrt{1-\beta^2
       c_{12}^2}-\beta s_{12}}\right) }{ 64 \beta^8 s_{12} x (x
   (y-1)+1) (x y-1) \sqrt{1-\beta^2 c_{12}^2}} \nonumber \\ & \times
 \Big ( \beta^6 x \big[x^2 (y-1) y+x \left(-4 y^2+4
 y-5\right)+5\big] +\beta^4 \big [x^2 \left(54 y^2-54 y-17\right)
\label{eq:G5text}
\\
& -21 x^3 (y-1) y+55 x-38 \big ]
+5 \beta^2 \big [x^2 \left(-24 y^2+24 y+5\right)
\nonumber \\
& +11 x^3 (y-1) y-17 x+12 \big]
-35 (x-2) \left(x^2 (y-1) y+x-1\right)
\Big  ).
\nonumber 
\end{align}
We note that $G_5$ is
integrable at $\beta = 1$ but not at $\beta = 0$.
Hence, we can set
$\beta_{\rm max}$ to  $1$ in $I_5$,  but we need to 
consider the regulated integral
\begin{equation}
I^{\rm reg}_{5} = \int \limits_{\beta_{\rm min}}^{1} {\rm d} \beta \;
G_5(\beta,x,y).
\label{eq:i5reg}
\end{equation}
With a slight abuse of notation, we will drop the ``reg'' superscript in
what follows. 

The function $G_5$ contains two logarithms of $\beta$, two
$\beta$-dependent square roots and a rational function of $\beta$.  To
integrate over $\beta$,  it is convenient to introduce an integral
representation for the two logarithms in eq.~(\ref{eq:G5text}). We
write them as
\begin{equation}
\begin{split} 
& \ln \frac{1+\beta}{1-\beta} = 2 \beta \int \limits_{0}^{1}
  \frac{{\rm d} r} {1 - r^2 \beta^2}, \\ & \ln
  \left(\frac{\sqrt{1-\beta^2 c_{12}^2}+\beta s_{12}}{\sqrt{1-\beta^2
      c_{12}^2}-\beta s_{12}}\right) = 2 \beta s_{12} \sqrt{1 -
    \beta^2 c_{12}^2} \int \limits_{0}^{1} \frac{{\rm d} \xi }{ 1-
    \beta^2 \Delta^2},
\end{split}
\label{eq:logrep}
\end{equation}
where $\Delta$ is a function of $\xi$, $\Delta = \sqrt{c_{12}^2 +
  s_{12}^2 \xi^2}$. After this transformation, eq.~\eqref{eq:i5reg} has the
structure
\begin{equation}
I_5 = \int \limits_{0}^{1} {\rm d} r\; {\rm d} \xi \; \int
\limits_{\beta_{\rm min}}^{1} \; {\rm d} \beta \; \sqrt{1-\beta^2} \;
R(\beta^2,r^2,\xi^2),
\label{eq2.12}
\end{equation}
where $R(\beta^2,r^2,\xi^2)$ is a \emph{rational function} of its
variables whose dependence on $x$ and $y$ is suppressed. We note that,
thanks to the integral representations shown in  eq.~\eqref{eq:logrep}, one of the
two $\beta$-dependent square roots has disappeared from the integrand
in eq.~(\ref{eq2.12}).
To remove the second root,  we  write $\beta = \sin \ourphi$ and
obtain
\begin{equation}
I_5 = \int \limits_{0}^{1} {\rm d} r\; {\rm d} \xi \; \int
\limits_{\ourphi_{\rm min} }^{\pi/2} \; {\rm d} \ourphi \;
\cos^2(\ourphi) \; R(\sin^2(\ourphi),r^2,\xi^2),
\end{equation}
where $\ourphi_\textrm{min} = \arcsin(\beta_{\rm min})$.

Since the integrand  depends on squares of $\cos \ourphi$ and $\sin
\ourphi$, it is convenient to change variables one more time and write
$\ourphi = \ourphi_1/2$, so that $ 0 < \ourphi_1 < \pi$,
$\cos^2(\ourphi_1/2) = (1+\cos \ourphi_1 )/2$, $\sin^2(\ourphi_1/2) =
(1-\cos \ourphi_1 )/2$.  We find
\begin{equation}
I_5 = \frac{1}{2} \int \limits_{0}^{1} {\rm d} r\; {\rm d} \xi \; \int
\limits_{2 \ourphi_{\rm min}}^{\pi} \; {\rm d} \ourphi_1 \; R_1(\cos
\ourphi_1 ,r^2,\xi^2),
\end{equation}
where $R_1$ is another rational function of its arguments.
To proceed further, we perform the partial fractioning of
$R_1$ with respect to
$\cos \ourphi_1 $ and obtain 
\begin{equation}
R_1 = \sum \limits_{i=-3}^{-1} \frac{P_i(r^2,\xi^2)}{(1-\cos \ourphi_1)^i}
+
\frac{1}{\Delta^2 - r^2} \left [
  \frac{P_\Delta(r^2,\xi^2)}{(2 - \Delta^2 + \Delta^2 \cos \ourphi_1 )} 
+
\frac{P_r(r^2,\xi^2)}{(2 - r^2 + r^2 \cos \ourphi_1 )}
\right ], 
\label{eq:R1}
\end{equation}
where $P_{-3,-2,-1}(r^2,\xi^2)$ and $P_{\Delta,r}(r^2,\xi^2)$ are
polynomials in $r^2$ and $\xi^2$.  As indicated in eq.~(\ref{eq:R1})
these polynomials only contain even powers of $r$ and $\xi$, a property that will be important for what follows. 

Integrating over $\ourphi_1$ is straightforward for all of the five
terms shown in eq.~(\ref{eq:R1}). In fact, as far as the first three
terms are concerned, we perform a trivial
integration over $\ourphi_1$ and expand in $\beta_{\rm min}$.  The
resulting expressions can be integrated over $\xi$ and $r$ in a
straightforward way.  We do not show the results of such an
integration since they are not very illuminating; the important point,
however, is that they are easy to obtain. 
We then focus on the last two terms in eq.~(\ref{eq:R1}) and 
study
\begin{equation}
  I_5^{45} = \frac{1}{2} \int \limits_{0}^{1} {\rm d} r\; {\rm d} \xi
  \; \int \limits_{2 \ourphi_{\rm min}}^{\pi} \; 
  \frac{{\rm d} \ourphi_1}{\Delta^2 - r^2} \left [
  \frac{P_\Delta(r^2,\xi^2)}{2 - \Delta^2 + \Delta^2 \cos \ourphi_1 } 
+
\frac{P_r(r^2,\xi^2)}{2 - r^2 + r^2 \cos \ourphi_1 }
\right ].
  \label{eq:i45}
\end{equation}
Both terms in eq.~\eqref{eq:i45}
are integrable at $\ourphi_1 = 0$ which corresponds to $\beta = 0$.
Hence, we can set $\varphi_{\rm min} \to 0$. 
Using
\begin{equation}
\int \limits_{0}^{\pi} \frac{ {\rm d} \ourphi }{a - b \cos \ourphi}
= \frac{\pi}{\sqrt{a^2 - b^2}},
\label{eq:eq2.16}
\end{equation}
we obtain the following result 
\begin{equation}
\begin{split} 
 I_5^{45} = \frac{ \pi}{4} \int \limits_{0}^{1} \frac{ {\rm d} r \;
   {\rm d} \xi}{\Delta^2 - r^2} \left [ \frac{P_\Delta(r^2,\xi^2)}{
     \sqrt{1-\Delta^2}} + \frac{P_r(r^2,\xi^2)}{\sqrt{1-r^2}} \right
 ].
\label{eq:i45az}
\end{split} 
\end{equation}
An apparent feature of the integrand in eq.~(\ref{eq:i45az}) is the
singularity at $\Delta = r$. This singularity is fake; indeed, using
the explicit form of the two polynomials $P_\Delta(r^2,\xi^2)$ and
$P_r(r^2,\xi^2)$, one can check that the expression in the square
brackets in eq.~(\ref{eq:i45az}) vanishes at $\Delta = r$.

Although the singularity at $\Delta = r$ and the way it is regulated
in eq.~(\ref{eq:i45az}) suggest that one has to integrate both terms in
eq.~(\ref{eq:i45az}) at once, it turns out to be beneficial to integrate
them separately.  This requires introducing a regulator, and we do
this by moving the pole at $\Delta = r$ away from the real axis,
i.e. we write $1/(\Delta^2 - r^2) \to 1/(\Delta^2 - r^2 +i 0)$.
Once the regulator is introduced, we can deal with the two terms in eq.~(\ref{eq:i45az}) separately.
We focus on the first one to illustrate the next step.
Changing
variables $\xi = {\rm th}(u), r = {\rm th}(w)$, we find
\begin{equation}
\begin{split}
& \int \limits_{0}^{1} \frac{ {\rm d} r \; {\rm d} \xi}{\Delta^2 -
    r^2+i0} \frac{P_\Delta(r^2,\xi^2)}{ \sqrt{1-\Delta^2}} =
  \frac{1}{s_{12}} \int \limits_{0}^{\infty} \frac{ {\rm d} u \; {\rm
      d} w \; {\rm ch}(u) \; P_\Delta({\rm th}^2(u),{\rm th}^2(w))}{
    {\rm ch}^2(u) - s^2_{12} {\rm ch}^2(w) + i0 }.
\end{split}
\end{equation}
We observe that the integrand in the above equation is an even
function of $u$. Hence, we extend the $u$-integration region to the
whole real axis, change the integration variable $u \to z = {\rm
  sh}(u)$ and arrive at an integral that can be readily evaluated
using Cauchy's residue theorem. We are then left with a one-dimensional
integral in $w$. An analogous calculation can be done for the second
term of eq.~\eqref{eq:i45az}, only in this case we integrate over
$w$ using Cauchy's theorem and we are left with a one-dimensional
integral in $u$.

We then map the integration regions
of the two remaining one-dimensional
integrals on the interval $[s_{12},1]$ by performing the change of
variable $t=1/{\rm ch}(w)$ for the first term of eq.~\eqref{eq:i45az}
and $t=1/{\rm ch}(u)$ for the second one. Combining the two terms,
we obtain a result of the form
\begin{equation}
I_5^{45} = \int\limits_{s_{12}}^1 \frac{\mathd t\;
  t^2}{\sqrt{1-t^2}{\sqrt{t^2-s_{12}^2}}}
\left[
  \alpha(x,y)+\beta(x,y)t^2 + \gamma(x,y)t^4 + \delta(x,y)t^6
  \right],
\label{eq:intt}
\end{equation}
where we have re-introduced the dependence on $x,y$ to stress that
$I_5^{45}$ depends non-trivially on the underlying Born kinematics.
It is straightforward to integrate over $t$ in eq.~(\ref{eq:intt}).
The result can be
expressed through the two complete elliptic integrals
 \begin{equation}
\begin{split} 
&  K(z) = \int \limits_{0}^{1}
  \; \frac{{\rm d} t}{\sqrt{(1-t^2)(1-z t^2)}},\;\;\;\;
  E(z) = \int \limits_{0}^{1}
 \; \frac{{\rm d} t \; \sqrt{1-z t^2} }{\sqrt{1-t^2}}.
\end{split}
\label{eq:EK}
 \end{equation}
 We do not present this result here since it is not very illuminating; in fact
 it is significantly more complex than the result for the full
 function $I_C$ that we will show below.

Summarising the key steps,
we note that we were able to obtain this relatively simple result by $a)$ using an
 integral representation for the logarithms in eq.~\eqref{eq:G5text},
 which also removes one of the square roots; $b)$ performing a change of
 variables to linearise the other square root in
 eq.~\eqref{eq:G5text}; $c)$ using a \emph{different} integration strategy
 for the two terms in eq.~\eqref{eq:i45az}, which required introducing
 a regulator at intermediate stages.

 All other contributions $I_i$ of eq.~\eqref{eq:Iidef} can be computed
 along similar lines. Interestingly, none of the results for the other
 contributions contains elliptic functions. When all the different
 contributions are put together, we obtain a rather cumbersome result
 for $I_C$ which is function of the underlying Born kinematics and
 $\beta_{\rm max}$. However, we observe that ${\cal O}(\lambda)$ terms
 simplify dramatically.  In fact, we find that for a generic three-jet
 configuration all the terms that do not contain elliptic integrals
 cancel and the result assumes the remarkably simple form
 \begin{equation}
   {\cal T}_\lambda \left [ I_C \right ] = \frac{15}{128\pi}
   \frac{s_{12}^3}{1-\zthree}
   \left(\frac{\lambda}{q}\right) \left [
     \frac{(1+\zthree)}{2} K\left(c_{12}^2\right) -(1-\zone \ztwo)
     E\left(c_{12}^2\right) \right ].
   \label{eq:tlIc}
 \end{equation}
 We note that when writing eq.~(\ref{eq:tlIc}), we switched back
 to the $z_{1,2,3}$ variables, defined in
eq.~(\ref{eq:zdef}), and used
  \begin{equation}
s_{12}^2 = \frac{\zthree}{(1-\zone)(1-\ztwo)}.
  \end{equation}
  
  Using eq.~\eqref{eq:mastereqsimp} we can immediately translate
  this result into a shift in the cumulant distribution for a generic
  three-jet configuration
\begin{equation}
  \begin{split}
    \label{eq:shifC_nice}
          {\cal T}_\lambda \left [ \Sigma(c,\lambda) \right ] =& 
          \int\mathd\sigma^{\rm b}(\Phi_{\rm b})\delta(\V(\Phi_{\rm b})-\v)
          \times
          \\
          &
          \alpha_s C_F  \frac{45\pi}{16}
          \frac{s_{12}^3}{1-\zthree} \left
               [ \frac{(1+\zthree)}{2} K\left(c_{12}^2\right) -(1-\zone
                 \ztwo) E\left(c_{12}^2\right) \right ]
               \left(\frac{\lambda}{q}\right).
  \end{split}
\end{equation}
In the two-jet limit, $c_{12}=0$, $\ztwo \to 0$ and $\zone+\zthree\to
1$ (or, equivalently $\zone\to 0$ and $\ztwo+\zthree\to1$), and we
reproduce the well-known result~\cite{Dasgupta:1999mb,Smye:2001gq}
\begin{equation}
  \frac{{\cal T}_\lambda \left [\Sigma(0,\lambda) \right ]}
  {\mathd\sigma/\mathd C|_{c=0}}= -
  \frac{15}{16}\pi^2 \left(\frac{\lambda}{q}\right)\alpha_s.
  \label{eq:c2j}
\end{equation}
At the symmetric point, when the three jets are produced with equal energies,
$\zone=\ztwo=\zthree=1/3$, $c_{12}=1/2$ and
we obtain
\begin{equation}
\frac{{\cal T}_\lambda \left [\Sigma(3/4,\lambda) \right
]}{\mathd\sigma/\mathd C|_{c=3/4}} = \frac{15}{32} \sqrt{3} \pi \big[3
K\left({1}/{4}\right)-4 E\left({1}/{4}\right)\big]
\left(\frac{\lambda}{q}\right)\alpha_s,
\label{eq:c3j}
\end{equation}    
which agrees with the result of ref.~\cite{Luisoni:2020efy}, adapted
to our $\gamma^*\to q+ \bar q + \gamma$ case and corrected for the $n_f\to \infty$ limit.
We note that  we have used
\begin{equation}
\int\mathd\sigma\,\delta(\V-\v) = \frac{\mathd\sigma}{\mathd\V},
\end{equation}
in the above equations.

It is interesting to note that in ref.~\cite{Luisoni:2020efy} the
calculation of the power correction was performed within the so-called
Milan-factor approach, i.e. by multiplying the result of a simplified
computation that only involves the emission of a single massless gluon
by a universal factor to correct for gluon splitting.  It is clear
that the above computation neither explains the simplicity of the
final result, nor its relation to
the Milan-factor approach. In
order to find an explanation, we need to change the way we approach
the integration in  eq.~\eqref{eq:mastereqsimp}, as we discuss in what
follows.

\subsection{Factorised form of the power correction to the $C$-parameter}
\label{sec:fact}
Similar to the previous section, our starting point is
eq.~\eqref{eq:Iv} with $\V=C$. However, at variance with
what we did before, we now remove the $\theta$-function that provides
the upper limit for the gluon energy and consider
\begin{eqnarray}
  I^{\rm unreg}_{C}(\{\pt\},\lambda) &=&
  \int [\mathd k] \frac{J^\mu
      J^\nu}{\lambda^2} 
  \int
      [\mathd \lone][\mathd\ltwo] (2\pi)^4 \delta^{(4)}(k-\lone-\ltwo)
      \nonumber\\
          &&\times{\rm
      Tr}\left[\hat \lone \gamma^\mu \hat \ltwo
      \gamma^\nu\right]\left[
        C(\{\pt\},\lone,\ltwo)-C(\{\pt\})\right].
      \label{eq:IcUn}
\end{eqnarray}
As in the previous section, we will replace
$\pt_i\to p_i$ since there is no room for ambiguity. We note that,
since we removed the gluon energy cut-off, eq.~(\ref{eq:IcUn}) is
ill-defined.  In what follows, we will introduce a suitable
regularisation that, on the one hand, makes it finite and, on the
other hand, allows for its straightforward computation.

Before discussing the regularisation, we simplify eq.~\eqref{eq:IcUn} by
computing the trace and inserting the definition
of the $C$-parameter. We obtain
\begin{equation}
I_C^{\rm unreg} = -24 \int [{\rm d} k] \frac{J_\mu J_\nu}{\lambda^2}
\int [{\rm d} \lone] [{\rm d} \ltwo] \; 
(2\pi)^4 \delta^{(4)}(k - \lone - \ltwo)
\tilde C^{\alpha \beta} \frac{\lone^\alpha \lone^\beta}{(\lone q)}
\left [ -2 \lone^\mu \lone^\nu - g^{\mu \nu} \frac{\lambda^2}{2} \right ],
\label{eq:eq42}
\end{equation} 
where we have dropped the arguments of $I_C^{\rm unreg}$ to shorten the
notation. To obtain eq.~\eqref{eq:eq42}, we have used $\ltwo = k-\lone$ and
$k_\mu J^\mu=0$ to discard terms proportional to $k$ that originate from the
trace.  We have also defined a new rank-two 
tensor $\tilde C_{\alpha \beta}$ that reads 
\begin{equation}
\tilde C_{\alpha \beta} = \sum \limits_{i=1}^{3} \frac{ p_i^\alpha p_i^\beta }{(p_iq)},
\label{eq:eq3.2}
\end{equation}
and accounted for the fact that the contributions of the quark and the
antiquark to the $C$-parameter are equal.

To proceed further, we find it convenient to use $p_{1,2}$ as the basis vectors 
and employ  a Sudakov parametrisation for $l$
\begin{equation}
  \exlone^\mu = \exloneperpmod e^{\etal} \frac{p_1^\mu}{\sqrt{s} } +
  \exloneperpmod e^{-\etal} \frac{p_2^\mu}{\sqrt{s}} +
  \exloneperp^\mu, ~~~~~~~~ [\mathd \exlone] = \frac{\mathd \eta\,
    \mathd^2\vec\exlone_\perp}{2(2\pi)^3},
  \label{eq:eq3.4}
\end{equation}
where $s=2(p_1 p_2)$ and $\exloneperpmod = |\exloneperp|$. In
terms of these variables, 
the quark-antiquark phase space reads 
\begin{equation}
  [\mathd \exlone][\mathd\ltwo](2\pi)^4\delta^{(4)}(k-\exlone-\ltwo)
  = 2\pi[\mathd\exlone]\delta_+\left((k-l)^2\right) =
  \frac{1}{8\pi^2}\mathd\eta\,\exloneperpmod\mathd\exloneperpmod\mathd\ourphi
  \delta\left(\lambda^2-2(kl)\right),
\end{equation}
where $\ourphi$ is the azimuthal angle of $\vec\exlone_\perp$.
We would like to remove the $\delta$-function in the above equation by
integrating over the absolute value of the quark transverse momentum
$\exloneperpmod$.  Since, according to eq.~(\ref{eq:eq3.4}),
$\exlone^\mu$ is proportional to $\exloneperpmod$, it is straightforward
to do so. 
We introduce the rescaled
vector
\begin{equation}
  \ltone^\mu = \frac{\lone^\mu}{l_t} = \frac{p_1^\mu}{\sqrt{s}}e^{\eta}
  +
  \frac{p_2^\mu}{\sqrt{s}}e^{-\eta} + n^\mu,
\end{equation}
where $n^\mu = \exloneperp^\mu/\exloneperpmod$, and find 
\begin{equation}
  [\mathd\lone][\mathd\ltwo](2\pi)^4\delta^{(4)}(k-\lone-\ltwo) =
  \frac{1}{8\pi^2}\mathd\eta\mathd\ourphi\frac{\lambda^2}
       {(2k\ltone)^2},~~~~~~~~ \exloneperpmod = \frac{\lambda^2}{2(k\ltone)}.
       \label{eq:pslt}
\end{equation}
Using eq.~\eqref{eq:pslt}, we write the function $I_C$ given in 
eq.~\eqref{eq:eq42} in the following way
\begin{equation}
  I_C^{\rm unreg} = W_C\times \lambda\FM(p_1,p_2,\ltone),
  \label{eq:eq45}
\end{equation}
where
\begin{equation}
  W_C = -3
  \int \frac{  {\rm d} \etal {\rm d} \ourphi }{2 (2\pi)^3}
  \; \tilde C_{\alpha \beta} \frac{\ltone^\alpha \ltone^\beta }{(\ltone q)},
  \label{eq:wc}
\end{equation}
and
\begin{equation}
 \FM(p_1,p_2,\ltone) = 16  \pi \int [{\rm d} k]  \frac{J_\mu J_\nu}{\lambda^3} \left \{
-2 \ltone^\mu \ltone^\nu \frac{\lambda^8}{(2 k \ltone)^5}
- \frac{g^{\mu \nu} \lambda^6}{2 (2 k \ltone)^3} \right  \}.
\label{eq:eq46}
\end{equation}
The function $I_C$ can be
written as a product of two terms: one term, $\FM$, that is
observable-independent and involves the integration over the gluon
momentum, and another term that involves the integration over the rapidity and
azimuthal angle of the quark with momentum $\exlone$ and contains the
dependence on the observable. As indicated in eq.~(\ref{eq:eq46}), $\FM$ may depend upon $p_1,p_2$ and $\ltone$. However, as we will show in
the next section it is actually a constant.

We conclude this section by stressing that
eqs.~(\ref{eq:eq45}, \ref{eq:eq46}) are ill-defined. However, as we
will explain below, it is possible to introduce a regularisation
scheme that allows us to compute $\FM$ and $I_C$ in a relatively
straightforward way.  As we will show, the
\emph{observable-independent} constant $\FM$ can be associated with
the Milan factor of refs.~\cite{Dokshitzer:1997iz,Dokshitzer:1998pt}.

\subsubsection{The observable-independent factor $F$}
\label{sec:F}
We now discuss how to compute function $\FM$ in eq.~(\ref{eq:eq46}).
We  introduce a Sudakov parametrisation of the gluon momentum $k$
\begin{equation}
k^\mu = m_t e^{\eta_k} \frac{p_1^\mu}{\sqrt{s}}+
m_t e^{-\eta_k} \frac{p_2^\mu}{\sqrt{s}} +
k_\perp^\mu,
\label{eq:sudk}
\end{equation}
where we have defined $k_t=|k_\perp|$ and $m_t =
\sqrt{k_t^2+\lambda^2}$.  Using eq.~\eqref{eq:eq3.4} and
eq.~\eqref{eq:sudk}, it is straightforward to obtain
\begin{equation}
  \begin{gathered}
      k_\mu \ltone^\mu = m_t\;{\rm ch}( \eta_{kl} )-k_t\cos(\ourphi_{kl}),\\
    J_\mu J^\mu = -\frac{2 (p_1 p_2)}{(p_1 k) (p_2 k)} = -\frac{4}{m_t^2},
    ~~~~~
   ( J_\mu \ltone^\mu )^2 = \frac{4 \; {\rm sh}^2(\eta_{kl})}{m_t^2},
\end{gathered} 
\end{equation}
where $\eta_{kl}=\eta_k-\eta$, $\ourphi_{kl}=\ourphi_k-\ourphi$. These
relations allow us to write  the function $\FM$ defined   in eq.~(\ref{eq:eq46})
as follows
\begin{equation}
  \begin{split}
    &\FM = \frac{\lambda^3}{4\pi^2}\int \mathd k_t\,k_t \,\mathcal I_M,
  \\
  &\mathcal I_M = \int \frac{ {\rm d} \eta_k
   \,{\rm d} \ourphi_k }{m_t^2} \Bigg \{
\frac{-\lambda^2 {\rm sh}^2(\eta_{\kl}) }{ \left [m_t\, {\rm
      ch}(\eta_{\kl}) - k_t \cos ( \ourphi_{\kl} ) \right ]^5} + \frac{1
}{ \left [m_t\, {\rm ch}(\eta_{\kl}) - k_t \cos ( \ourphi_{\kl} ) \right
  ]^3} \Bigg \}.
\end{split}
\label{eq:eq3.16}
\end{equation}

\subsubsection{Regularisation procedure for $F$} 
\label{sec:regF}
To regulate $\FM$, we multiply the integrand by the factor 
$k_t^{2\epsilon}$ and write 
\begin{equation}
  \FM = \frac{\lambda^3}{4\pi^2}\int \limits_{0}^{\infty} \mathd k_t\,k_t^{1+2\epsilon}
  \,\mathcal I_M.
  \label{eq:regkt}
\end{equation}
We will explain why this
regularisation can be chosen at the end of this section;
for now, we assume  that it
does indeed regulate all divergences present in eq.~\eqref{eq:eq3.16}. 
Before explicitly computing $\FM$, we note that the integrand in
eq.~\eqref{eq:eq3.16} only depends on the difference of two rapidities
$\eta_{\kl}$ and on the difference of two azimuthal angles
$\ourphi_{\kl}$.  Since we integrate over all possible gluon
rapidities and over all azimuthal angles, $\FM$ is independent of
$\ltone$. Hence, as pointed out at the end of the previous section,
$\FM$ is a constant.

To compute $\FM$, we change variables $\eta_k \to \eta_{\kl} = x$
and $\ourphi_k \to \ourphi_{\kl}$.  It is straightforward to integrate
over $\ourphi_{\kl}$. We use eq.~(\ref{eq:eq2.16}) and compute
an appropriate number of derivatives with respect to $a$, to obtain an
expression for the integrals of $1/(a-b \cos ( \ourphi ) )^5$ and $1/(a-b
\cos ( \ourphi ))^3$.  We find the following result
\begin{equation}
\FM = \frac{\lambda^{2 \epsilon}}{128 \pi} \; \int
\limits_{-\infty}^{\infty} {\rm d} x \int \limits_{0}^{\infty} {\rm d}
\xi \; \frac{ \; \xi^{\epsilon} \; f(\xi,x) }{ (\xi+1) \left [ (\xi+1)
    {\rm ch} ^2(x)-\xi \right ]^{9/2}},
\end{equation}
where $\xi = k_t^2/\lambda^2$ and the function $f(\xi,x)$ reads
\begin{equation}
\begin{split} 
f(\xi,x) & = \xi (\xi+1)^2 {\rm ch} (6 x)+4 (1-2 \xi) (\xi+1) {\rm ch} (4 x)
\\
&  - \left(9 \xi^3+8 \xi^2-23 \xi-16\right) {\rm ch} (2 x)+2 \left(4 \xi^3+7 \xi^2+14 \xi+6\right).
\end{split} 
\end{equation}
To integrate over $\xi$, we change variables $\xi \to r$, with $\xi =
[r - {\rm ch}^2(x)]/[{\rm ch}^2(x) - 1]$, and find
\begin{equation}
\FM = \frac{ \lambda^{2\epsilon}}{256 \pi} \; \int
\limits_{-\infty}^{\infty} \frac{ {\rm d} x }{{\rm sh}^{2+
    2\epsilon}(x)} \; \int \limits_{{\rm ch}^2(x)}^{\infty} {\rm d} r
\; \frac{ \left(r - {\rm ch}^2(x)\right)^{\epsilon} \; \tilde f(x,r)}{(r-1) \;
  r^{9/2}},
\label{eq:3.37}
\end{equation}
where
\begin{equation}
\begin{split} 
\tilde f(x,r) & = \left (-8 r^2+40 r
-35\right) {\rm ch} (4 x)+  4 \left(8 r^3-44 r^2+70 r-35\right) {\rm ch} (2 x)
\\
& +64 r^3-192 r^2+240 r-105.
\end{split} 
  \end{equation}
The integration over $r$ can now be performed in a straightforward way and
the result can be written in terms of hypergeometric functions.  We
obtain 
 \begin{equation}
\begin{split} 
 \FM & = \frac{\lambda^{2\epsilon} }{1890 \pi^{3/2} } \Gamma
 \left(\frac{3}{2}-\epsilon\right) \Gamma (\epsilon+1) \int
 \limits_{0}^{\infty} \frac{ {\rm d} x}{ { \rm sh}^{2 + 2\epsilon}(x)
   \; {\rm ch}^{5-2 \epsilon}(x) } \Bigg \{ \\ & 630\; {\rm ch}^2(x) \;
 ({\rm ch} (2 x)+2) \,
 F_{21}\left(1,\frac{3}{2}-\epsilon,\frac{5}{2};\frac{1}{{\rm
     ch}^2(x)} \right) + \frac{63}{8} (2 \epsilon-3) \Bigg [ \\ &
   \left(16 \epsilon^2+8 (2 \epsilon+3) {\rm ch} (2 x)+3 {\rm ch} (4
   x)+21 \right ) \, F_{21}
   \left(1,\frac{5}{2}-\epsilon;\frac{7}{2},\frac{1}{{\rm
       ch}^2(x)}\right) \\ & +4 {\rm sh}^2(x) ( 4 \epsilon+{\rm ch} (2
   x)+3 ) \,
   F_{21}\left(2,\frac{5}{2}-\epsilon;\frac{7}{2},\frac{1}{{\rm
       ch}^2(x)} \right) \Bigg ] \Bigg \},
\end{split}
\label{eq:eq86}
\end{equation}
where we used the fact that the integrand is a symmetric
function of $x$ to restrict the integration region to the positive
real axis.

The integrand of the above expression has a quadratic singularity at
$x = 0$. We need to extract this singularity before integrating over
$x$. We do this by subtracting a suitable limiting form of the
integrand in the $x \to 0$ limit, computed for finite $\epsilon$, and
add it back. The difference is integrable and can be expanded in
$\epsilon$ while the subtracted term is simple enough to be integrated
for arbitrary $\epsilon$.
Following this  discussion, we write $\FM$ as the sum of  two terms 
\begin{equation}
  \FM = \FM^{(a)} + \FM^{(b)}.
\end{equation}
To obtain the subtraction term $F^{(a)}$, we let ${\rm ch}(x)\to 1$ in
eq.~\eqref{eq:3.37}, keeping the $1/{\rm
  sh}^{2+2\epsilon}(x)$ term as it is and find
\begin{equation}
  \begin{split}
  F^{(a)} &\equiv \frac{ \lambda^{2\epsilon}}{256 \pi} \; \int
  \limits_{-\infty}^{\infty} \frac{ {\rm d} x }{{\rm sh}^{2+
      2\epsilon}(x)} \; \int \limits_{1}^{\infty} {\rm d} r
  \; \frac{ \left(r - 1\right)^{\epsilon-1} \; \tilde f(0,r)}{ \;
    r^{9/2}} \\
  &=
  \frac{ \lambda^{2\epsilon}  \Gamma (1+ \epsilon) \Gamma \left(\frac{3}{2}-\epsilon\right)}{6 \pi^{3/2}}
  \big[  3 - \epsilon (1-2 \epsilon) \big]
  \int \limits_{0}^{\infty} \frac{ {\rm d} x }{ {\rm sh}^{2+2\epsilon}(x) }.
  \end{split}
  \label{eq:fa}
\end{equation}
The regular piece $ \FM^{(b)}$ is given by the difference between
$\FM$ in eq.~(\ref{eq:eq86}) and $\FM^{(a)}$.  Since this difference is
integrable at $x=0$, we can expand it in $\epsilon$. Moreover, it is
convenient to change the integration variable from $x$ to $t = e^{x}$,
$ 1 < t < \infty$.  We obtain 
  \begin{equation}
\begin{split} 
  \FM^{(b)} & = \frac{1}{ \pi} \int \limits_{1}^{\infty} {\rm d} t \;
  \left [ -\frac{(3 + t^2)(1 + 3 t^2)}{32 t^3} \ln \frac{t+1}{t-1} +
    \frac{(3+6 t + 14 t^2 + 6 t^3 + 3 t^4)}{16 t^2(1+t)^2} \right ]
  \\ & = \frac{1}{\pi} \left ( - \frac{5 \pi^2}{64} + \frac{1}{4}
  \right ).
  \label{eq:eq79}
  \end{split} 
\end{equation}
It remains to compute $\FM^{(a)}$. In order to do that, we 
change the integration variable using $x = \ln t$  and find
\begin{equation}
\int \limits_{0}^{\infty} \frac{ {\rm d} x }{ {\rm
    sh}^{2+2\epsilon}(x) } = 2^{2+2\epsilon} \int \limits_{1}^{\infty}
\frac{{\rm d} t \; t^{1+2\epsilon}}{(t^2-1)^{2+2\epsilon}} = 2^{1+2
  \epsilon} \frac{\Gamma(-1-2 \epsilon)
  \Gamma(1+\epsilon)}{\Gamma(-\epsilon)}.
\label{eq:sinh}
\end{equation}
We note that despite the original integral being ill-defined for $\epsilon=0$,
this result has a smooth $\epsilon\to 0$ limit.
We then use eq.~\eqref{eq:sinh} in the expression for $\FM^{(a)}$ given in 
eq.~\eqref{eq:fa}, take the $\epsilon \to 0$ limit and obtain
\begin{equation}
\FM^{(a)} = -\frac{1}{4 \pi}.
\label{eq:eq81}
\end{equation}
Combining the results for $\FM^{(a)}$ and $\FM^{(b)}$ in eq.~(\ref{eq:eq79})
and eq.~(\ref{eq:eq81}), we derive  the final result for the factor $\FM$
\begin{equation}
\FM = -\frac{5 \pi}{64}.
\label{eq:eq82}
\end{equation}

We conclude this section with a discussion of the regulator that we
employed to regularise the large-$k_t$ divergence, cf.
eq.~\eqref{eq:eq3.16}.  To justify the introduction of the analytic
regulator, we note that the large-$k_t$ region in
eq.~(\ref{eq:eq3.16}) does not lead to linear terms in
$\lambda$. Since the result with the analytic regulator should be
proportional to $\lambda^{1+2\epsilon}$ for dimensional reasons, terms
independent of $\lambda$ are forced to vanish. Hence, the analytic
regulator ``projects'' the integral in eq.~\eqref{eq:eq3.16} on ${\cal
  O}(\lambda)$ terms which are of interest to us, and removes all
$\lambda$-independent terms that exhibit divergences but are
irrelevant for the discussion of power corrections.

\subsubsection{Alternative procedure to evaluate $\FM$}
\label{sec:AltF}
We now examine an alternative procedure to evaluate $F$ that,  besides showing that
$F$  is a constant, allows us  to express it in terms of a convergent integral
amenable to  a direct numerical evaluation.\footnote{That we did in fact carry out as
  a further check of the analytic  result shown in eq.~(\ref{eq:eq82}). }

If we look at $\mathcal I_M$ in
eq.~\eqref{eq:eq3.16}, we notice that for small values of $\lambda$
both terms in the square bracket diverge when $\eta_{\kl}$ and
$\ourphi_{\kl}$ become small at the same time. We are expecting this
divergence, since for small $\lambda$ there is a collinear singularity
when the quark and gluon momenta are parallel to each other.

For  $\eta_{\kl}\approx \ourphi_{\kl} \approx 0$ the following equation holds 
\begin{equation}
  \frac{1}{m_t{\rm ch}(\eta_{kl})-k_t\cos\ourphi_{kl}}=
  \frac{2}{m_t}\times\frac{1+\mathcal O(\eta_{kl}^2,\ourphi^2_{kl})}
  {\frac{2(m_t-k_t)}{m_t}+\frac{k_t}{m_t}\ourphi_{kl}^2+\eta_{kl}^2}.
\end{equation}
Since the integral in  eq.~\eqref{eq:eq3.16} is dominated by the region
where $\eta_{kl}$ and $\ourphi_{kl}$ are of order $\lambda/m_t$, for
large $k_t$ we can further substitute in the integrand
\begin{equation}
  \frac{1}{m_t{\rm ch}(\eta_{kl})-k_t\cos\ourphi_{kl}}\to
  \left(\frac{2}{k_t}\right)
  \times\frac{1}{{\lambda^2}/{k_t^2}+\ourphi_{kl}^2+\eta_{kl}^2},
\end{equation}
as neglected terms only lead to $\mathcal O(\lambda^2/k_t^2)$
corrections. Using this and changing variables
$\{\eta_{kl},\ourphi_{kl}\} \to
\{r_{\eta\ourphi}\cos\theta_{\eta\ourphi},
r_{\eta\ourphi}\sin\theta_{\eta\ourphi}\}$, we can write 
\begin{equation}
  \mathcal I_M \approx \int r_{\eta\ourphi} \mathd r_{\eta\ourphi}
  \mathd\theta_{\eta\ourphi}\left[
    -32\left(\frac{\lambda^2}{k_t^7}\right)
    \frac{r_{\eta\ourphi}^2\cos^2\theta_{\eta\ourphi}}
         {\left({\lambda^2}/{k_t^2} +r_{\eta\ourphi}^2\right)^5}+
         \frac{8/k_t^5}{\left({\lambda^2}/{k_t^2}
           +r_{\eta\ourphi}^2\right)^3}
         \right]=\frac{8\pi}{3\lambda^4}\times\frac{1}{k_t},
\end{equation}
up to
$\mathcal O(\lambda^2/k_t^2)$ corrections. 
It follows that the expression
\begin{equation}
   \FM^{\rm (reg)} = \frac{\lambda^3}{4 \pi^2} \int \limits_{0}^{\infty}  \mathd k_t\;\left(
   k_t\, \mathcal I_M- \frac{8\pi}{3\lambda^4} \right),
\label{eq3.16-bis2}
\end{equation}
yields a convergent $k_t$-integration. In fact, the integral in eq.~(\ref{eq3.16-bis2})  can
be computed  numerically confirming the analytic result in eq.~(\ref{eq:eq82}).

Notice that using eq.~(\ref{eq:eq82}) we can provide yet another
  argument to justify the use of the regularisation procedure employed in the previous
  section.
In fact, since the integral in $\FM^{{\rm (reg)}}$ is convergent, we can introduce
the  analytic regulator $k_t^{2\epsilon}$ in the integrand of eq.~(\ref{eq3.16-bis2})
and treat the two terms separately.  Thanks to the properties of the analytic
regulator, the subtraction term vanishes 
\begin{equation}
\int  \limits_{0}^{\infty}  {\rm d} k_t k_t^{1+2\epsilon} = 0,
\end{equation}
and the first term in eq.~(\ref{eq3.16-bis2}) yields the analytically-regulated
expression, cf. eq.~(\ref{eq:regkt}),
which was used  in this section to compute the constant $F$. 

\subsubsection{The observable-dependent part}
\label{sec:3jetcaseW}
We now discuss the observable-dependent term $W_C$ defined in eq.~\eqref{eq:wc}.
To make the integral well-behaved, we introduce the analytic regulator
$e^{-\epsilon|\eta-\eta_q|}$ where $\eta_q$ is the rapidity of $q$ in the
dipole rest frame and write 
\begin{equation}
  W_C = -3 \int \frac{ {\rm d}\etal {\rm d} \ourphi }{2 (2\pi)^3} \;
  \frac{\tilde C_{\alpha \beta} \ltone^\alpha \ltone^\beta }{(\ltone q)}
  e^{-\epsilon|\etal - \eta_q|}.
  \label{eq:eq44}
  \end{equation}
A justification for this procedure is provided further down
in this section.

Given the structure of the tensor $\tilde C_{\alpha \beta}$,
cf. eq.~(\ref{eq:eq3.2}), $W_C$ in eq.~(\ref{eq:eq44}) naturally
splits into the sum of three terms. They read 
\begin{equation}
  W_C = -\frac{3}{8 \pi^3}\sum_{i=1}^{3} \frac{W^{(i)}_C}{(p_i q)},
  ~~~~~~~\textrm{with}~~ W_C^{(i)} = \int {\rm d}\etal {\rm d} \ourphi
  \; \frac{(p_i \ltone)^2 }{2 (\ltone q)} e^{-\epsilon|\etal - \eta_q|}.
\label{eq:eq3.31}
\end{equation}
We first discuss $W_C^{(3)}$. We write
$p_3 = q - p_{12}$, where $p_{12} = p_1 + p_2$ and obtain
\begin{equation}
  \begin{split}
    W_C^{(3)} &= \int \frac{ {\rm d}\etal {\rm d} \ourphi}{ 2 (\ltone
      q) } \left[(q \ltone) - (p_{12} \ltone)\right]^2
    e^{-\epsilon|\etal - \eta_q|} \\ &= \int \frac{ {\rm d}\etal {\rm
        d} \ourphi}{ 2 (\ltone q) } \left[ ( \ltone q )^2 - 2 ( p_{12}
      \ltone ) (\ltone q) + (p_{12} \ltone)^2
      \right]e^{-\epsilon|\etal - \eta_q|}.
  \end{split}
\label{eq:eq47}
\end{equation}
Among the three terms that appear in the last integral, the first two
involve a scalar product $(\ltone q)$ that cancels the $1/(\ltone q)$
factor; we will now show that the resulting integrals do not
contribute to the final result. Indeed, upon averaging over the
azimuthal angle $\ourphi$, both of these terms involve integrals of
the form
\begin{equation}
\int \limits_{-\infty}^{\infty} {\rm d} \etal
e^{-\epsilon|\etal-\eta_q|} e^{\pm \etal} =
e^{\pm\eta_q}\left(\frac{1}{1+\epsilon}-\frac{1}{1-\epsilon}\right)
=\mathcal O(\epsilon).
\end{equation}
Hence, taking the $\epsilon \to 0$ limit, we find that the two
integrals without $1/(\ltone q)$ in eq.~(\ref{eq:eq47}) do not contribute
to $W_C^{(3)}$.  We, therefore, re-write $W_C^{(3)}$ as
\begin{equation}
W_C^{(3)} = W_C^{(1)} + W_C^{(2)} + \Delta W_C^{(3)},
\label{eq:eq3.36}
\end{equation}
with
\begin{equation}
\Delta W_C^{(3)} = \int \frac{ {\rm d}\etal {\rm d} \ourphi \;
  e^{-\epsilon|\etal-\eta_q|} }{ 2 (\ltone q) } \; 2 (p_1 \ltone) (p_2
\ltone)=
\frac{s}{2}\int \frac{ {\rm d}\etal {\rm d} \ourphi \;
  e^{-\epsilon|\etal-\eta_q|} }{ 2 (\ltone q) }.
\end{equation}
In the last step, we have used $2(p_1\ltone)(p_2\ltone)=s/2$. To
proceed, we introduce a Sudakov decomposition for $q$ in the $p_1 p_2$
dipole rest frame
\begin{equation}
  q^\mu = m_{t,q}  \frac{p_1^\mu}{\sqrt{s}}e^{\eta_q}
  + m_{t,q}  \frac{p_2^\mu}{\sqrt{s}} e^{-\eta_q}
  +q^\mu_\perp,
\end{equation}
where $m_{t,q} = \sqrt{q^2+q_t^2}$ and $q_t=|q_\perp|$.
Writing $\Delta W_C^{(3)}$ in terms of
these variables and of the azimuthal angle $\ourphi_q$ of $q_\perp$,
we find
\begin{equation}
\begin{split} 
  \Delta W_C^{(3)} & = 
  \frac{s}{4} \int \frac{ {\rm d}\etal {\rm d} \ourphi \;
    e^{-\epsilon|\etal-\eta_q|} }{ m_{t,q} {\rm
      ch} (\etal - \eta_q) - \qtran \cos (\ourphi -\ourphi_q) } 
  = \frac{s \pi}{2} \int \frac{ {\rm d}\etal \;
    e^{-\epsilon|\etal-\eta_q| }}{ \sqrt{ m_{t,q}^2 {\rm
        ch}^2(\etal - \eta_q) - q_t^2 }}
  \\
  &
  = \frac{s \pi}{2
    \sqrt{q^2+\qtran^2} } \int \limits_{-\infty}^{\infty} \frac{ {\rm
      d}x} { \sqrt{ {\rm ch}^2(x) - c_{12}^2 }} =
  \frac{s \pi}{
    \sqrt{q^2+\qtran^2} } \int \limits_{0}^{\infty} \frac{ {\rm
      d}x} { \sqrt{ {\rm ch}^2(x) - c_{12}^2 }},
\end{split} 
 \end{equation}
where $c_{12}\equiv \qtran/\sqrt{q^2+\qtran^2}$ and in the second line
we have set $\epsilon = 0$ because the integral is finite.\footnote{We
note that $c_{12}$ defined here is indeed the cosine of the half angle
between the direction of $p_1$ and $p_2$ in the $q$ rest frame,
cf. sec.~\ref{sec:dirint}. Indeed, the cosine of the half-angle can
be computed in a frame-independent way using $c_{12}^2 = 1-(p_1
p_2)q^2/(p_1q)(p_2q)/2$, which evaluates to $q_t^2/m_{t,q}^2$ in the
dipole rest frame.} We then change the integration
variable $x={\rm arcch}(1/\xi)$ and use
\begin{equation}
\int\limits_{0}^{\infty} \frac{ {\rm
    d}x} { \sqrt{ {\rm ch}^2(x) - c_{12}^2 }} =
\int\limits_{0}^{1} \frac{ {\rm
    d}\xi} { \sqrt{(1-\xi^2)(1-c_{12}^2\xi^2)}} = K(c_{12}^2),
\end{equation}
where $K$ is the complete elliptic integral of the first kind. The
final result for $\Delta W_C^{(3)}$ reads
\begin{equation}
\Delta W_C^{(3)}=  \frac{s \pi}{
  \sqrt{q^2+\qtran^2} } K(c_{12}^2).
 \label{eq:eq3.41}
\end{equation}

We continue with the calculation of $W_C^{(1)}$. Proceeding in the
same way as in the calculation of $\Delta W_C^{(3)}$, we obtain
\begin{equation}
  W_C^{(1)} = \frac{s}{4} \int \frac{ {\rm d}\etal {\rm d} \ourphi \;
    e^{-\epsilon|\etal-\eta_q|} \; e^{-2 \etal} }{ 2 (\ltone q) } =
  \frac{s \pi }{4 \sqrt{q^2 + \qtran^2} } \int
  \limits_{-\infty}^{\infty} \frac{ {\rm d}\etal \;
    e^{-\epsilon|\etal-\eta_q|} \; e^{-2 \etal} }{ \sqrt{ {\rm
        ch}^2(\etal - \eta_q) - c_{12}^2 }}.
  \label{eq:eq3.46}
\end{equation}
The integral in eq.~(\ref{eq:eq3.46}) diverges. To extract the
divergence, we change the integration variable
$\etal = \eta_q - x$ and map the
integration region over $x$ to the positive semi-axis.  We find
\begin{equation}
  W_C^{(1)} = \frac{s \pi e^{-2\eta_q} }{4 \sqrt{q^2 + \qtran^2} } \int
  \limits_{0}^{\infty} \; \frac{ {\rm d}x \; e^{-\epsilon x} \; \left (
    e^{2 x} + e^{-2x} \right ) }{ \sqrt{ {\rm ch}^2(x) - c_{12}^2 }}.
\end{equation}
We now perform the same change of variables as before,
$x={\rm arcch}(1/\xi)$,
and obtain
\begin{equation}
  W_C^{(1)} = \frac{s \pi e^{-2\eta_q} }{4 \sqrt{q^2 + \qtran^2} }
  \int\limits_0^1
  \frac{\mathd\xi\,\xi^{\epsilon-2}\left(1+\sqrt{1-\xi^2}\right)^{-\epsilon}
  (4-2\xi^2)}{\sqrt{(1-\xi^2)(1-c_{12}^2\xi^2)}}.
\end{equation}
The above integral has a power divergence at $\xi=0$.  However, there
is no logarithmic divergence at $\xi = 0$ 
since the Taylor expansion of the integrand at small $\xi$ proceeds in
powers of $\xi^2$.  This implies that no $1/\epsilon$ contributions
are generated upon integration over $\xi$ and that, therefore,  the
factor $\left ( 1 + \sqrt{1-\xi^2} \right )^{-\epsilon}$ can be
dropped as it changes the result only at ${\cal O}(\epsilon)$.  Hence,
we write
\begin{equation}
\int \limits_{0}^{1} \; \frac{ {\rm d} \xi \; \xi^{\epsilon - 2} \left
  ( 1 + \sqrt{1-\xi^2} \right )^{-\epsilon} \; \left ( 4 -2 \xi^2
  \right ) }{ \sqrt{(1-\xi^2)(1- c_{12}^2 \xi^2) }} =
Z_1^a+Z_1^b,
   \label{eq:splitX1X2int1}
\end{equation}
where
\begin{equation}
\begin{split} 
& Z_1^a = \int \limits_{0}^{1} \frac{ {\rm d} \xi \; \xi^{\epsilon -
      2} \; \left ( 4 - 4 \xi^2 \right ) }{ \sqrt{(1-\xi^2)( 1-
      c_{12}^2 \xi^2) }}, \\
  & Z_1^b = \int \limits_{0}^{1} \frac{
    {\rm d} \xi \; \xi^{\epsilon - 2} \; \left ( 2 \xi^2 \right ) }{
    \sqrt{(1-\xi^2)(1- c_{12}^2 \xi^2) }} = 2 \int
  \limits_{0}^{1} \frac{ {\rm d} \xi }{ \sqrt{(1-\xi^2)(1-
      c_{12}^2 \xi^2) }} = 2 K(c_{12}^2),
\end{split} 
\label{eq:splitX1X2int2}
\end{equation}
and we set $\epsilon=0$ in $Z_1^b$ because the corresponding integral
is finite and does not require a regulator.

To compute $Z_1^a$, we integrate by parts, recognise that boundary
terms do not contribute and obtain
\begin{equation}
\begin{split} 
Z_1^a & = \frac{\xi^{\epsilon-1}}{\epsilon-1} \frac{ \left ( 4 - 4
  \xi^2 \right ) }{ \sqrt{(1-\xi^2)(1- c_{12}^2 \xi^2) }}
\Bigg |_{\xi=0}^{\xi=1} - \frac{1}{\epsilon-1} \int \limits_{0}^{1}
      {\rm d} \xi \; \xi^{\epsilon - 1} \frac{{\rm d} }{{\rm d} \xi}
      \frac{ \left ( 4 - 4 \xi^2 \right ) }{ \sqrt{(1-\xi^2)(
          1- c_{12}^2 \xi^2) }} \\ & = -4 (1-c_{12}^2) \int
      \limits_{0}^{1} \frac{ {\rm d} \xi}{ \sqrt{1-\xi^2} \; ( 1-
        c_{12}^2 \xi^2 )^{3/2} }= -4 E(c_{12}^2),
\end{split}
\label{eq:splitX1X2int3}
\end{equation}
where $E$ is the complete elliptic integral of the second kind. 
Putting everything together, we find 
\begin{equation}
W_C^{(1)} = \frac{s \pi e^{-2\eta_q} }{2 \sqrt{q^2 + \qtran^2} }
\left ( K(c_{12}^2) - 2 E(c_{12}^2) \right ).
\label{eq:eq3.48}
\end{equation}

It is clear  that $W_C^{(2)}$ is described by a similar formula, except that  the factor $e^{-2\eta_q}$ should be replaced
with $e^{2 \eta_q}$. Therefore 
\begin{equation}
W_C^{(2)} = \frac{s \pi  e^{2\eta_q} }{2 \sqrt{q^2 + \qtran^2} } \left (  K(c_{12}^2) - 2 E(c_{12}^2) \right ).
\label{eq:eq3.49}
\end{equation}
It is straightforward to combine eqs.~(\ref{eq:eq3.31},
\ref{eq:eq3.36}, \ref{eq:eq3.41}, \ref{eq:eq3.48}, \ref{eq:eq3.49}) to
obtain the full result for $W_C$. Written in terms of the $z_i$
variables introduced in eq.~\eqref{eq:zdef}, it reads
 \begin{equation}
   W_C = -\frac{3  }{ 2 \pi^2 q}
   \frac{s_{12}^3}
   {(1-\zthree)} \left [ \frac{(1+\zthree)}{2}
 K(c_{12}^2) - (1-\zone \ztwo) E(c_{12}^2) \right ].
 \label{eq:eq74}
 \end{equation}
 Multiplying $W_C$ and $\FM$, given in eqs.~\eqref{eq:eq74} and  \eqref{eq:eq82} respectively,
 we reproduce the result of the  direct calculation shown in 
eq.~\eqref{eq:tlIc}.

Before concluding this section, we note that $W_C$ in eq.~(\ref{eq:wc})
has an interesting interpretation. Indeed, consider the emission of a
soft massless gluon in the process $\gamma^*(q) \to q(p_1) + \bar q(p_2) +
\gamma(p_3)$ and compute  its contribution to the $C$-parameter.
Denoting the momentum of the soft gluon as $l$, writing it using a
Sudakov decomposition with  $p_{1,2}$ as two basis vectors,  and considering
contributions up to some value of the transverse momentum $l_t^*$,
we obtain
\begin{equation}
  G_C(\exlone_t^*) = -3 \int \limits_{0}^{\exlone_t^*} \frac{
    \exlone_t {\rm d} \exlone_t {\rm d} \eta {\rm d} \ourphi }{2
    (2\pi)^3} \; J_\mu J_\nu (-g^{\mu \nu}) \frac{\tilde C_{\alpha
      \beta} l^\alpha l^\beta }{(l q)}.
  \label{eq:Gc}
\end{equation}
Using 
\begin{equation}
  J_\mu J_\nu g^{\mu \nu} =   -\frac{4}{\exloneperpmod^2 },
\end{equation}
we find
\begin{equation}
  \frac{{\rm d} G_C(\exlone_t^*)}{ {\rm d} \exlone_t^*} \Bigg |_{\exlone_t^*
    = 0} = -12 \int \frac{ {\rm d}\eta {\rm d} \ourphi }{2 (2\pi)^3}
  \; \frac{\tilde C_{\alpha \beta} {\tilde l}^\alpha {\tilde l}^\beta
  }{{(\tilde l} q)}.
\end{equation}
Hence, although $\ltone$ in eq.~(\ref{eq:wc}) is the quark momentum,
we can compute the kinematic dependence of the power corrections,
including the gluon splitting to quarks, by considering the
contribution of a \emph{massless} soft gluon to the $C$-parameter at
zero transverse momentum since
\begin{equation}
W_C = \frac{{\rm d} G_C(\exlone_t^*)}{ 4 {\rm d} \exlone_t^*} \Bigg
|_{\exlone_t^* = 0}.
\end{equation}
In other words, we can use a gluon
with a small transverse momentum as a technical proxy
for our calculation, even though this replacement does not have any
particular physical significance in our formalism. We note that
this replacement only works
if an 
observable is linear in the momentum of the soft emissions.
We will
return  to this
point in the next section, when we will discuss how to generalise our
analysis to a broader class of observables.

\section{Generalisation of the factorised approach}
\label{sec:general}
 In the previous section we have focused on
the $C$-parameter. This was done in the interest of clarity but it is clear that many features of the
factorised approach discussed there are applicable to a
broader class of observables.
In this section we identify the properties
that a shape variable should satisfy for the factorisation result
to be applicable. 

As we will show, this generalisation bears striking
similarities to the Milan-factor approach of
refs.~\cite{Dokshitzer:1997iz,Dokshitzer:1998pt}, which was
successfully applied to study non-perturbative corrections near
singular kinematic configurations.  We explore the connections between
our formalism and the one of
refs.~\cite{Dokshitzer:1997iz,Dokshitzer:1998pt} in sec.~\ref{sec:mil},
where we show that our findings both confirm and generalise the Milan-factor
approach.

\subsection{Recoil, observables and factorised
  form of linear power corrections}
\label{sec:oncollinear}
We now study in more detail the
role of recoil effects on shape variables. As shown in
ref.~\cite{Caola:2021kzt} and reviewed in sec.~\ref{sec:gen}, to compute
linear power correction to an observable $\V$ we need to consider the difference
$\V(\{p\},\lone,\ltwo)-\V(\{\pt\})$, that we now examine including all recoil effects. We 
find
\begin{align}
  \V(\{p\},\lone,\ltwo)-\V(\{\pt\}) &= \V(\{\pt\},\lone,\ltwo) -
  \V(\{\pt\})+\frac{\partial \V(\{\pt\})}{\partial \pt_i^\mu}
  \Tmap^\mu_{i,\nu}(\{\pt\})\, k^\nu \nonumber \\ &+\left\{
  \frac{\partial \V(\{\pt\},\lone,\ltwo)}{\partial \pt_i^\mu} -
  \frac{\partial \V(\{\pt\})}{\partial \pt_i^\mu} \right\}
  \Tmap^\mu_{i,\nu}(\{\pt\})\, k^\nu+{\cal
    O}(k_0^2)\,, \label{eq:SimplifiedSfactor1}
\end{align}
where we assume summation over the repeated indices $\mu$, $\nu$ and
$i$.  The term in the curly bracket on the right-hand side of
eq.~(\ref{eq:SimplifiedSfactor1}) is suppressed in the soft limit,
and, since it is multiplied by $k^\nu$, can be dropped. Many
interesting observables are linear with respect to soft
emissions.\footnote{The definition of linearity implies
  $V(\{p\},l,\bar l) = V(\{p\}) + V_2^\mu(\{p\},l)\, l_\mu +
  V_3^\mu(\{p\},\bar l)\,\bar l_\mu$.  } In this case, we have
\begin{align}\label{eq:splittingdeltaS}
  \V(\{\pt\},\lone,\ltwo) - \V(\{\pt\}) &= \V(\{\pt\},\lone,\ltwo) -
  \V(\{\pt\},\lone) + \V(\{\pt\},\lone) - \V(\{\pt\}) \nonumber
  \\ &\approx \V(\{\pt\},\ltwo) - \V(\{\pt\}) + \V(\{\pt\},\lone) -
  \V(\{\pt\}),
\end{align}
leading to
\begin{align}
  \V(\{p\},\lone,\ltwo)-\V(\{\pt\})&\approx \V(\{\pt\},\lone) -
  \V(\{\pt\}) + \frac{\partial \V(\{\pt\})}{\partial \pt_i^\mu}
  \Tmap^\mu_{i,\nu}(\{\pt\})\, \lone^\nu \, \nonumber \\ &+
  \V(\{\pt\},\ltwo) - \V(\{\pt\}) + \frac{\partial \V(\{\pt\})}{\partial
    \pt_i^\mu} \Tmap^\mu_{i,\nu}(\{\pt\})\, \ltwo^\nu
  \,, \label{eq:SimplifiedSfactor2}
\end{align}
were we have neglected terms of higher order in $l$.\footnote{For ease of notation, we omit to
  indicate these higher order terms in the following.}
If the observable $\V$ and the mapping procedure
satisfy the linearity condition eq.~(\ref{eq:SimplifiedSfactor2}),
the effect of the emission of two soft massless partons (arising
from the decay of a virtual gluon) splits into the sum of two equivalent
contributions, one for each massless parton, where in each one of them
the recoil effect is computed as if only one parton was emitted.

In what follows, we assume that our observable is linear in the soft
limit -- in the sense of eq.~\eqref{eq:SimplifiedSfactor2} --  and thus focus
on modifications to the shape variable due to the emission
of a single massless parton.  Calling $\exlone$ the momentum of the
emitted soft parton, and $\ourphi$, $\eta$, $\exlone_\perp$ its
azimuthal angle, rapidity and transverse momentum in the emitting dipole rest
frame, we expect that the shape variable is  described
by the following equation at small $l_t= |\exlone_\perp|$ 
\begin{equation}\label{eq:genericetaphidep}
  \V(\{p\},\exlone) - \V(\{\pt\}) = \frac{\exlone_t}{q}\,
  h_\V(\eta,\ourphi).
\end{equation}
This expression obviously vanishes
in the soft limit. In order for it to vanish also in the collinear
limit, the function $h_\V$ should be bounded for large $\eta$ and
arbitrary $\ourphi$. In fact, since the rapidity is limited by a
logarithm of $Q/l_t$, an exponential behaviour in $\eta$ may 
ultimately lead to powers of $Q/\exlone_t$ (where $Q$ is the hard
scale in the process) thus canceling the $\exlone_t$ suppression.
For our purposes, however, this is not sufficient.
We work  under the assumption that no linear power
corrections can arise from hard collinear divergences (see
ref.~\cite{Caola:2021kzt} for more details). Thus, $h_\V$ must also be
suppressed for large $\eta$.\footnote{Not all shape variables satisfy
this requirement, jet-broadening being one such case.}  In the case of
emission from a $q\bar{q}$ dipole, the collinear divergence does not
depend upon the azimuth of the emitted parton, so that we may rely
upon the weaker condition that $\int \mathd \ourphi h_\V(\eta,\ourphi)$
vanishes for large $\eta$. When generalising our result to the
emission from $q g$ dipoles, however, we may have to worry about
azimuthal-dependent collinear divergences, caused by terms
proportional to $\exlone_\perp^\mu \exlone_\perp^\nu$ in the splitting
function. These terms are even in $\ourphi$, i.e. they are invariant
under the transformation $\ourphi\to \pi-\ourphi$.  We thus make the
slightly stronger requirement that $h_\V(\eta,\ourphi)+h_\V(\eta,\pi-\ourphi)$
must be integrable for $-\infty<\eta<\infty$.

The conditions highlighted above are sufficient to generalise the
factorised approach of the previous section to a wider class of
observables. More specifically, if the observable and mapping are such
that for soft emissions the linearity condition
eq.~\eqref{eq:SimplifiedSfactor2} is satisfied; and the observable
modification due single emission can be cast in the form
eq.~\eqref{eq:genericetaphidep}, and
$h_\V(\eta,\ourphi)+h_\V(\eta,\pi-\ourphi)$ is integrable for
$-\infty<\eta<\infty$; then an expression for linear power corrections
similar to eq.~\eqref{eq:eq45} can be obtained along the lines
described in the previous section for the
$C$-parameter.\footnote{Examples of such observables are the thrust,
  the heavy jet mass, the jet mass difference, the sum of the jet
  masses, and the wide jet broadening. The narrow jet and total
  broadening do not satisfy the condition of integrability in $\eta$,
  since they are proportional to the absolute value of the transverse
  momentum relative to the thrust axis, with no rapidity
  suppression. Clustering algorithms that have a tendency to cluster
  the soft particles together, like the Durham algorithm, may violate
  the linearity condition in multiple soft emissions, and thus may
  lead to a modification of the Milan factor~\cite{Dasgupta:2009tm}.}  
We write
\begin{equation}
  I_\V^{\rm unreg} = W_{\V}\times\lambda F,
  \label{eq:fact}
\end{equation}
where 
\begin{equation}
  W_\V =
  \int\frac{\mathd\eta\mathd\ourphi}{2(2\pi)^3}\frac{h_\V(\eta,\ourphi)}{q},
  \label{eq:Wv}
\end{equation}
and  $h_\V$ is  defined  in eq.~\eqref{eq:genericetaphidep}.
As we have shown in
sec.~\ref{sec:F}, the singularities in $F$
can be dealt with an analytic
regularisation as in sec.~\ref{sec:regF}, or by subtraction as in sec.~\ref{sec:AltF}.
Under the assumptions listed above, the $W_\V$ integral
of eq.~(\ref{eq:Wv}) is convergent and no regularisation is needed. Using
eq.~\eqref{eq:mastereqsimp} we arrive at our final formula for the
linear power corrections to the cumulant
\begin{equation}
  \begin{split}
  \mathcal T_\lambda\left[\Sigma(v;\lambda)\right]&=
  \int\mathd\sigma^{\rm b}(\tilde\Phi_{\rm b})
  \delta\big(\V(\tilde\Phi_{\rm b})-\v\big)
  \; \frac{\lambda}{q} \; \left[\mathcal N \; F \; (qW_\V)
    \right]
  \\
  &=
  \int\mathd\sigma^{\rm b}(\tilde\Phi_{\rm b})
  \delta\big(\V(\tilde\Phi_{\rm b})-\v\big)
  \;  \frac{\lambda}{q} \;
  \left[
    -\frac{15}{64}\as\pi\Cf
    \int\mathd\eta\frac{\mathd\ourphi}{2\pi}h_\V(\eta,\ourphi)
    \right].
  \end{split}
  \label{eq:finres}
\end{equation}

  We now observe that rather than using the full expression for the shape variable, including recoil
  effects, we can simply compute
\begin{equation}
  \V(\{\pt\},\exlone)-\V(\{\pt\}) \equiv \frac{\exlone_t}{q}\, \hat{h}(\eta,\ourphi).
\end{equation}
We now know that $\exlone_t\hat{h}$ must differ from $\exlone_th$ by terms linear in the components of $l$, i.e.
we must have
\begin{equation}
  \hat{h}(\eta,\ourphi)-h(\eta,\ourphi)=A e^\eta + B e^{-\eta} +C \cos\ourphi +D \sin \ourphi,
\end{equation}
where $A$ and $B$ must be the same for all acceptable mappings, and $C$ and $D$ at the end do not contribute.
Thus, we can complete the calculation of the shape variable contribution by replacing
\begin{equation}\label{eq:hregulated}
  \hat{h}(\eta,\ourphi) \to \frac{1}{2}\left[\hat{h}(\eta,\ourphi)+ \hat{h}(\eta,\pi-\ourphi)\right]- A e^\eta -B e^{-\eta},
\end{equation}
with $A$ and $B$ chosen to cancel the $\eta \to +\infty$ and $\eta \to - \infty$ behaviour
of the first term.
Alternatively, if we use the analytic regulator introduced in eq.~(\ref{eq:eq44}) these subtraction terms give a vanishing
contribution, thus yielding a further justification to the procedure introduced there.

As a last observation, we remark that in ref.~\cite{Caola:2021kzt} it
was pointed out that underlying Born mappings that are not linear in
$k$, but that are such that the non-linear term has the form $k_\perp
g(\eta)$ (e.g. the Catani-Seymour dipole scheme~\cite{Catani:1996vz}), share
the same feature of linear schemes as far as the absence of linear
power corrections is concerned. These schemes, however, do not satisfy
the property of eq.~(\ref{eq:splittingdeltaS}), and therefore are not
suitable for the analytic arguments that we presented here. We stress
that this does not imply that such schemes are pathological in any
sense, and indeed we have used them for the numerical checks of
sec.~\ref{sec:validation}.

\subsubsection{The example of the $C$-parameter} 
\label{sec:Cexample}
We now illustrate  the construction of  the previous section
by  considering the case of the $C$-parameter.
Its variation caused  by  the emission of one soft massless parton is given
by
\begin{equation}
  \delta C=-3\sum_{i>j=1,3}\frac{(p_i p_j)^2}{(p_i q)(p_j q)}
  +3\sum_{i>j=1,3}\frac{(\pt_i \pt_j)^2}{(\pt_i q)(\pt_j q)}
  - 3\sum_{j=1,3}\frac{(\pt_j \exlone)^2}{(\pt_j q)(\exlone q)} + {\cal O}(l_0^2)\,,
  \label{eq:deltaC}
\end{equation}
where we introduced the notation
$\delta C = C(\{p\},l)-C(\{\pt\})$. 
We adopt a dipole-local mapping defined for small $l$ as follows
\begin{align}
  p_1 &= \pt_1-\frac{(\pt_2 \exlone)}{(\pt_1 \pt_2)}
  \pt_1-\frac{1}{2} \exlone_\perp, \nonumber\\
  p_2 &=
  \pt_2-\frac{(\pt_1 \exlone)}{(\pt_1 \pt_2)} \pt_2-\frac{1}{2}
  \exlone_\perp, \label{eq:dipolemap}\\
  \nonumber p_3 &= \pt_3.  
\end{align}

We note that the mapping~(\ref{eq:dipolemap}), besides being accurate
up to terms of order $l_0^2$, is also accurate in the hard collinear
region up to terms of order $l_t^2$. In fact, it fully satisfies
momentum conservation, and it preserves the on-shell properties of
$p_1$ and $p_2$ up to terms of order $l_t^2$.  Defining
\begin{align}
  x_i & =\frac{2(\pt_i q)}{q^2},\nonumber \\ r_3 & =\frac{ |\pt_{3,\perp}|
  }{\sqrt{q^2}}=\sqrt{\frac{(1-x_1)(1-x_2)}{1-x_3}},\nonumber \\ l & =
  l_t\frac{\pt_1}{\sqrt{s}}\alpha+l_t\frac{\pt_2}{\sqrt{s}}\beta+l_\perp
  \,,\nonumber
\end{align}
with $\alpha=\exp(\eta)$ and $\beta=\exp(-\eta)$ and $s=2(\pt_1\pt_2)$,
a straightforward but tedious computation leads to the following result:
\begin{eqnarray}
  \delta C &=& \frac{l_t}{q} h_C(\eta,\ourphi),\nonumber
  \\ h_C(\eta,\ourphi) &=&\frac{6}{x_3\,x_1^2\,x_2^2}\Bigg\{ - 2r_3^2
  \frac{ (1 - x_3)^{\frac{5}{2}}(x_1 + x_2 - x_1 x_2) \cos^2
    \ourphi}{\beta x_2 + \alpha x_1 + 2 \cos \ourphi\, r_3 \sqrt{1 -
      x_3}} \nonumber
  \\ &&\phantom{\frac{6}{x_3\,x_1^2\,x_2^2}\Bigg\{}+ r_3 \frac{(x_1 -
    x_2) (1 - x_3)^2(\beta x_2-\alpha x_1) \cos \ourphi}{\beta x_2 +
    \alpha x_1 + 2 \cos \ourphi \,r_3 \sqrt{1 - x_3}} \nonumber \\ &&
  \phantom{\frac{6}{x_3\,x_1^2\,x_2^2}\Bigg\{}+ \frac{ x_1x_2(1 -
    x_3)^{\frac{3}{2}}(x_1 + x_2 - 2 x_1 x_2)}{\beta x_2 + \alpha x_1
    + 2 \cos \ourphi \,r_3 \sqrt{1 - x_3}} \;\Bigg\}\,.
  \label{eq:fullC}
\end{eqnarray}
This has the form of eq.~(\ref{eq:genericetaphidep}).  We can see that
the first and third term in the curly bracket vanish exponentially for
large $|\eta|$. This is not the case of the central term, that for
large $|\eta|$ is proportional to $\cos \ourphi$. However
$h_C(\eta,\ourphi)+h_C(\eta,\pi-\ourphi)$ is suppressed for large $\eta$,
so our convergence requirements are satisfied, and formula~(\ref{eq:fullC})
  can be inserted in eq.~(\ref{eq:Wv}), yielding a finite integral that
  can be performed numerically.

It is interesting to compute separately the recoil
contribution to $h_C$, i.e.  the contribution coming from the first
two terms of eq.~\eqref{eq:deltaC}. It reads
\begin{equation}
  \begin{split}
  h^{\rm rec}_C(\eta,\ourphi) = \frac{3}{x_1^2 x_2^2 x_3}\bigg\{ 2 r_3 (1
  - x_1) (1 - x_2) \big[3 x_1 x_2 - x_1 (1 - x_1) - x_2 (1 -
    x_2)\big]\cos\ourphi \\ + \frac{x_1 x_2}{\sqrt{1-x_3}}\big[\beta
    x_1 (1 - x_1)^2 + \alpha x_2 (1 - x_2)^2 + x_3 (1 - x_3)^2 (\alpha
    + \beta)\big]\bigg\}.
  \end{split}
  \label{eq:dcrec}
\end{equation}
While this contribution does not lead to linear power
corrections~\cite{Caola:2021kzt}, it is not suppressed at large
rapidities. Hence, if we neglect recoil when computing the
shape-variable modification, we are 
left with the quantity $h^{\rm no rec}_C=h_C-h^{\rm rec}_C$ that is
ill-behaved at large rapidities. On the other hand,  the terms
proportional to $\cos\ourphi$ drop out of $h^{\rm no rec}_C$ after azimuthal integration, and we can get rid of the terms growing with $\eta$ using eq.~(\ref{eq:hregulated}), with the coefficients $A$ and $B$
tuned to exactly cancel the large $\eta$ behaviour.
Alternatively, as shown in sec.~\ref{sec:3jetcaseW}, we can
 regulate the rapidity
integral with the analytic regulator 
\begin{equation}
  \mathd\eta \to \mathd\eta \,e^{-\epsilon|\eta-\eta_q|},
  \label{eq:anreg}
\end{equation}
where $\eta_q$ is the rapidity of $q$ in the
emitting-dipole rest frame. It is straightforward to see that with
this regulator the contribution of the recoil eq.~\eqref{eq:dcrec} vanishes.

\subsection{Relation to the Milan-factor approach}
\label{sec:mil}
In this section, we investigate the connections between our approach and
results already available in the literature, namely the Milan-factor
formalism of refs.~\cite{Dokshitzer:1997iz,Dokshitzer:1998pt}.

Historically, the calculation of linear corrections to shape variables
was first formulated under the assumption that it was sufficient  to
compute corrections due to the emission of a massive gluon, neglecting
the effect of gluon splitting into a $q\bar q$
pair~\cite{Webber:1994cp,
  Dokshitzer:1995zt,Akhoury:1995sp,Manohar:1994kq}. This was very
early proved to be incorrect~\cite{Nason:1995np}, and in
ref.~\cite{Dokshitzer:1997iz} it was shown how to include the effects
of gluon splitting. The result was expressed in terms of a correction factor,
dubbed \emph{Milan factor}, to be applied to previous calculations
where the splitting was ignored. The universality of this procedure
was investigated in ref.~\cite{Dokshitzer:1998pt}, and an analytic form for
the Milan factor was extracted from explicit calculations for the
$C$-parameter in the two-jet limit in
refs.~\cite{Dasgupta:1999mb,Smye:2001gq}. Until recently, the
Milan-factor approach has only been used to compute non-perturbative
corrections near the two-jet limit. In ref.~\cite{Luisoni:2020efy}
it was also applied to the study of the $C$-parameter near the
symmetric three-jet configuration $c=3/4$. 

To make a connection between the Milan-factor approach and our
formalism, we now briefly sketch the main features of the former.
According to refs.~\cite{Dokshitzer:1997iz,Dokshitzer:1998pt}, one can
obtain linear power corrections to the cumulant of an observable $\V$
by first computing its  modification induced by
the emission of a soft massless gluon, dubbed
``gluer'',
\begin{equation}\label{eq:gluereffect}
  \delta \Sigma(v)= - \int\mathd\sigma^{\rm b}
  \delta\big(\V(\Phi_{\rm b})-\v\big)
  \int
  \frac{\mathd l_t}{l_t} \left[4 \frac{\as(l_t) C_F}{2\pi} \right]
  \frac{l_t}{q} \int \mathd \eta\, \frac{\mathd \ourphi}{2\pi}
  h_\V(\eta,\ourphi)\,.
\end{equation}
In eq.~\eqref{eq:gluereffect},
$l_t$ is the transverse momentum of the gluer in the
emitting-dipole rest frame.\footnote{If there is more than one emitting dipole,
contributions from each dipole are summed.}
To account for gluon splitting, one then ``corrects''
eq.~\eqref{eq:gluereffect} by multiplying it by the Milan factor $\mathcal M$,
which in
the large-$\nf$ limit reads~\cite{Dasgupta:1999mb,Smye:2001gq}
\begin{equation}
\mathcal M = \frac{15\pi^2}{128}.
\end{equation}
In refs.~\cite{Dokshitzer:1997iz,Dokshitzer:1998pt} it was argued that
this is sufficient to obtain linear power corrections in regions where
recoil effects are strongly suppressed, i.e. where different recoil
prescriptions lead to modifications of the observable which are at
least quadratic in the gluer momentum.  This requirement
implies that the Milan-factor approach cannot be naively applied to
generic three-jet configurations, since in the non-degenerate three-jet region
recoil will typically induce a \emph{linear} dependence on the gluer momentum.

We now compare the Milan-factor formalism to our approach for
obtaining linear power corrections. For observables 
satisfying the requirements listed in sec.~\ref{sec:oncollinear}, it 
is summarised by eq.~\eqref{eq:finres}, which can be written as
\begin{equation}
  \mathcal T_\lambda\left[\Sigma(v;\lambda)\right] =
  -\int\mathd\sigma^{\rm b}\delta\big(\V(\Phi_{\rm b})-\v\big)
  \left[\mathcal M
    \times
    4\frac{\as\Cf}{2\pi}
    \right] \frac{\lambda}{q}
  \int\mathd\eta\frac{\mathd\ourphi}{2\pi}h_\V(\eta,\ourphi).
  \label{eq:masterform1}
\end{equation}
We note that, in  the context of large-$n_f$ formalism, we obtain the final   result
from eq.~(\ref{eq:masterform1}) by replacing $\lambda$ with the integral
of an effective coupling (cf. eq.~(C.1) in
ref.~\cite{Caola:2021kzt}, and eq.~(3.83) in ref.~\cite{Beneke:1998ui}).
In the Milan-factor approach, the non-perturbative correction is also
modeled as the integral over an effective coupling. We then
immediately see that in the two-jet and symmetric three-jet limit our result 
is formally equivalent to the Milan-factor one.\footnote{In this
spirit, the calculation of sec.~\ref{sec:F} can be viewed as an
analytical derivation of the Milan factor from first principles  which, 
at variance with  the original
computations~\cite{Dasgupta:1999mb,Smye:2001gq}, does not depend on any
specific observable or kinematic configuration.} However,
our approach is valid for generic three-jet configurations, at least in
the large-$\nf$ limit and for observables of the
type discussed in sec.~\ref{sec:oncollinear}.  It follows  that our
large-$\nf$ derivation both confirms and generalises the Milan-factor
formalism.

However, we would like to stress  that the formulation of our result is
quite different from the one in
refs.~\cite{Dokshitzer:1997iz,Dokshitzer:1998pt}.
First, we note that the standard presentation of the Milan-factor
approach as a correction to a naive soft gluon result is perhaps
justified from the historical perspective, but did  cause
misunderstanding in the past.  In fact, although it is quite clear
from ref.~\cite{Dokshitzer:1998pt} that the Milan factor is to be
applied to the computation of the non-perturbative effect due to the
emission of a massless gluon, it is sometimes understood as the
correction to be applied to calculations performed using a massive
gluon.  From our calculation, it is apparent that there are no
contributions to the power corrections that arise from the emission of
a massive gluon.\footnote{In ref.~\cite{Caola:2021kzt} we have
presented a calculation of the linear correction to the $C$-parameter
induced by the emission of a gluon with mass $\lambda$. However, that
calculation had only illustrative purposes, and physical results
including the $g\to q\bar q$ splitting were instead computed numerically.}
Furthermore, the effect of a massive gluon emission also depends upon
ambiguities that arise when the definition of the shape variable is extended to the case of
massive partons, see footnote~\ref{foot:amb}.  What determines
the power correction is the behaviour of the shape variable under the
emission of a soft massless parton. The reason why this is the case
has to do with the fact that for shape variables of the kind discussed
in sec.~\ref{sec:oncollinear} the variation of the observable under
the emission of two massless partons splits into the sum of two
independent contributions, one for each emission, that can be computed
in terms of the differential distribution of just one of the two
partons arising from gluon splitting. This distribution turns out to
be invariant (in the radiating dipole rest frame) under boosts along
the dipole direction, and even under the azimuthal angle of the
emitted parton. This is sufficient  to guarantee that the result has the
same form \emph{as if} the parton was just a soft (massless!) gluon
emitted by the radiating dipole. However, in our formalism there is no
compelling reason to express this result as a correction factor to the
effect computed with the emission of a massless gluon.

\section{Applications of the factorised approach}
\label{sec:app}
In this section, we will apply the factorised formalism developed in
sec.~\ref{sec:general} to the computation of linear
power corrections to observables other than the $C$-parameter in the
three-jet region. Specifically, in sec.~\ref{sec:thrust} we will study the
thrust distribution in the three-jet region, which is of great
phenomenological interest.

In sec.~\ref{sec:njet} we will instead show that
our results for the power corrections to the $C$-parameter and thrust can be generalised to
final states with $N > 3$ jets with relative ease. 
This  case is less relevant 
phenomenologically  than the three-jet one, because the
requirement of having at least four unresolved jets restricts the
available phase space for the $C$-parameter and thrust to the regions
$c>3/4$, $1-t < 1/3$, respectively.\footnote{We note however that
four- and five-jet  regions are  also used for $\alpha_s$ determinations, see
e.g. refs.~\cite{Kluth:2006bw,OPAL:2005vad,Schieck:2006tc}.} Nevertheless, we present the corresponding 
calculation here since we believe that the study of the $N$-jet case illustrates the
flexibility and robustness of our approach.

\subsection{Power corrections to thrust in the three-jet region}
\label{sec:thrust}
We consider linear power corrections to the thrust distribution. 
If all final-state particles are massless, the thrust shape variable
$T$ is defined as 
\begin{equation}
  T = \max_{\vec{n}}\sum_{i}\frac{|\vec n \cdot \vec p_i|}{q},
  \label{eq:Tdef}
\end{equation}
where $\vec{n}$ is a unit vector, $i$ runs over all final-state
particles and $p_i$ is the momentum of the $i$-th particle.  For ease
of notation, we use ${\vec n}_{\rm m}$ to denote the vector which maximises
eq.~(\ref{eq:Tdef}). From the definition eq.~\eqref{eq:Tdef} it follows
that in the two-jet limit $T=1$. Since we are not interested in the two-jet
region we find it convenient
to consider the distribution in $\overline T = 1-T$. We then
study the cumulant 
\begin{equation}
  \Sigma\left(\ot;\lambda\right) = \sum_F \int {\rm d} \sigma_F \; \theta \left(\overline T(\Phi_F)-\ot\right),
\end{equation}
see eq.~\eqref{eq:CumulantDefinition}.

It is easy to see that thrust satisfies all the requirements presented in 
sec.~\ref{sec:oncollinear}. Therefore, linear power corrections to the cumulant
distribution can be written as (cf. eq.~(\ref{eq:finres}))
\begin{equation}
  \mathcal T_\lambda\left[\Sigma(\ot;\lambda)\right]=
  \int\mathd\sigma^{\rm b}\,\delta\big(\,\overline T(\tilde\Phi_{\rm b})-\ot\,\big)
  \; \frac{\lambda}{q}\;
  \left[
    \frac{15}{8}\as\Cf\pi^3 q W_T\right],
  \label{eq:Tt}
\end{equation}
where
\begin{equation}
  W_T = \frac{1}{q}\int\frac{\mathd\eta\mathd\ourphi}{2(2\pi)^3}
  \left|\nmv\cdot\vec\ltone\,\right|.
  \label{eq:Wt}
\end{equation}
 
As in
sec.~\ref{sec:oncollinear}, $l$ denotes the momentum of a massless
soft parton, and $\eta$, $\ourphi$, $\lone_t$ are its rapidity,
azimuthal angle  and absolute value of transverse momentum in the emitting
dipole rest frame. Also, $\ltone=\lone/\lone_t$. The different overall
sign in eq.~\eqref{eq:Tt} with respect to eq.~\eqref{eq:finres} is
because we consider the distribution in $\overline T=1-T$.
In eq.~\eqref{eq:Wt}, the analytic regularisation of eq.~\eqref{eq:anreg}
is implicitly assumed. 

To compute eq.~\eqref{eq:Wt}, we find it convenient to introduce
a Sudakov parametrisation for the four-vector $\nm = (0,\nmv)$.
We write 
\begin{equation}
  \nm = \alpha p_1 + \beta p_2 + \nmperp,
\end{equation}
where, as usual, $(p_{1,2} \nmperp) = 0$.
Since
$\nm$ is a unit space-like vector in the rest frame of $q$, $\nm^2 = -1$
in an arbitrary frame. This implies
\begin{equation}
  \nmt = |\nmperp|=  \sqrt{1+\frac{(2p_1t)(2p_2t)}{(2p_1p_2)}}.
  \label{eq:nmtdef}
\end{equation}
In terms of Sudakov variables, the scalar product
between $\nm$ and $\ltone$ reads
\begin{equation}
  -2\left(\nmv\cdot \vec\ltone\,\,\right) = 2(\nm\ltone) =
  \sqrt{s}e^{-\eta}\left(\alpha+\beta e^{2\eta}
  - \frac{2\nmt}{\sqrt{s}}e^\eta\cos\philn\right),
  \label{eq:scalprod0}
\end{equation}
where $s=2(p_1p_2)$ and $\philn = \ourphi-\ourphi_{\nm}$. Introducing
the notation $\omega = e^\eta$, the scalar product can be written as
\begin{equation}
  2(\nm\ltone) = \frac{\sqrt{s}}{\omega} P(\omega),
  ~~~{\rm with}~~~P(\omega) = \beta(\omega-\omega_+)(\omega-\omega_-),
  \label{eq:scalprod1}
\end{equation}
and
\begin{equation}
  \omega_\pm = \frac{\nmt \cos\philn
    \pm\sqrt{1-\nmt^2\sin^2\philn}}{\sqrt{s}\beta}.
  \label{eq:roots}
\end{equation}
In deriving this expression, we have used $\nm^2 = \alpha\beta s - \nmt^2=-1$.

To proceed further, we note that the analytically-regulated measure, introduced in eq.~\eqref{eq:anreg}, can be written in terms of $\omega$ as
\begin{equation}
  \mathd\eta\,e^{-\epsilon|\eta-\eta_q|} = \frac{\mathd \omega_{\rm
      reg}}{\omega}, ~~~~\mathd\omega_{\rm reg} = \mathd\omega
  \left[\left(\frac{\omega}{\omega_q}\right)^{-\epsilon}
    \theta(\omega-\omega_q)+
    \left(\frac{\omega}{\omega_q}\right)^{\epsilon}
    \theta(\omega_q-\omega)\right],
  \label{eq:domegaq}
\end{equation}
where $\omega_q=e^{\eta_q}$. Using eqs.~(\ref{eq:scalprod0},
\ref{eq:scalprod1}, \ref{eq:domegaq}), we obtain
the following result for $W_T$ as defined in eq.~\eqref{eq:Wt}
\begin{equation}
  W_T = \frac{\sqrt{s}}{32\pi^3q}\int
  \mathd \omega_{\rm reg\,}\mathd\ourphi \,\omega^{-2}|P(\omega)|.
  \label{eq:Wt2}
\end{equation}
From eq.~\eqref{eq:roots} it follows that to explicitly write
  $|P(\omega)|$ we need to consider two different cases: case A
with $\nmt<1$ and case B with $\nmt>1$. In case A, the roots
of $P(\omega)$ are real-valued for all values of $\philn$. Also, in
this case
\begin{equation}
  \sqrt{1-\nmt^2\sin^2\philn} > |\nmt\cos\philn|,
\end{equation}
so that $\omega_+>0$ and $\omega_-<0$ for all values of $\philn$. In case
B on the other hand the two roots are real-valued only for
$|\sin\philn|<1/\nmt$. Also,
\begin{equation}
  \sqrt{1-\nmt^2\sin^2\philn} < |\nmt\cos\philn|,
  \label{eq:rootB}
\end{equation}
provided that a real-valued root exists. Eq.~\eqref{eq:rootB} implies that
if $\cos\philn$ is negative, both roots $\omega_\pm$ are negative and
if $\cos\philn$ is positive, both roots are positive. 

We now write eq.~\eqref{eq:Wt2} as
\begin{equation}
W_T =
  \frac{\sqrt{s}}{32\pi^3q}\int
  \mathd\ourphi \,X_T,
~~~{\rm with}~~~
  X_T = \int\limits_0^\infty\mathd\omega_{\rm reg}\,\omega^{-2}|P(\omega)|,
  \label{eq:WfromX}
\end{equation}
and study $X_T$ in cases A and B.

For case A, we write
\begin{equation}
  X^{A}_T = \int\limits_{\omega_+}^\infty \mathd\omega_{\rm
    reg}\,\omega^{-2}P(\omega) -
  \int\limits_0^{\omega_+} \mathd\omega_{\rm
    reg}\,\omega^{-2}P(\omega),
\end{equation}
or, equivalently, 
\begin{equation}
  \begin{split}
  X^{A}_T &= 2\int\limits_{\omega_+}^\infty \mathd\omega_{\rm
    reg}\,\omega^{-2}P(\omega)-
  \int\limits_0^{\infty} \mathd\omega_{\rm
    reg}\,\omega^{-2}P(\omega)
  \\
  &=
  -2\int\limits_0^{\omega_+} \mathd\omega_{\rm
    reg}\,\omega^{-2}P(\omega)+
  \int\limits_0^{\infty} \mathd\omega_{\rm reg}\,\omega^{-2}P(\omega).
  \end{split}
  \label{eq:XtA}
\end{equation}
For case B we find
\begin{align}
  &X^{B}_T = \int\limits_0^\infty \mathd\omega_{\rm
    reg}\,\omega^{-2}P(\omega), &{~}\cos\philn<0,
  \nonumber
  \\
  &X^{B}_T = -2\int\limits_{\omega_-}^{\omega_+} \mathd\omega_{\rm
    reg}\,\omega^{-2}P(\omega)+
  \int\limits_0^{\infty} \mathd\omega_{\rm
    reg}\,\omega^{-2}P(\omega), &{~}\cos\philn>0.
  \label{eq:XtB}
\end{align}
A straightforward calculation leads to
\begin{equation}
  \int\limits_0^\infty\mathd\omega_{\rm reg}\; \omega^{-2} \; P(\omega) =
  -\frac{2\beta(\omega_++\omega_-)}{\epsilon}+\mathcal O(\epsilon).
\end{equation}
We note that the $\mathcal O(1/\epsilon)$ term in the above expression vanishes
upon azimuthal integration since $\omega_+ + \omega_-\propto\cos\philn$.
Hence, we can drop  all terms from eqs.~(\ref{eq:XtA}, \ref{eq:XtB}) where
the $\omega$ integration is unrestricted. As a consequence, in case B we
only need to consider the situation $\cos\philn>0$. 

We now consider case A. If $\omega_q<\omega_+$, we use the representation
in the first line of eq.~\eqref{eq:XtA}. If $\omega_q>\omega_+$ we use
instead the one in the second line. In either case, integrating over
$\omega$ and expanding in $\epsilon$ we obtain
\begin{equation}
  X_T^{A} = -2\beta
  \left[\omega_+-\omega_-
    \pm\frac{\omega_++\omega_-}{\epsilon}
    -(\omega_++\omega_-)\ln\frac{\omega_+}{\omega_q}
    \right],
\label{eq5.19}
\end{equation}
where the $\pm$ refers to $\omega_q<\omega_+$ and $\omega_q>\omega_+$,
respectively. Neglecting terms that vanish after azimuthal integration,
we can re-write eq.~(\ref{eq5.19}) as
\begin{equation}
  \begin{split}
&  X_T^{A} \to -2\beta
  \left[\omega_+-\omega_- -(\omega_++\omega_-)\ln\omega_+
    \right]\to
  \\
  &
  -\frac{4}{\sqrt{s}}
  \left[
    \sqrt{1-\nmt^2\sin^2\philn}-\nmt\cos\philn
    \ln\left(\nmt\cos\philn+\sqrt{1-\nmt^2\sin^2\philn}\right)
    \right].
  \end{split}
\end{equation}
Inserting this result into eq.~\eqref{eq:WfromX}, changing variables
from $\ourphi$ to $\philn$ and performing a straightforward azimuthal
integration we obtain
\begin{equation}
  W_T|_{\rm case~A} = \frac{\sqrt{s}}{32\pi^3q}
  \int\limits_0^{2\pi}\mathd\philn X_T^A = 
  -\frac{1}{2\pi^3q}\big[2 E(\nmt^2)-K(\nmt^2)\big].
  \label{eq:Wta}
\end{equation}

Case B can be dealt with in a similar way. As we mentioned earlier,
we only
need to consider the situation $\cos\philn>0$, i.e. the second line of
eq.~\eqref{eq:XtB}. After dropping the second term on the right hand side,
we note that the integral is convergent so we can set $\epsilon\to0$.
After performing the $\omega$ integration, we find
\begin{equation}
  \begin{split}
    &X_T^B = -2\beta\left[
    2(\omega_+-\omega_-)-(\omega_+ + \omega_-)\ln\frac{\omega_+}{\omega_-}
    \right] = \\ & -\frac{4}{\sqrt{s}} \left[ 2
    \sqrt{1-\nmt^2\sin^2\philn}-\nmt\cos\philn
    \ln\frac{\nmt\cos\philn+\sqrt{1-\nmt^2\sin^2\philn}}
            {\nmt\cos\philn-\sqrt{1-\nmt^2\sin^2\philn}}
    \right].
  \end{split}
\end{equation}
Inserting $X_T^B$ into eq.~\eqref{eq:WfromX}, changing variable
$\ourphi\to\philn$ and recalling  the reality condition $|\sin\philn|<1/\nmt$,
we obtain
\begin{equation}
  W_T|_{\rm case~B} = \frac{\sqrt{s}}{32\pi^3q}
  \int\limits_{-\ourphi_{\rm max}}^{\ourphi_{\rm max}}\mathd\philn
  X_T^B=
  \frac{\sqrt{s}}{16\pi^3q}
  \int\limits_{0}^{\ourphi_{\rm max}}\mathd\philn
  X_T^B,
\end{equation}
with $\ourphi_{\rm max}=\arcsin(1/\nmt)$.
The azimuthal integration is straightforward, and yields
\begin{equation}
  W_{T}|_{\rm case~B} = -\frac{ \nmt}{\pi^3 q } \left [
    E \left (
    \frac{1}{\nmt^2} \right )
    -
    \frac{2 \nmt^2 -
      1}{2 \nmt^2 } K \left ( \frac{1}{\nmt^2} \right )  \right ].
  \label{eq:Wtb}
\end{equation}

To understand when the results for the A,B cases are applicable, we note
that in a three-jet event the thrust axis in the $q$ rest frame coincides
with the three-momentum of the most energetic particle in the event.
If $\nm$ is aligned with the momentum of either $p_1$ or $p_2$, then
${\rm min}(z_1,z_2,z_3) \ne z_3$, where $z_i$ is the variable introduced
in eq.~\eqref{eq:zdef}, and
\begin{equation}
  \nmt^2 = \frac{z_1 z_2}{z_3} < {\rm max}(z_1,z_2) < 1.
\end{equation}
This implies that if $\nmv\propto \vec p_1$ or $\nmv \propto \vec p_2$,
case A applies. If on the other hand $\nmv\propto \vec p_3$, then
${\rm min}(z_1,z_2,z_3) = z_3$ and
\begin{equation}
  \nmt^2 = \frac{z_1 z_2(1+z_3)^2}{z_3(1-z_3)^2} >
  \frac{z_3^2(1+z_3)^2}{z_3(1-z_3)^2} > 1,
\end{equation}
so case B applies.

We note that in the three-jet case
$\nmt\ne1$. Indeed, $\nmt=1$ only if $\nmv$ is orthogonal to either
$\vec p_1$ or $\vec p_2$, see eq.~\eqref{eq:nmtdef}. However, in the
three-jet case the thrust axis is aligned with the direction of the
most energetic particle, hence momentum conservation implies that none
of the two remaining particles can be orthogonal to it. Summarising our
findings, we can write
\begin{equation}
  \begin{split}
  \label{eq:shiftThtot}
  \mathcal T\left[\Sigma(\ot;\lambda)\right]
  =& 
  \int\mathd\sigma^{\rm b}\,\delta\big(\,\overline T(\Phi_{\rm b})-\ot\,\big)
  \times
  \\
  &\as \Cf \lambda \frac{15 \pi^3}{8}\times
  \begin{cases}
    & W_{T}|_{\rm case ~ A}, \qquad \mbox{ if } \min(\zone,\ztwo,\zthree)\neq \zthree  \\
    & W_{T}|_{\rm case ~ B}, \qquad \mbox{ if } \min(\zone,\ztwo,\zthree)=\zthree.
  \end{cases}
  \end{split}
\end{equation}
In the two-jet limit, we can assume for example $\min(\zone,\ztwo,\zthree)=\zone=0$ and $\ztwo+\zthree=1$. This yields the well-known result
\begin{equation}
  \frac{\mathcal
    T\left[\Sigma(0;\lambda)\right]}{\mathd\sigma/\mathd\overline T|_{\ot=0}} =
  -\frac{5\pi}{8}\left(\frac{\lambda}{q}\right)\as.
  \label{eq:t2j}
\end{equation}
In  the symmetric limit  $\zone=\ztwo=\zthree=1/3$, so that the thrust axis is not uniquely defined. 
In this case,  the result is obtained by averaging over the three possible alignments of the thrust axis, \emph{i.e.}
\begin{equation}
  \begin{split}
    \frac{\mathcal
      T\left[\Sigma(1/3;\lambda)\right]}{\mathd\sigma/\mathd\overline
      T|_{\ot = 1/3}} &= \as \Cf \lambda \frac{15 \pi^3}{8} \times
    \left(\frac{2}{3} W_{T}|_{\rm case~A}+ \frac{1}{3}W_{T}|_{\rm
      case~B} \right) \\ & = \left[
      \frac{5}{6}K\left(\frac{1}{3}\right)
      -\frac{5}{3}E\left(\frac{1}{3}\right)
      +\frac{25}{24\sqrt{3}}K\left(\frac{3}{4}\right)
      -\frac{5}{3\sqrt{3}}E\left(\frac{3}{4}\right) \right]
    \left(\frac{\lambda}{q}\right)\as.
  \end{split}
\end{equation}

\subsection{Power corrections to shape variables for $N$-jet
  final states}
\label{sec:njet}
In this section, we present results for the $C$-parameter
and the thrust in a generic $N$-jet final state. We note that in order
to avoid the contribution of singular configurations with one
or more unresolved partons, an appropriate cut must be imposed upon
the shape variable. For example, for $N=4$ we must require $c>3/4$ for
the $C$-parameter and $\bar t > 1/3$ for thrust.

Similar to the three-jet case, we start with 
eq.~\eqref{eq:IcUn}
\begin{eqnarray}
  I_{\V}(\{\pt\},\lambda) &=&
  \int [\mathd k] \frac{J^\mu
      J^\nu}{\lambda^2} 
  \int
      [\mathd \lone][\mathd\ltwo] (2\pi)^4 \delta^{(4)}(k-\lone-\ltwo)
      \nonumber\\
      &&
      \times{\rm
      Tr}\left[\hat \lone \gamma^\mu \hat \ltwo
      \gamma^\nu\right]\left[
        \V(\{\pt\},\lone,\ltwo)-\V(\{\pt\})\right],
      \label{eq:NI}
\end{eqnarray}
where $V=C$ or $V=1-T$ and the regularisation discussed in
sec.~\ref{sec:fact} is assumed. The radiation of a
soft gluon with momentum $k$ off a final state with $N$ hard partons
with momenta $p_i$, $i=1,...,N$ is described by the current
\begin{equation}
J^{\mu}=\sum_{i=1}^N Q_i \frac{p_i^{\mu}}{(p_i k)},
\end{equation}
where $Q_i$ is the gauge charge of particle $i$.\footnote{For photon
emission in QED,
$Q_i$ is the electric charge of the outgoing
particle/antiparticle $i$. To properly define gauge charges in QCD, we
need to use the colour-space formalism reviewed for example in
ref.~\cite{Catani:1996vz}.}  Charge neutrality of
the final state implies $\sum\limits_{i=1}^N Q_i=0$.
Using this, we can write the product of the two soft
currents $J^\mu J^\nu$ in eq.~\eqref{eq:NI}
as a sum over $N(N-1)/2$ dipoles
\begin{equation} 
J^{\mu}J^{\nu}=-\sum_{i\neq j}^N \frac{Q_i
  Q_j}{2}\left[-\frac{2p_i^{\mu}p_j^{\nu}}{\left(p_i k\right)\left(p_j
    k\right)}+\frac{p_i^{\mu} p_i^{\nu}}{\left(p_i
    k\right)^2}+\frac{p_j^{\mu} p_j^{\nu}}{\left(p_j
    k\right)^2}\right].
\end{equation}
Repeating the steps that led to eq.~\eqref{eq:eq45}, we
 obtain
\begin{equation}
  I_\V = -\lambda\FM \sum_{i\ne j}^{N} \frac{Q_iQ_j}{2} W_\V^{ij},
\end{equation}
where $\FM$ is the universal factor given in eq.~\eqref{eq:eq82} and $W_\V^{ij}$
is a function that describes the contribution to the observable $V$ due to an 
emission by an $(i-j)$ dipole, similar to  the
one in eq.~\eqref{eq:Wv}. 
We stress that to compute $W_\V^{ij}$ for a given dipole, we need to
perform the Sudakov decomposition of (all) particles' momenta with
respect to the four-momenta of $p_i$ and $p_j$. Explicitly, we write
\begin{equation}
  \begin{split}
    &\ltone^\mu = \frac{p_i^\mu}{\sqrt{2(p_ip_j)}} e^\eta
    +\frac{p_j^\mu}{\sqrt{2(p_ip_j)}} e^{-\eta} + n^\mu,
    \\
    & p_k^\mu = p_{k,t}\frac{p_i^\mu}{\sqrt{2(p_ip_j)}} e^{\eta_k}
    +p_{k,t}\frac{p_j^\mu}{\sqrt{2(p_ip_j)}} e^{-\eta_k} + p_{k,\perp}^\mu,\;\;\; k \in N, \; k \ne i,j,
    \\
    & q^\mu = m_{t,q}\frac{p_i^\mu}{\sqrt{2(p_ip_j)}} e^{\eta_q}
    +m_{t,q}\frac{p_j^\mu}{\sqrt{2(p_ip_j)}} e^{-\eta_q} + q_{\perp}^\mu,
  \end{split}
  \label{eq:genSudakov}
\end{equation}
where $(n p_{i,j}) = 0$, $p_{k,t}=|p_{k,\perp}|$,
$m_{t,q} = \sqrt{q^2+q_t^2}$ and $q_t = |q_\perp|$. Linear power corrections
are then written as
\begin{equation}
  \frac{\mathcal
    T_\lambda\left[\Sigma(v;\lambda)\right]}{\mathd\sigma^{\rm
      b}/\mathd V} =
  -\frac{15\pi\as}{64}\frac{\lambda}{q}
  \int\mathd\sigma^{\rm b}\,\delta\big(\,\V(\Phi_{\rm b})-\v\,\big)
  \times\left[
    -\frac{Q_iQ_j}{2}\sum_{i\ne j}^{N} q W_\V^{ij}
    \right],
  \label{eq:finres_N}
\end{equation}
see eq.~\eqref{eq:finres}. Eq.~\eqref{eq:finres_N} holds for any
observable that satisfies the requirements described in
sec.~\ref{sec:oncollinear}.\footnote{We remind the reader that we are
not yet able to deal with the full non-abelian case in a rigorous
way. As a consequence, technically our result is solid only if all emitting
dipoles are $q\bar q$ ones.  We will discuss in sec.~\ref{sec:numerics} a
conjectured generalisation to processes involving $gg$ and $qg$
dipoles.} In the rest of this section, we will present results for $V=C$
and $V=1-T$.

It is straightforward to write $W_C^{ij}$ for the $C$-parameter. It reads
\begin{equation}
  W_C^{ij} = -\frac{3}{8\pi^3}\sum_{k=1}^{N}
  \frac{W_C^{ij,(k)}}{(p_kq)}, ~~~{\rm with~} W_C^{ij,(k)} =
  \int\mathd\eta\mathd\ourphi\frac{(p_k\ltone)^2}{2(\ltone
    q)}e^{-\epsilon|\eta-\eta_q|}.
  \label{eq:5.36}
\end{equation}
We stress that both $\eta$ and $\ourphi$ in this equation implicitly
depend on the choice of $i$ and $j$ through the Sudakov decomposition
eq.~\eqref{eq:genSudakov}.  Although $W_C^{ij,(k)}$ looks similar to
$W_C^{i}$ of eq.~\eqref{eq:eq3.31}, we cannot directly use the results
of sec.~\ref{sec:3jetcaseW}. Indeed, there we
used momentum conservation to simplify the calculation, cf. the
discussion around eq.~\eqref{eq:eq47}. This simplification does not
work anymore in the $N$-jet case and, for this reason, we have to
consider the general case carefully. We describe the calculation in
detail in appendix~\ref{sec:cn}. Here we limit ourselves to
the presentation of the 
final result, which is remarkably compact. Introducing the variables
\begin{equation}
  s_{ij} = \frac{q}{m_{t,q}} = \sqrt{\frac{q^2 (2p_ip_j)}{(2p_iq)(2p_jq)}},
  ~~~~~
  c_{ij} = \frac{q_t}{m_{t,q}} = \sqrt{1-s_{ij}^2},
  \label{eq:5.37}
\end{equation}
and 
\begin{equation}
  n_3^{i(j)} = \frac{q^2}{(2qp_{i(j)})}\sum_{k=1}^{N}
  \frac{\left(p_{i(j)}p_k\right)^2} {(qp_{i(j)})(qp_k)},
  ~~~~~
  n_4^{ij} = \frac{q^2}{2}\sum_{k=1}^{N}\frac{(p_ip_k)(p_jp_k)}
  {(qp_i)(qp_j)(qp_k)},
  \label{eq:5.38}
\end{equation}
we obtain
\begin{equation}
  \begin{split}
  W_C^{ij} = -\frac{3s_{ij}}{4\pi^2 c_{ij}^4 q}\bigg\{
  K(c_{ij}^2)\left[s_{ij}^4-s_{ij}^2\left(3-n_3^{\langle ij\rangle}-3
    n_4^{ij}\right) + \left(n_3^{\langle ij\rangle} - n_4^{ij}\right)
    \right] \\ +2 E(c_{ij}^2) \bigg[ 1+n_3^{\langle ij\rangle}c_{ij}^2
    -2 n^{ij}_4-\frac{n_3^{\langle ij\rangle}-n_4^{ij}}{s_{ij}^2}
    \bigg]\bigg\},
  \end{split}
  \label{eq:Wcij}
\end{equation}
with
\begin{equation}
  n_3^{\langle ij\rangle} = \frac{n_3^i+n_3^j}{2}.
\end{equation}
Given the result in  eq.~\eqref{eq:Wcij}, one may worry that
$W_C^{ij}$ diverges in kinematics configurations where $c_{ij}\to 0$ or
$s_{ij}\to 0$. However, it is simple to check that this is not the
case and eq.~\eqref{eq:Wcij} actually yields a finite result for any kinematic
configuration. It is also easy to verify that in the $N=3$ case 
eq.~\eqref{eq:Wcij} reduces to eq.~\eqref{eq:eq74}.

We continue with the thrust variable.
In this case, the observable-dependent function
$W_T^{ij}$ reads
\begin{equation}
  W_T^{ij} = \frac{1}{q}\int\frac{\mathd\eta\mathd\ourphi}{2(2\pi)^3}
  \left|\nmv\cdot\vec\ltone\,\right|,
  \label{eq:Wtij}
\end{equation}
and the only dependence on $i$,$j$ is through the definition of $\eta$ and
$\ourphi$. We note that eq.~\eqref{eq:Wtij} is formally identical to
eq.~\eqref{eq:Wt}. We can then immediately read-off the $N$-jet result
from eqs.~(\ref{eq:Wta}, \ref{eq:Wtb}):
\begin{equation}
W_T^{ij}=\begin{cases} W_{T}^{ij}|_{\nmt^{ij}<1} & \text{when }\left(2
p_i \nm\right)\left(2 p_j \nm\right)<0, \\
W_{T}^{ij}|_{\nmt^{ij}>1} &
\text{when }\left(2 p_i \nm\right)\left(2 p_j \nm\right)>0,
\end{cases}
\label{eq:WT2casesNjet}
\end{equation}
where
\begin{equation}
  \begin{split}
    &
    W^{ij}_T|_{\nmt^{ij}<1} = -\frac{1}{2\pi^3q}\left[2
      E\left(\nmt^{ij,2}\right)-K\left(\nmt^{ij,2}\right)\right],
    \\
    &
    W^{ij}_T|_{\nmt^{ij}>1} = -\frac{ \nmt^{ij}}{\pi^3 q } \left [
    E \left (
    \frac{1}{\nmt^{ij,2}} \right )
    -
    \frac{2 \nmt^{ij,2} -
      1}{2 \nmt^{ij,2} } K \left ( \frac{1}{\nmt^{ij,2}} \right )  \right ].
  \end{split}
  \label{eq:Wtgen}
\end{equation}
and
\begin{equation}
  \nmt^{ij} = \sqrt{1+\frac{(2p_i\nm)(2p_j\nm)}{(2p_ip_j)}}.
  \label{eq:nmtgen}
\end{equation}
If multiple thrust axes exist in a given kinematic configuration, the
calculation is repeated for each of them and then averaged over their
number to obtain the effective power correction to thrust, in analogy to
what we did in sec.~\ref{sec:thrust} for the symmetric three-jet limit. 

The expressions for $W^{ij}_T$ in eq.~\eqref{eq:Wtgen} diverge 
when $\nmt^{ij}=1$. From eq.~\eqref{eq:nmtgen} it is apparent that this
can only happen if the thrust axis is orthogonal to the direction of
one of the external particles. We now show  that this cannot  occur  in a 
generic $N$-jet case. 
We work in the event
center of mass frame and write
\begin{equation}
  T = \max_{\vec{n}} \sum_{i}\frac{|\vec n \cdot \vec p_i|}{q} =
  \max_{\vec{n}}\sum_{i}\frac{E_i}{q} f_i(\theta,\ourphi) =
  \max_{\vec{n}}\left[T_n(\theta,\phi)\right].
\end{equation}
Here, $E_i$ is the energy of particle $i$ and
\begin{equation}
  f_i(\theta,\phi) = \sqrt{(\cos\theta\cos\theta_i + \cos(\ourphi-\ourphi_i)
    \sin\theta\sin\theta_i)^2},
\end{equation}
where $(\theta_i,\ourphi_i)$ are the polar and azimuthal angles of parton
$i$ and $(\theta,\varphi)$ are angular variables for  the vector $\vec n$.  We now
assume that the vector $\vec n$ is orthogonal to $\vec p_1$.  Without
loss of generality, we choose our axis such that
$\vec p_1 = E_1 \hat z$ and $\vec n = \hat x$, i.e. we set
$\theta_1 = 0$ and $(\theta,\phi) = (\pi/2,0)$. We find 
\begin{equation}
T_n(\pi/2,0) = \sum_{i=2}^N\frac{E_i}{q}
\sin\theta_i\left|\cos\ourphi_i\right|.
\end{equation}
We now investigate what happens if we move away from the $\vec n\perp \vec p_1$
configuration. More precisely, we consider the case
$(\theta,\ourphi) = (\pi/2 -\delta\theta,\ourphi)$, with $|\delta\theta|\ll 1$.
We then write
\begin{equation}
  T_n(\pi/2-\delta\theta,0) - T_n(\pi/2,0) =
  \frac{E_1}{q}|\delta\theta| +
  \delta\theta\sum_{i=2}^{N}\frac{E_i}{q}\cos\theta_i\frac{|\cos\ourphi_i|}
                  {\cos\ourphi_i}
  + \mathcal O(\delta\theta^2).
  \label{eq:difft}
\end{equation}
Keeping in mind that $E_i>0$, it is clear that the first term in eq.~\eqref{eq:difft} is positive and independent of the sign of $\delta\theta$. As for the second term, the coefficient of $\delta\theta$ can take either positive or negative values. We can then make a choice of $\delta\theta$ with appropriate sign such that the second term is positive as well leading to an increase in thrust compared to the orthogonal case.
Hence, we conclude that the $\vec n\perp\vec
p_1$ configuration does not maximise an expression for  $T_n$, which implies that the thrust axis
$\nm$ cannot be orthogonal to $\vec p_1$ and, since parton $i=1$ plays no special role in our analysis,
to any other hard parton as well.

\section{Phenomenological predictions in the three-jet region}
\label{sec:numerics}
We now apply the formalism described in this paper  to obtain predictions for
linear power corrections to the $C$-parameter and thrust cumulative
distribution in the three-jet region. Here, we follow
ref.~\cite{Caola:2021kzt} and conjecture that our results can be
 extended to processes involving gluons at the Born level
by modifying QCD charges of the corresponding dipoles.
This allows us to consider the phenomenologically interesting
$e^+ e^-\to q\bar q g$ process.

The rest of this section is organised as follows. 
In sec.~\ref{sec:validation} we
validate the analytic results obtained in this paper
against  numerical ones obtained with
the formalism of ref.~\cite{Caola:2021kzt}. 
We find that the analytic
results allow us to obtain phenomenological
predictions in a very efficient way, eliminating the practical
shortcomings of the numerical approach.  In
sec.~\ref{sec:cpar_pheno}, we compare our results to the ones of 
ref.~\cite{Luisoni:2020efy}. In particular, we show that our formalism
resolves an ambiguity present there, and sheds light on its origin. 

\subsection{Comparison of analytic and numerical results for $C$-parameter and thrust}
\label{sec:validation}
We  validate the analytic results derived in this paper
by comparing them with 
the numerical ones obtained using the method of
ref.~\cite{Caola:2021kzt}. 

Non-perturbative corrections are often presented as a shift with respect
to the perturbative result~\cite{Dokshitzer:1997ew}. More precisely, 
one writes the full, ``hadronic'' cumulant as~\cite{Dokshitzer:1997ew}
\begin{equation}
\widetilde\Sigma^{\rm had}(\v) = \widetilde\Sigma\big(\v-\delta_{\rm
  NP}(\v)\big) \approx \widetilde\Sigma(\v)-
\frac{1}{\sigma}\frac{\mathd\sigma}{\mathd\V}\delta_{\rm NP}(\v),
\end{equation}
where the perturbative cumulant is defined as
\begin{equation}
  \widetilde \Sigma(\v) = \frac{1}{\sigma}\int_0^c
  \mathd \V \frac{\mathd \sigma}{\mathd\V}.
\end{equation}
Taking into account the  definition of $\Sigma$,
cf. eq.~\eqref{eq:CumulantDefinition}, and the fact that the total
cross section $\sigma$ is free from linear power corrections, we
obtain
\begin{equation}
  \delta_{\rm NP}(\v) = \frac{\mathcal
    T_\lambda\left[\Sigma(\v;\lambda)\right]}{\mathd\sigma/\mathd\V}.
\end{equation}
Since we are mostly interested in
the dependence of non-perturbative corrections
on kinematics,
we parameterise $\delta_{\rm NP}$ as
\begin{equation}
  \delta_{\rm NP}(\v) = h\zeta(\v),~~~~~~ h \equiv\delta_{\rm NP}(0).
  \label{eq:zeta}
\end{equation}
It follows that $\zeta(0)=1$. In what follows, we will
discuss the function $\zeta(\v)$. 

\begin{figure}[t]
  \includegraphics[width=0.48\textwidth]{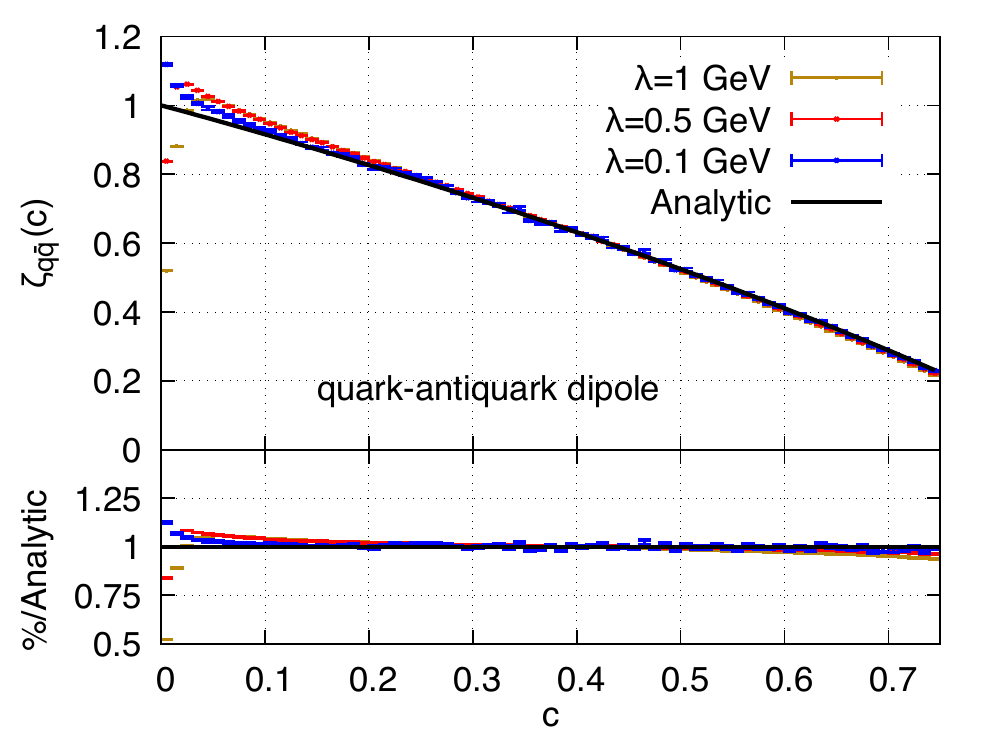}
  \includegraphics[width=0.48\textwidth]{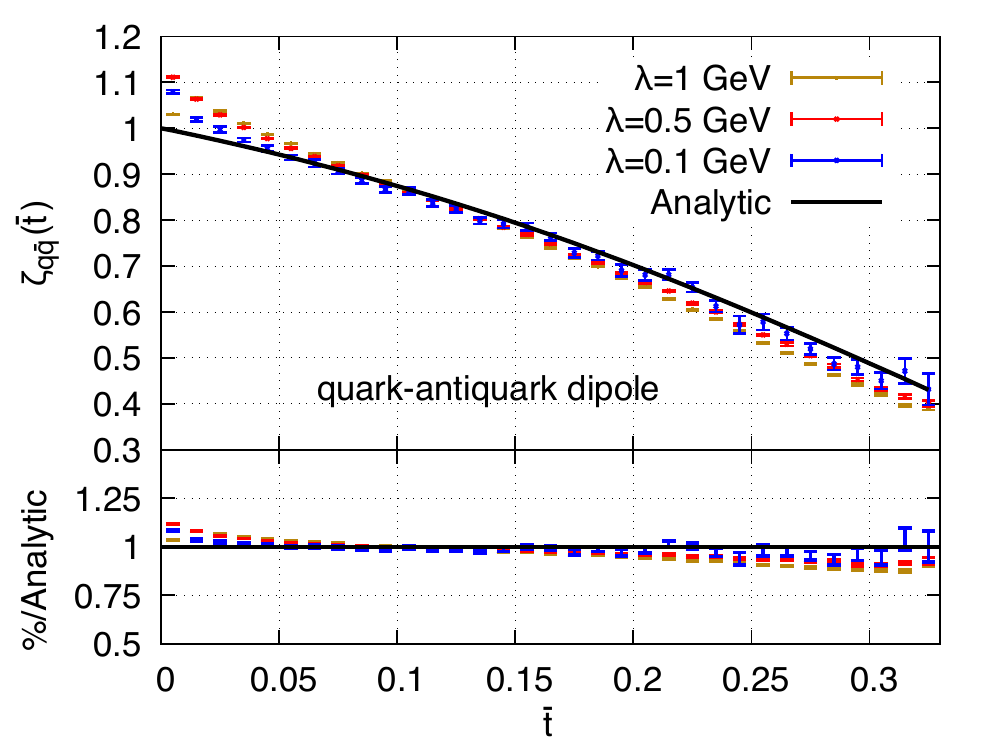}
  \caption{\label{fig:zetaqq} The function $\zeta_{q\bar{q}}(\v)$ for
    the $C$-parameter (left panel) and the thrust $\overline T=1-T$ (right
    panel), evaluated numerically using $q=100$ GeV and
    $\lambda=0.1,~0.5,~1$ GeV, compared to the analytic computation
    (solid line). In the lower panels
    we reported the ratio plot of each numerical curve with respect to the
    analytic one.}
\end{figure}

We start by considering the process $\gamma^*(q)\to q(p_1)+\bar
q(p_2)+\gamma(p_3)$, as in the previous sections. In this case
the only emitting dipole is the $q\bar q$ system. We then write
\begin{equation}
  \delta_{\rm NP}^{q\bar q\gamma} = h \zeta_{q\bar q}(\v),
\end{equation}
with $\zeta_{q\bar q}(0) = 1$ as before. We note that, since for $\v=0$
the $q\bar q\gamma$ configuration approaches a two-parton $q\bar q$
configuration, $h$ must correspond in this case to the non-perturbative
shift computed in previous literature for the two-jet configuration.
Thus, for the $C$-parameter, we have 
$h=-(\lambda/q)\as\times 15\pi^2/16$, while for the thrust
$h=-(\lambda/q)\as\times 5\pi/8$, see eqs.~(\ref{eq:c2j}) and
(\ref{eq:t2j}). 

In  fig.~\ref{fig:zetaqq}   analytic and numerical results~\cite{Caola:2021kzt} are compared
for $\zeta_{q\bar q}$, for both the $C$-parameter and the thrust. 
We take  $q = 100~{\rm GeV}$. 
For the numerical results, we consider three different values of
the gluon mass $\lambda = 1,~0.5,~0.1~{\rm GeV}$. We
observe that the numerical
result converges to the analytic one,
to a good approximation. However, it is apparent from these figures that the agreement
is not perfect and that even smaller values of $\lambda$ should be chosen in numerical computations
to improve the situation. Unfortunately,  it  is quite difficult to do so.
This is because the numerical results contain extra quadratic terms
in $\lambda$ that can also be enhanced by powers of $\ln\lambda$. We also
note 
that the analytic and numerical results tend to depart from each other
at small $\v$, for both the $C$-parameter and the thrust. This is because,
in this region, the effective hard scale is reduced, so that
the expansion parameter for power corrections is no longer $\lambda/q$,
but rather $\lambda$ divided by the reduced scale. In contrast, the analytic
calculation only contains linear power corrections, and is not affected
by these numerical issues.

\begin{figure}[t]
  \includegraphics[width=0.48\textwidth]{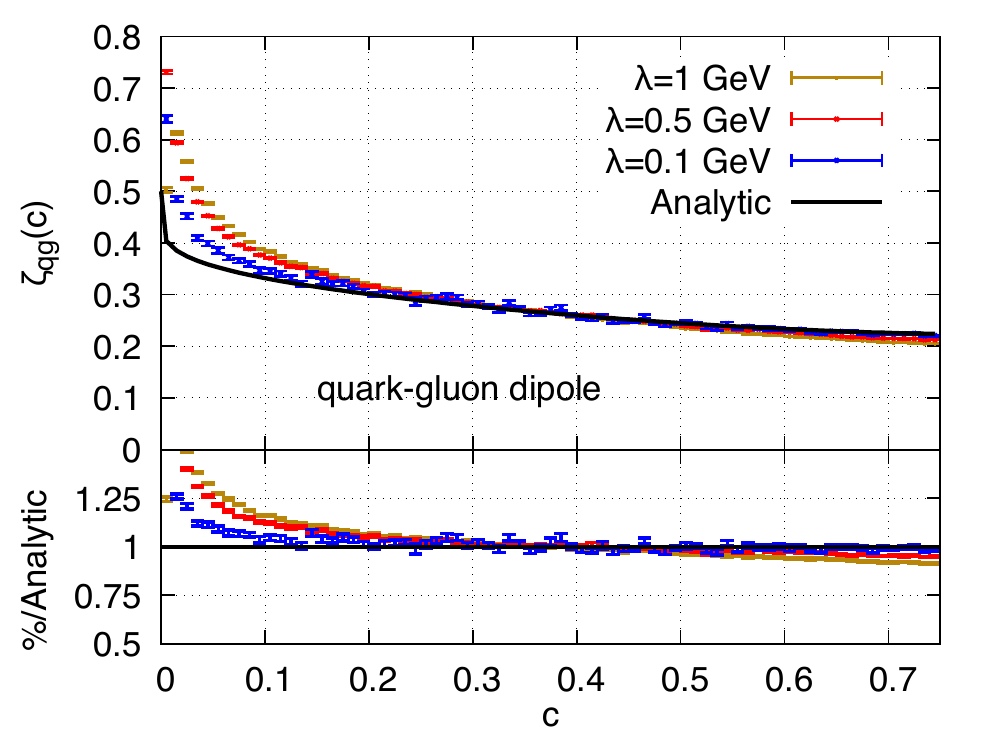}
  \includegraphics[width=0.48\textwidth]{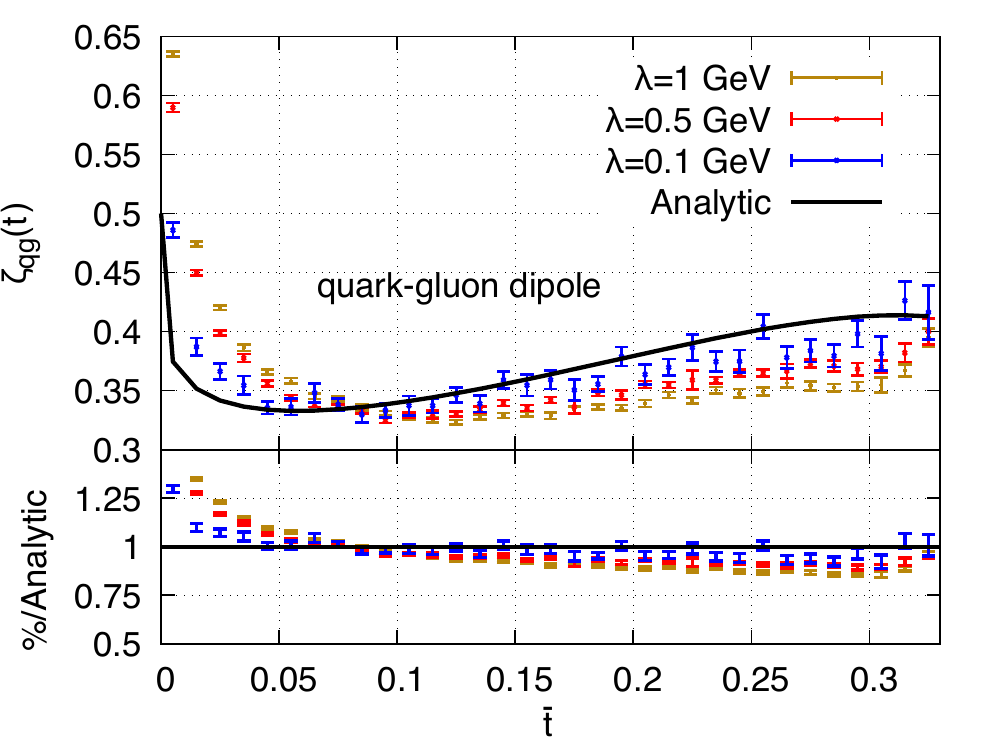}
  \caption{\label{fig:zetaqg} Same as fig.~\ref{fig:zetaqq}, but for
    the $qg$ dipole.  }
\end{figure}

To generalise the analytic result to the QCD case, 
$\gamma^*(q)\to q(p_1)+\bar q(p_2) + g(p_3)$, 
we consider radiation off the three dipoles $q \bar q$, $q g$ and $\bar q g$   and account for the relevant
color factors, see
ref.~\cite{Caola:2021kzt} for more detail. We  write 
\begin{equation}
  \delta_{\rm NP}^{q\bar q g}(\v) = h \zeta^{q\bar q g}(\v),
  ~~~~~~~~~~~\zeta^{q\bar q g}(\v) = \zeta_{q\bar q}(\v)
  \frac{\Cf-\Ca/2}{\Cf} + \zeta_{qg}(\v)\frac{\Ca}{\Cf},
  \label{eq:ansatz}
\end{equation}
where we have assumed that the $qg$ and $\bar q g$ dipoles contribute
equally. In eq.~\eqref{eq:ansatz}, $\zeta_{q\bar q}$ is identical to 
the contribution of
the $q\bar q$-dipole correction considered in this paper. The function
$\zeta_{q g}$ is  as $\zeta_{q \bar q}$   except that  the gluon plays the
role of an anti-quark. We note that it follows that $\zeta_{qg}(0)
= 1/2$~\cite{Caola:2021kzt} and, hence,  $\zeta^{q\bar qg}(0)=1$.

In order to get a more realistic prediction, the value of the constant
$h$ should be corrected to include also the effects of the gluon
splitting into two gluons, as it is commonly done in the dispersive
model~\cite{Dokshitzer:1997iz,Dokshitzer:1998pt,Smye:2001gq}, but this
is irrelevant for us since we only report results for the $\zeta$
function.

We compare analytic and numerical
results for  $\zeta_{qg}$
in fig.~\ref{fig:zetaqg}. We observe  that also in this case the numerical
result converges towards the analytic one, and that  features
discussed in connection
with   fig.~\ref{fig:zetaqq} are also present for the $qg$ dipole. 
We note that it is evident from fig.~\ref{fig:zetaqg} 
that
the numerical result for the $qg$ dipole is less stable  than the
result
for the $q\bar q$ one, so that the availability of the analytic computation in this case is
especially welcome.

\begin{figure}[t]
  \includegraphics[width=0.48\textwidth]{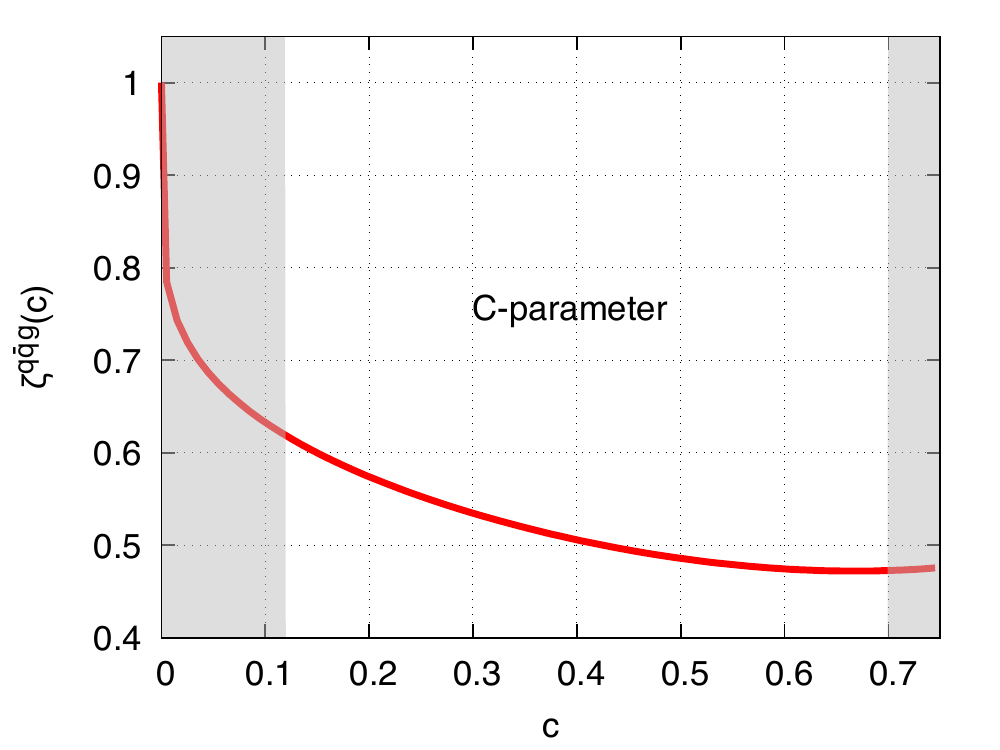}
  \includegraphics[width=0.48\textwidth]{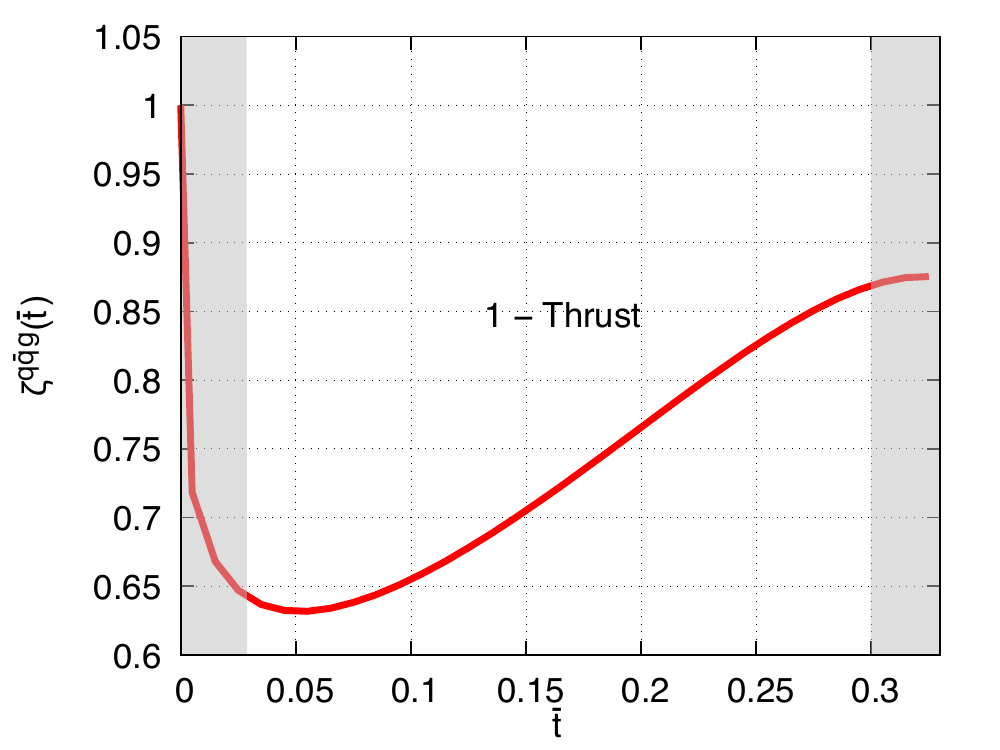}
  \caption{\label{fig:zetatot} Same as fig.~\ref{fig:zetaqq}, but for
    the sum of all the dipoles, where each contribution is
    supplemented with the proper colour factor, as in
    eq.~\eqref{eq:ansatz}. The grey shading shows the region which
    is usually excluded from $\alpha_s$
    determinations~\cite{Luisoni:2020efy}.}
  \end{figure}

  \subsection{Results for the $C$-parameter and the thrust in the three-jet region and
   comparison with existing literature}
\label{sec:cpar_pheno}
Having validated the analytic result against the numerical
ones of ref.~\cite{Caola:2021kzt}, we can compare our
predictions to the results in the literature. In
fig.~\ref{fig:zetatot}, we show our prediction for $\zeta^{q\bar qg}$
for both the $C$-parameter (left) and the thrust (right)
distributions using eq.~\eqref{eq:ansatz}. In those
figures the grey shaded areas show the kinematic regions 
that are typically excluded from  high-precision extractions of $\alpha_s$~\cite{Luisoni:2020efy}. We observe that for both the
$C$-parameter and the thrust the shape of non-perturbative corrections
in the bulk of the three-jet region is non-trivial.

It is interesting to compare our result for the $C$-parameter with 
the predictions of ref.~\cite{Luisoni:2020efy}.  Using
eqs.~(\ref{eq:c2j},~\ref{eq:c3j}), it is straightforward to check that
our results agree with those of
ref.~\cite{Luisoni:2020efy} at the endpoints
$c=0,3/4$.\footnote{We note that our definition of $\zeta^{q\bar q g}$
and $\zeta$ in ref.~\cite{Luisoni:2020efy} differ by a factor $3\pi$.}
However, in the bulk of the three-jet region the
formalism of ref.~\cite{Luisoni:2020efy} does  not lead to  an unambiguous
prediction. Instead, the authors of ref.~\cite{Luisoni:2020efy}
adopted an agnostic approach and obtained different results depending
on the assumption they made for the recoil due to the  emission of
a soft massless gluon, see their fig.~3.
It is interesting to note that most of the recoil schemes considered in
ref.~\cite{Luisoni:2020efy} (Catani-Seymour~\cite{Catani:1996vz},
PanLocal (antenna variant) and PanGlobal~\cite{Dasgupta:2020fwr}) gave
the same result. It turns out that this result is also compatible with our
prediction based on eq.~\eqref{eq:ansatz}.\footnote{We are grateful to
the authors of ref.~\cite{Luisoni:2020efy} for providing the
input data for their fig.~3.}  The FHP (Forshaw-Holguin-Pl\"atzer)
scheme of ref.~\cite{Forshaw:2020wrq} however led to a different prediction.

Our formalism provides a clear explanation of why this is the
case. Indeed, the Catani-Seymour, PanLocal, and PanGlobal schemes all
satisfy the smoothness requirement in the soft limit that is needed in
order for the recoil effects not to contribute to linear power
corrections~\cite{Caola:2021kzt}. In other words, if one uses these
schemes, then a naive soft analysis leads to the correct result for
linear power corrections to the $C$-parameter, without additional
contributions. This is not the case for the FHP scheme, which does not
satisfy the smoothness requirements.\footnote{We however stress
  that this does not impact the logarithmic accuracy of a dipole
  shower based on such a recoil.}  We elaborate on this remark in
what follows.

We start by analysing the PanLocal (antenna) mapping. It is
dipole-local, which means that it preserves the four-momentum of the radiating
dipole. It is defined by
\begin{align}
  p_1=\alpha_1 \pt_1+\beta_1 \pt_2 - f l_\perp, \qquad
  p_2=\alpha_2 \pt_1+\beta_2 \pt_2 - (1-f) l_\perp, \label{eq:panlocal1}
\end{align}
where $l$ is the momentum of the radiated soft gluon and 
$\alpha$ and $\beta$ are specified by the momentum-conservation
and on-shell requirements
\begin{align}
  p_1+p_2+l=\pt_1+\pt_2,\qquad\qquad
  p_{1/2}^2= 2\alpha_i\beta_i\,(p_1 p_2)+{\cal O}(l_\perp^2)=0\,.
\end{align}
These two equations can be satisfied only if either $\beta_1=\alpha_2=0$
or $\beta_2=\alpha_1=0$ up to terms of order $l_\perp^2$.
Assuming the first assignment, eq.~(\ref{eq:panlocal1}) implies
\begin{align}
  \alpha_1=1-\frac{(l \pt_2)}{(\pt_1 \pt_2)}\,, \qquad\qquad
  \beta_2=1- \frac{(l \pt_1)}{(\pt_1 \pt_2)}\,,
\end{align}
so, in the soft limit the PanLocal scheme has the form
\begin{align}
  p_1=\left(1-\frac{(l \pt_2)}{(\pt_1 \pt_2)}\right)\,\pt_1-f\,l_\perp,
  \qquad
  p_2=\left(1-\frac{(l \pt_1)}{(\pt_1 \pt_2)}\right)\,\pt_2-(1-f)\,l_\perp.
\end{align}
The PanLocal choice for $f$ is
\begin{equation}
  f=\frac{e^{2\bar{\eta}_k}}{1+e^{2\bar{\eta}_k}}\,,
  \label{eq:f}
\end{equation}
where $\bar{\eta}_k$ is a rapidity-type variable, defined as
\begin{equation}
\bar{\eta}_l = \frac{1}{2}\ln \frac{(l\tilde{p}_2)( \tilde{p}_1
  q)}{(l\tilde{p}_1)(\tilde{p}_2  q)}.
\label{eq:etabar}
\end{equation}
Thus the PanLocal mapping is non-linear in the soft limit. However,
the non linear term is proportional to $l^\mu_\perp$ with an
azimuthally-independent coefficient.  Hence, when computing recoil
effects using this mapping, the non-linear term always cancels after
azimuthal integration, and the PanLocal mapping satisfies the
smoothness criteria~\cite{Caola:2021kzt}. The Catani-Seymour mapping
is also non-linear, but again the non-linear term vanishes after
azimuthal integration for a similar reason.

The PanGlobal mapping is defined as follows. First, one introduces
the intermediate variables
\begin{align}
  \bar{p}_1=\left(1-\frac{(l\pt_2)}{(\pt_1\pt_2)}\right)\,\pt_1,
  \qquad\bar{p}_2=\left(1-\frac{(l\pt_1)}{(\pt_1 \pt_2)}\right)\,\pt_2\,.
\end{align}
Then one finds a boost $B$ followed by a rescaling $R$ to be applied
to $\bar{p}_1,\bar{p}_2,p_3$ such that
\begin{align}
  p_{1/2}=R B \bar{p}_{1/2}, \qquad
  p_3= R B \pt_3,\qquad
  l' &= R B l.
\end{align}
In the small-$l$ limit, it is easy to see that this yields the mapping
\begin{align}
p_1 &\approx \pt_1 -\frac{(\pt_2 l)}{(\pt_1 \pt_2)}\pt_1-\frac{(\pt_1 q)}{q^2}\, l_\perp-\frac{(l_\perp q)}{q^2} \,\pt_1,\nonumber \\
p_2 &\approx \pt_2 -\frac{(\pt_1 l)}{(\pt_1 \pt_2)}\pt_2-\frac{(\pt_2 q)}{q^2}\, l_\perp-\frac{(l_\perp q)}{q^2} \,\pt_2, \\
p_3 &\approx \pt_3 -\frac{(\pt_3 q)}{q^2}\,l_\perp + \frac{\pt_3\cdot l_\perp}{q^2}\, q -\frac{(l_\perp q)}{q^2} \,\pt_3\,.\nonumber
\end{align}
This is fully linear, and it satisfies the
requirements of  ref.~\cite{Caola:2021kzt}. As we have said, this implies that
with these recoil schemes a naive soft analysis completely captures 
linear power corrections. 

The FHP mapping is similar to the PanGlobal one, but only one side of
the dipole is rescaled. In practice, one parametrises the intermediate momenta
\begin{align}
    \bar{p}_1=\left(1-\frac{(l \pt_2)}{(\pt_1 \pt_2)}\right)\,\pt_1, \qquad
    \bar{p}_2=\pt_2\,,
\end{align}
with probability $f(\bar{\eta_l})$, where $f$ and $\bar{\eta}_l$ are
given in eqs.~(\ref{eq:f}) and~(\ref{eq:etabar}) and coincide with the
PanLocal/PanGlobal ones. The parametrisation
\begin{align}
    \bar{p}_1=\pt_1, \qquad
    \bar{p}_2=\left(1-\frac{(l \pt_1)}{(\pt_1 \pt_2)}\right)\,\pt_2\,,
\end{align}
is instead selected with probability $1-f(\bar{\eta_l})$.  Then the kinematic
reconstruction proceeds as in PanGlobal, where all the momenta are
rescaled to preserve the mass of the total system, and then boosted in
the original event frame.
It is thus clear that the soft limit is non-linear in the
longitudinal components of the momenta of the radiated parton and, therefore,
it does not satisfy the requirements~\cite{Caola:2021kzt}. As a
consequence, an analysis
of soft emissions is not sufficient 
to compute the linear power corrections if one were to use this recoil
scheme. This is exactly what is seen in fig.~3 of
ref.~\cite{Luisoni:2020efy}.  We conclude this section by noting
 that at the endpoints $c=0$, $c=3/4$ all recoil schemes yield
the same result.  This is expected since in these regions the
sensitivity of the shape variable to recoil effects is strongly
suppressed~\cite{Luisoni:2020efy}.

\section{Conclusions}\label{sec:conc}
The goal of this paper was to continue exploration of  linear power corrections to
shape variables in $e^+e^-$ annihilation to hadrons that we started in  ref.~\cite{Caola:2021kzt}.
We performed explicit  analytic computations for the
$C$-parameter and the thrust distributions and showed that  our considerations apply  to a large class
of shape variables.

Our results  can be briefly summarised  as follows:  linear power
corrections which affect   cumulants of  shape variables in the
three-jet region are  proportional to a universal, constant factor times
 a function which  characterises  the behaviour of the shape
variable under the emission of a single soft  massless parton.  Our
result applies \emph{a fortiori} to the two-jet region where, however,  an
equivalent result has been formulated long
ago~\cite{Dokshitzer:1998pt}.

Our calculations are performed in the  large-$n_f$ model of
QCD,  but with a final state $q \bar q \gamma$ rather than $q \bar q g$. 
The linear renormalon contribution to shape variables  associated
with the power correction can be exactly calculated in this framework. It 
arises from the emission  of a soft virtual gluon that can both fluctuate into
virtual $q \bar q$ pairs and eventually decay into one of them. 

Since renormalons can be studied using soft emissions,
we can speculate that, in the realistic case of a
$q\bar{q} g$ final state, linear power corrections can be obtained
by summing up the soft emissions off all three colour dipoles,
$q\bar{q}$, $qg$ and $\bar{q}g$, multiplied by the corresponding
colour factor. By doing this, we obtain  a theoretical prediction that can be confronted
with data. 

These assumptions are essentially the same ones that
underlie the so called \emph{dispersive model} of power
corrections~\cite{Dokshitzer:1997iz,Dokshitzer:1998pt}. This model is fully consistent with the large-$n_f$ limit of QCD, and
it was recently applied to the calculation of power corrections to the
$C$-parameter near the three-jet symmetric point~\cite{Luisoni:2020efy}.

It is useful to briefly summarise the reasons that allow for the
computation of power corrections in the two-jet region, in the three-jet
symmetric limit for the $C$-parameter~\cite{Luisoni:2020efy} and in the generic three-jet
configuration as described in this paper.
In general, we can write the
cumulant of a shape variable as
\begin{equation}\label{eq:illustration}
  \Sigma(\v) = \int \mathd \sigma(\{\pt\}) \theta\big(\V(\{\pt\})-\v\big) +
  \int \mathd \sigma(\{p,k\}) \theta\big(\V(\{p,k\})-\v\big),
\end{equation}
where $\{\pt\}$ denote the hard final state momenta when no soft
particles are emitted, $p$ denote the hard momenta in the final state
that are accompanied by the emission of a soft gluon, and by $k$ we
denote generically the momenta of the final state particles arising from the
radiated soft gluon. Furthermore, we assume that $\sigma(\{\pt\})$
also includes the virtual soft gluon corrections, so that the
left-hand  side of eq.~(\ref{eq:illustration}) includes all relevant
corrections to the cumulant. By $\V(\{\pt\})$ and $\V(\{p,k\})$ we denote
the value of the shape variable computed for the hard partons final
state (without soft gluon emission) and for the final state including
the gluon emission with its decay products.

In the two-jet region $v\to 0$, the final state consists of two back-to-back
partons. Thus $\V(\{\pt\})$ is constant, and we can manipulate
eq.~\eqref{eq:illustration} as follows
\begin{eqnarray}\label{eq:illustration1}
  \Sigma(\v) &=& \left[\int \mathd \sigma(\{\pt\})+ \int \mathd
    \sigma(\{p,k\})\right] \theta\big(\V(\{\pt\})-\v\big) \nonumber
  \\ &+& \int \mathd \sigma(\{p,k\}) \left[\theta\big(\V(\{p,k\})-\v\big) -
    \theta\big(\V(\{\pt\})-\v\big) \right]\,.
\end{eqnarray}
We can see now that the first line of eq.~(\ref{eq:illustration1}) is
proportional to the total cross section, and thus is free from linear
infrared renormalons. In the second term, on the other hand, the factor
in the square bracket vanishes in the soft limit, so that the soft
approximation to $\mathd \sigma(\{p,k\})$ is all what is needed for
the computation, that can then be carried out along the lines of
refs.~\cite{Dokshitzer:1997iz,Dokshitzer:1998pt}.

If we consider the $C$-parameter near the three-jet symmetric point,
the crucial observation is that the contribution to $C$ from the
hard partons is given by~\cite{Luisoni:2020efy}
\begin{equation}\label{eq:cparsymm}
    C_{\rm hard}=\frac{3}{4}-\frac{81}{16}(\epsilon_q^2+\epsilon_q\epsilon_{\bar q}
    +\epsilon_{\bar q}^2)+{\cal O}(\epsilon^3),
\end{equation}
where $\epsilon_{q(\bar q)}=2E_{q(\bar{q})}/E-2/3$ and $E=\sqrt{q^2}$
is the total energy.  If we apply eq.~(\ref{eq:illustration}) for $c$
very near $3/4$, the momenta $\{\pt\}$ will be forced to approach the
symmetric limit, and $C\{\pt\}$ is nearly constant there. So, we can
again carry out the manipulation that led to
eq.~(\ref{eq:illustration1}) with $\V=C$.  Again, the first term is
proportional to a cross section that is inclusive in the radiation of
the soft gluon, and can be assumed to be free of linear
renormalon~\cite{Caola:2021kzt}. Because of eq.~\eqref{eq:cparsymm},
we can replace $C(\{p(\pt,k),k\})-c) \to C(\{\pt,k\})-c)$ in the
square bracket, and because of the ensuing soft suppression, we can
again resort to the soft approximation to perform the computation, as
it was done in ref.~\cite{Luisoni:2020efy}. Hence, we see that the
derivation in both the two-jet and symmetric three-jet cases discussed
above crucially relies on recoil effects being strongly suppressed. As
a consequence, it is not immediately clear how to generalise these
constructions to the three-jet region beyond the symmetric point.

Let us now see how this problem is solved in this paper. Assuming,
for the sake of discussion, that the soft emission originates from  a
single dipole, we introduce a mapping $p(\pt,k)$ that, for small $k$,
is linear in $k$
and that is collinear-safe when the gluon is collinear
to the partons constituting the dipole. Then, for a generic shape
variable and for generic momenta we write eq.~(\ref{eq:illustration})
as
\begin{eqnarray}\label{eq:illustration2}
  \Sigma(\v) &=& \int \big[ \mathd \sigma(\{\pt\})+ \mathd
    \sigma(\{p(\pt,k),k\})\big] \theta\big(\V(\{\pt\})-\v\big)
  \nonumber \\ &+& \int \mathd \sigma(\{p(\pt,k),k\})
  \big[\theta\big(\V(\{p(\pt,k),k\})-\v\big) -
    \theta\big(\V(\{\pt\})-\v\big) \big]\,.
\end{eqnarray}
Now the first term involves an inclusive integral of the cross section
at fixed underlying Born momenta, with the underlying Born defined by
a mapping that is linear in $k$ for small $k$. It was shown 
  in ref.~\cite{Caola:2021kzt} that
such integrals do not yield
linear power corrections. Hence, we can drop the first line in
eq.~(\ref{eq:illustration2}) and
consider only the second line, that again can be evaluated in the soft
approximation leading to the results discussed in this paper. 

Although the phenomenological consequences of our results are yet to
be fully explored, we can make a few statements using the results of
ref.~\cite{Luisoni:2020efy}.  There, in the framework of the
dispersive model for power corrections which employs a single
non-perturbative parameter related to the low-scale value of the
strong coupling constant, several interpolations of the
non-perturbative corrections to the $C$-parameter distribution between
the two- and three-jet symmetric points were considered, and for each
of them a fit of the strong coupling constant was performed. For some
of the extrapolations it was found that the fitted value of $\as$ was
larger than the one obtained using the conventional assumption that
non-perturbative corrections are constant and, therefore, are fixed by
their value in the two-jet region.  Our results clearly show that the
non-perturbative effects in the three-jet region are
kinematic-dependent.  In fact, our result agrees with the lowest
curves in fig.~3 of ref.~\cite{Luisoni:2020efy} that, in turn, are
below the line labeled as $\zeta_{b,3}$ there.  From table~2 in that
reference, we see that the use of $\zeta_{b,3}$ in the $\alpha_s$ fit
yields the value of $0.1167$ that is larger than the $\alpha_s =
0.1121$ value obtained by assuming that the non-perturbative
coefficient is constant.

In view of these observations, a conservative conclusion from the
results of the present paper is that the error on $\as$ determination
that employs constant non-perturbative corrections should be
considerably enlarged towards the upper side.  A more aggressive
conclusion would be that the central value should be shifted in the
upper direction. However, in the latter case, in view of the fact that
several assumptions underlie our findings, chief of which is an Abelian
nature of the large-$\nf$ calculation, a more refined theoretical
understanding and comparison with data should be needed to validate
the approach.

\acknowledgments We are grateful to the authors of
ref.~\cite{Luisoni:2020efy} for sharing with us the input data for
their fig.~3. We thank P. Monni and G. Salam for many interesting discussions.
We also thank C. Duhr and L. Tancredi for insightful exchanges on elliptic
functions. 
The
research of F.C. is supported by the ERC Starting Grant 804394 \textsc{hipQCD}
and the UK Science and Technology Facilities Council (STFC) under
grant ST/T000864/1. S.F.R.’s work is supported by the ERC Advanced
Grant 788223 PanScales. P.N. acknowledges support from Fondazione
Cariplo and Regione Lombardia, grant 2017-2070, from
INFN and from the Humboldt foundation. K.M. is
partially supported by the Deutsche Forschungsgemeinschaft (DFG,
German Research Foundation) under grant 396021762-TRR 257.

%%%Start of Appendix
\appendix

\section{Renormalon structure of the large-$\nf$ result}\label{app:renstruct}
The exact prediction for the cumulant $\Sigma (\v)$
eq.~\eqref{eq:CumulantDefinition} computed in the large-$\nf$
approximation reads
\begin{equation}
  \Sigma (\v) = \Sigma_{\rm b} (\v) -\frac{1}{\bnotf \as(\mu)}
  \int_0^\infty \frac{\mathd\lambda}{\pi} \frac{\mathd
    \Sigma(v;\lambda)}{\mathd \lambda} \arctan \frac{\pi \bnotf
    \as(\mu)}{1+\bnotf \as(\mu)\log \frac{\lambda^2e^{-C}}{\mu^2}},
  \label{eq:sigmalargenf}
\end{equation}
where $\Sigma_{\rm b} (\v)$ is the LO prediction, $\Sigma(v;\lambda)$
and $\bnotf$
are given in eqs.~(\ref{eq:Tolambda},~\ref{eq:bnotf}) and $C=5/3$. We note
that the factor $\as$ in the denominator in front of the integral cancels the
factor of $\as$ in $\Sigma(v;\lambda)$.
Eq.~(\ref{eq:sigmalargenf}) is well-known (see for example
ref.~\cite{Beneke:1998ui}) as far as the virtual corrections and the
inclusive real corrections are concerned, but we are not aware of a
reference before \cite{FerrarioRavasio:2018ubr} where also the
$q\bar{q}$ splitting term is cast in the same form. In appendix~B of
ref.~\cite{FerrarioRavasio:2018ubr} a complete derivation of
eq.~\eqref{eq:sigmalargenf} formula is also given.  Examples of real
and virtual Feynmann graphs contributing to
eq.~(\ref{eq:sigmalargenf}) are illustrated in
fig.~\ref{fig:largeNFgraphs},
\begin{figure}[t]
  \begin{center}
  \includegraphics[width=0.9\textwidth]{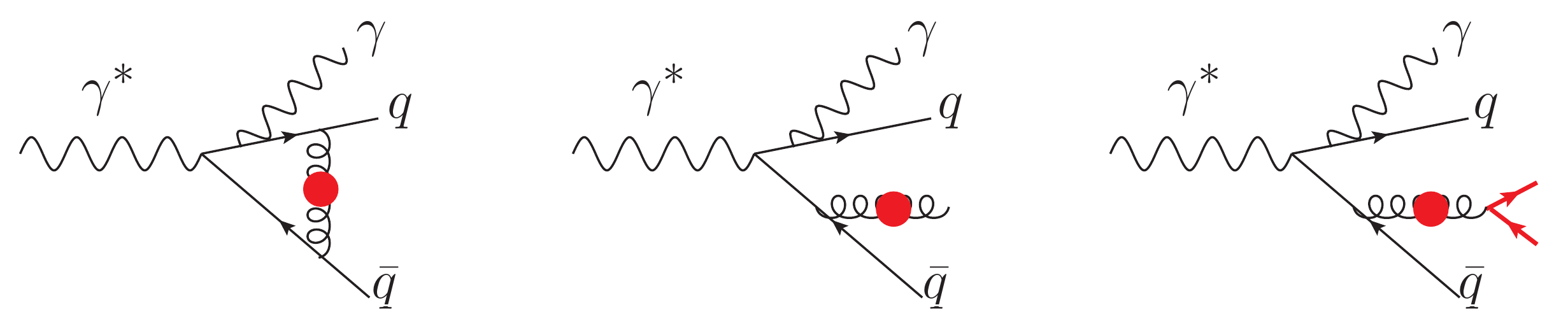}
  \end{center}
  \caption{\label{fig:largeNFgraphs} An example of a virtual, a real
    gluon emission and a real $q\bar{q}$ emission graph contributing to the
    large-$\nf$ correction to the process $\gamma^*\to q\bar{q} \gamma$.
    The right graph, with a $g\to q\bar{q}$ splitting, must be included, since
    the sum over final state quark flavour leads to a factor of $\nf$.
  }
\end{figure}
 where the solid blob insertion in the gluon propagator represents the inclusion of all corrections given by a fermion loop, as
represented by the recursive graphic equation
\begin{equation}
\raisebox{-0.5cm}{\includegraphics[width=0.8\textwidth]{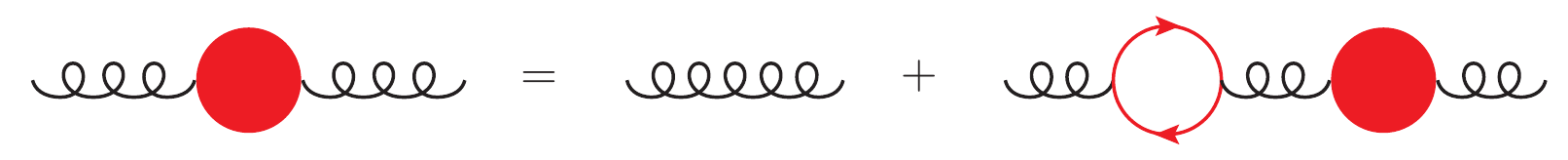}}.
\end{equation}

The linear power corrections arise in eq.~(\ref{eq:sigmalargenf}) from
the leading term in the small-$\lambda$ expansion of
$\Sigma(v;\lambda)$, that leads to
\begin{equation}\label{eq:renstruct1}
  \Sigma(v)\approx\Sigma_b(v)-
  \left[\frac{\mathd \Sigma(v;\lambda)}{\mathd \lambda} \right]_{\lambda=0}
  \left\{ \frac{1}{b_{0,f}\as(\mu)} \int_0^{\mu_C}\frac{d\lambda}{\pi} \arctan \frac{\pi \bnotf \as(\mu)}{1+\bnotf \as(\mu)\log \frac{\lambda^2}{\mu_C^2}}\right\},
\end{equation}
where $\mu_C=\mu e^{ C/2}$. The upper limit of integration is chosen for convenience, since the
renormalon structure arises from the region where the denominator in the arctangent vanishes, corresponding
to the Landau pole. Thus, any upper limit comprising the Landau pole is acceptable.

The renormalon structure of the expression in the curly bracket of
eq.~(\ref{eq:renstruct1}), according to the result presented in
appendix~A of ref.~\cite{FerrarioRavasio:2020guj}, is
\begin{equation}\label{eq:renstruct}
  \begin{split}
  & \frac{1}{b_{0,f}\as(\mu)} \int_0^{\mu_C}\frac{d\lambda}{\pi} \arctan
  \frac{\pi \bnotf \as(\mu)}{1+\bnotf \as(\mu)\log
    \frac{\lambda^2}{\mu_C^2}} = \\
  &  \frac{\mu_C}{b_{0,f}\as(\mu)}
  \left[ \text{P}\int_0^\infty \frac{\mathd t}{\pi}
    \frac{\exp\left(-\frac{t}{2\bnotf\as(\mu)}\right)}{1-t} + C_{\rm
      NP} \frac{\Lambda_{\rm QCD}}{\mu}\right]+ \mbox{analytic terms in
    $\as(\mu)$}.
  \end{split}
\end{equation}
In the last term, $\Lambda_{\rm QCD}$ is the QCD (large-$\nf$) scale,
defined by the equation
\begin{equation}
  \as(\mu)=\frac{1}{\bnotf \log \frac{\mu^2}{\Lambda_{\rm QCD}^2}},
\end{equation}
so that
\begin{equation}\label{eq:lamomu}
   \exp\left(-\frac{1}{2\bnotf\as(\mu)}\right) = \frac{\Lambda_{\rm QCD}}{\mu}.
\end{equation}
In ref.~\cite{FerrarioRavasio:2020guj} a further term of the form of
eq.~(\ref{eq:lamomu}), arising from the discontinuity of the integrand
in eq.~(\ref{eq:renstruct1}) at the Landau pole, is added to the
principal value part. If a suitable theta function is added to the
integrand in (\ref{eq:renstruct1}) such term is not present. The
important point, however, is that an ambiguity remains depending upon
the handling of the Landau pole.  Such ambiguity is related to the
residue of the integral at $t=1$, and thus has the form of
eq.~(\ref{eq:lamomu}).  In eq.~(\ref{eq:renstruct}) we parametrize
this ambiguity in terms of an (uncalculable) coefficient $C_{\rm NP}$.
We also note that the factorial growth in the asymptotic expansion of
the Borel sum in eq.~(\ref{eq:renstruct}) is due to the singularity at
$t=1$. Thus its $\mu$ dependence cancels against the $\mu_C$ factor,
and so does the $\mu$ dependence of the $C_{\rm NP}$ term. See
appendix~A in ref.~\cite{FerrarioRavasio:2020guj} for more details.

In some phenomenological approaches the expression in the curly
bracket of eq.~(\ref{eq:renstruct1}) is interpreted as an effective
coupling, that in the full theory should be replaced by a QCD
effective coupling that also includes non-perturbative
effects~\cite{Dokshitzer:1995qm}.

\section{The functions $G_{1,..,5}$}
\label{sec:funG5}

The functions $G_{1,..,5}$ were introduced in
eq.~\eqref{eq:Iidef}. They read
\begin{align}
G_1 & = \frac{\left(1-\beta^2\right)^{-3/2}}{48 \beta^6 (x (y-1)+1)^2
  (x y-1)^2} \Bigg [ \beta^6 \Bigg (2 x^3 y \left(-42 y^3+84 y^2-421
  y+379\right) \nonumber \\ & + 29 x^4 (y-1)^2 y^2 +x^2 \left(1706
  y^2-1706 y-271\right)+x \left(-948 y^2+948 y+542\right)-271 \Bigg )
  \nonumber \\ & +\beta^{4} \Bigg (-159 x^4 (y-1)^2 y^2 +2 x^3 y
  \left(187 y^3-374 y^2+1203 y-1016\right) \nonumber \\ & +x^2
  \left(-4542 y^2+4542 y+749\right) +2 x \left(1255 y^2-1255
  y-749\right)+749 \Bigg ) \\ & +5 \beta^2 \Bigg ( 47 x^4 (y-1)^2
  y^2+2 x^3 y \left(-50 y^3+100 y^2-271 y+221\right) \nonumber \\ &
  +165 x^2 \left(6 y^2-6 y-1\right)+x \left(-548 y^2+548
  y+330\right)-165 \Bigg ) \nonumber \\ & -105 \Bigg (x^4 (y-1)^2
  y^2-2 x^3 y \left(y^3-2 y^2+5 y-4\right)+3 x^2 \left(6 y^2-6
  y-1\right) \nonumber \\ & +x \left(-10 y^2+10 y+6\right)-3\Bigg )
  \Bigg ], \nonumber \\ G_2 & = \frac{\sqrt{1-\beta^2} (x-1) \ln
  ^2\left(\frac{1+\beta}{1-\beta}\right) }{32 \beta^8 x (x (y-1)+1)^2
  (x y-1)^2} \nonumber \\ & \times \Bigg [ \beta^4 \left(x^2
  \left(-117 y^2+117 y+34\right)+83 x^3 (y-1) y-53 x+19\right) \\ &
  -10 \beta^2 \left(x^2 \left(-29 y^2+29 y+8\right)+21 x^3 (y-1) y-11
  x+3\right) \nonumber \\ & +35 \left(x^2 \left(-7 y^2+7 y+2\right)+5
  x^3 (y-1) y-3 x+1\right) \Bigg ], \nonumber \\ G_3 & = \frac{\ln
  \left(\frac{\sqrt{1-\beta^2 c_{12}^2}+\beta s_{12}}{\sqrt{1-\beta^2
      c_{12}^2}-\beta s_{12}}\right)}{96 \beta^7
  \left(1-\beta^2\right)^{3/2} s_{12} x \sqrt{1-\beta^2 c_{12}^2} (x
  (y-1)+1) (x y-1)} \nonumber \\ & \times \Bigg [ \beta^8 x \left(29
  x^2 (y-1) y+x \left(-84 y^2+84 y+15\right)-15\right) \nonumber \\ &
  -2 \beta^6 \left(x^2 \left(-229 y^2+229 y+51\right)+94 x^3 (y-1)
  y-186 x+135\right) \\ & +\beta^4 \left(x^2 \left(-874 y^2+874
  y+232\right)+394 x^3 (y-1) y-758 x+526\right) \nonumber \\ & -10
  \beta^2 \left(x^2 \left(-71 y^2+71 y+25\right)+34 x^3 (y-1) y-78
  x+53\right) \nonumber \\ & +105 (x-2) \left(x^2 (y-1) y+x-1\right)
  \Bigg ], \nonumber \\ G_4 & = \frac{\ln
  \left(\frac{1+\beta}{1-\beta}\right)}{96 \beta^7
  \left(1-\beta^2\right)^{3/2} x (x (y-1)+1)^2 (x y-1)^2} \Bigg [
  \beta^8 x \Bigg (3 x^4 (y-1)^2 y^2 \nonumber \\ & -6 x^3 y \left(2
  y^3-4 y^2+57 y-55\right)+x^2 \left(758 y^2-758 y-113\right)
  \nonumber \\ & +x \left(-428 y^2+428 y+226\right) -113\Bigg ) -6
  \beta^6 \Bigg ( 11 x^5 (y-1)^2 y^2 \nonumber \\ & +x^4 y \left(-29
  y^3+58 y^2-479 y+450\right) +x^3 \left(1046 y^2-1046 y-171\right)
  \\ & +x^2 \left(-596 y^2+596 y+387\right)-261 x+45\Bigg ) +2 \beta^4
  \Bigg (114 x^5 (y-1)^2 y^2 \nonumber \\ & +x^4 y \left(-261 y^3+522
  y^2-3187 y+2926\right) +8 x^3 \left(848 y^2-848 y-139\right)
  \nonumber \\ & +x^2 \left(-3858 y^2+3858 y+2487\right)-1638
  x+263\Bigg ) -10 \beta^2 \Bigg (27 x^5 (y-1)^2 y^2 \nonumber \\ &
  +x^4 y \left(-57 y^3+114 y^2-607 y+550\right)+x^3 \left(1278
  y^2-1278 y-211\right) \nonumber \\ & +x^2 \left(-728 y^2+728
  y+475\right)-317 x+53\Bigg ) +105 \Big (x^5 (y-1)^2 y^2 \nonumber
  \\ & -2 x^4 y \left(y^3-2 y^2+10 y-9\right) +7 x^3 \left(6 y^2-6
  y-1\right)-8 x^2 \left(3 y^2-3 y-2\right)-11 x+2\Big ) \Bigg ],
\nonumber \\ G_5 & = \frac{\sqrt{1-\beta^2} \ln
  \left(\frac{1+\beta}{1-\beta}\right) \ln \left(\frac{\sqrt{1-\beta^2
      c_{12}^2}+\beta s_{12}}{\sqrt{1-\beta^2 c_{12}^2}-\beta
    s_{12}}\right) }{ 64 \beta^8 s_{12} x (x (y-1)+1) (x y-1)
  \sqrt{1-\beta^2 c_{12}^2}} \nonumber \\ & \times \Bigg [ \beta^6 x
  \left(x^2 (y-1) y+x \left(-4 y^2+4 y-5\right)+5\right) +\beta^4
  \Bigg (x^2 \left(54 y^2-54 y-17\right) \nonumber \\ & -21 x^3 (y-1)
  y+55 x-38 \Bigg ) +5 \beta^2 \Bigg (x^2 \left(-24 y^2+24 y+5\right)
  \\ & +11 x^3 (y-1) y-17 x+12 \Bigg ) -35 (x-2) \left(x^2 (y-1)
  y+x-1\right) \Bigg ].  \nonumber
\end{align}

\section{Elliptic integration}
\label{sec:ellint}
The goal of the discussion in this appendix is to explain how the
integrals in eq.~\eqref{eq:Iidef} can be calculated in a systematic
way using state-of-the-art technology. First, in
appendix~\ref{sec:esetup} we will show how the integration can be
recast so that the elliptic multiple
polylogarithms~\cite{Broedel:2017kkb,Broedel:2017siw} appear
naturally. This will allow us to obtain an analytic expression for
$I_i$ in eq.~\eqref{eq:Iidef}, albeit a quite complicated one.  To
simplify it, in appendix~\ref{sec:eisen} we will express it in terms of
the so-called $\tilde\Gamma$ integrals (see
refs.~\cite{Broedel:2017kkb,Broedel:2017siw,Broedel:2018iwv,Broedel:2018qkq,Broedel:2019hyg,Weinzierl:2022eaz})
and note that for the purposes of our calculation we only require
kernels evaluated at special, rational points. This will allow us to
rewrite our result in terms of iterated Eisenstein integrals, and
observe a significant simplification in the final expression.
In this appendix, we will assume that the reader is familiar with the
elliptic polylogarithms literature. For convenience, we will provide
a minimal background necessary to follow the relevant steps in
appendix~\ref{sec:brief}.

\subsection{Set-up of the calculation and result in terms of
  MPLs and eMPLs}
\label{sec:esetup}
The functions $G_{1,..,5}$ displayed in appendix~\ref{sec:funG5}
contain the integration variable $\beta$ under the square roots
$\sqrt{1-\beta^2}$ and $\sqrt{1-\beta^2 c_{12}^2}$. In particular,
$G_{1,2,4}$ only contains the first square root, while $G_{3,5}$ contain
both. We consider the integration of $G_{1,2,4}$ first. 

To rewrite the integrand in a form suitable for the integration with
multiple polylogarithmic kernels, we rationalise the square root with
a variable transformation \cite{Besier:2018jen,Besier:2019kco}
\begin{equation}
\beta= \frac{2 z}{1+z^2},
\label{eq:vartrans}
\end{equation}
so that
\begin{equation}
{\rm d} \beta  =  \frac{2 \left(1-z^2\right)}{\left(1+z^2\right)^2}\; {\rm d} z.
\end{equation}
The new integration limits 
are
\begin{equation}
0  < z < \sqrt{ \frac{\omega_{\rm max} - \lambda}{\omega_{\rm max} + \lambda} }.
\end{equation}
We note that each integral
\begin{equation}
I_i = \int \limits_{0}^{\beta_{\rm max} } {\rm d} \beta\, G_i,\;\;\; i=1,..,5,
\end{equation}
 is divergent at $\beta=0$
  but their sum is finite. Since we will compute  all the 
  different integrals separately,  we will need to introduce
  a lower integration boundary $\beta_{\rm min}$.  This implies that
  the lower integration boundary for the variable $z$ is 
\begin{equation}
  z_{\rm min} = \frac{\beta_{\rm min} }{1 + \sqrt{1-\beta_{\rm min}^2}}.
\end{equation}
We note that care has to be taken that the expansion in $z_{\rm min}$
is done properly for each individual expression up to finite terms.
Similarly, as we are interested only in ${\cal O}(\lambda)$
contribution to the $C$-parameter, we need to compute each term $G_i$
only up to $\mathcal{O}{\left(\lambda^0\right)}$ since we already have
a global factor $\lambda/q$ in front of the integral in
eq.~\eqref{eq:Iidef}. The change of variable in
eq.~\eqref{eq:vartrans} allows us to express the result for
$I_{1,2,4}$ in terms of the multiple polylogarithmic functions (MPLs)
of ref.~\cite{Goncharov:1998kja,Goncharov:2001iea}. After expanding
around $\beta_{\rm min}=0$ and $\lambda = 0$, we obtain
\begin{align}
  \label{eq:i1e}
I_1=&-\frac{\omega_{\rm max}}{\lambda}\frac{2 (x-1) \left(3 x^2 (y-1) y+x \left(-4 y^2+4 y+1\right)-1\right)}{3 (x (y-1)+1)^2 (x y-1)^2}
\nonumber \\
&-\frac{1}{\beta_{\rm min}^5}\frac{7}{16 (x (y-1)+1)^2 (x y-1)^2}\left[x^4 (y-1)^2 y^2-2 x^3 y \left(y^3-2 y^2+5 y-4\right)\right.
\nonumber \\
&\left.+3 x^2 \left(6 y^2-6 y-1\right)+x \left(-10 y^2+10 y+6\right)-3\right]
\nonumber \\
&+\frac{1}{\beta_{\rm min}^3}\frac{5}{288 (x (y-1)+1)^2 (x y-1)^2}\left[31 x^4 (y-1)^2 y^2+2 x^3 y \left(-37 y^3+74 y^2\right.\right.
 \\
&\left.\left.-227 y+190\right)+141 x^2 \left(6 y^2-6 y-1\right)+x \left(-466 y^2+466 y+282\right)-141\right]
\nonumber \\
&-\frac{1}{\beta_{\rm min}}\frac{1}{384 (x (y-1)+1)^2 (x y-1)^2}\left[27 x^4 (y-1)^2 y^2+2 x^3 y \left(-71 y^3+142 y^2\right.\right.
\nonumber \\
&\left.-1239 y+1168\right)+x^2 \left(5286 y^2-5286 y-817\right)
\nonumber \\
&\left. +x \left(-2950 y^2+2950 y+1634\right)-817\right],
\nonumber \\
\nonumber \\
\label{eq:i2e}
I_2=&\frac{1}{\beta_{\rm min}^5}\frac{7 (x-1) \left(5 x^3 (y-1) y+x^2 \left(-7 y^2+7 y+2\right)-3 x+1\right)}{8 x (x (y-1)+1)^2 (x y-1)^2}
\nonumber \\
&-\frac{1}{\beta_{\rm min}^3}\frac{5 \left((x-1) \left(217 x^3 (y-1) y+x^2 \left(-299 y^2+299 y+82\right)-111 x+29\right)\right)}{144 \left(x (x (y-1)+1)^2 (x y-1)^2\right)}
\nonumber \\
&+\frac{1}{\beta_{\rm min}}\frac{(x-1) \left(4121 x^3 (y^2-y)
  -x^2 \left(5875 (y^2-y)-1754\right)-2895 x+1141\right)}{576 x (x (y-1)+1)^2 (x y-1)^2}
 \\
&-\frac{\pi ^2 (x-1) \left(85 x^3 (y-1) y+x^2 \left(-127 y^2+127 y+42\right)-83 x+41\right)}{256 \left(x (x (y-1)+1)^2 (x y-1)^2\right)},
\nonumber \\
\nonumber \\
\label{eq:i4e}
I_4=&\frac{1}{\beta_{\rm min}^5}\frac{7}{16 x (x (y-1)+1)^2 (x y-1)^2}\left[x^5 (y-1)^2 y^2-2 x^4 y \left(y^3-2 y^2+10 y-9\right)\right.
\nonumber \\
&\left.+7 x^3 \left(6 y^2-6 y-1\right)-8 x^2 \left(3 y^2-3 y-2\right)-11 x+2\right]
\nonumber \\
&-\frac{1}{\beta_{\rm min}^3}\frac{5}{288 \left(x (x (y-1)+1)^2 (x y-1)^2\right)}\left[31 x^5 (y-1)^2 y^2
  -74 x^4 y \left(y^2(y-2)  \right.\right.
\nonumber \\
&\left.\left.  +12 y -11\right)+x^3 \left(1878 ( y^2- y)-305\right)-x^2 \left(1064 (y^2-y)-668\right)-421 x+58\right]
\nonumber  \\
&+\frac{1}{\beta_{\rm min}}\frac{1}{1152 x (x (y-1)+1)^2 (x y-1)^2}\left[81 x^5 (y-1)^2 y^2-2 x^4 y
  \left(213 y^2(y-2) \right.\right.
 \\
&\left.+7838 y-7625\right)+x^3 \left(35850 (y^2-y)-5959\right)-200 x^2 \left(103 (y^2- y) -71\right)
\nonumber \\
&\left.-10523 x+2282\right]
\nonumber \\
&+\frac{\omega_{\rm max}}{\lambda}\frac{4 (1-x) \left( (2 x^3 - 3 x^2)  (y^2-y)+(1-x)^2\right) (\log{\frac{\lambda}{2 \omega_{\rm max}}}+1)}{3 x (x y-1)^2 (x y-x+1)^2}
\nonumber \\
&-\frac{\pi ^2}{1024 x (x y-1)^2 (x y-x+1)^2}
\left[3 x^5 y^2(1-y)^2  -2 x^4 y^3 (y-2) -536 x^4 y^2
\right.
  \nonumber \\
&\left.  +534 x^4 y+(1278 x^3 -744 x^2)(y^2-y)-205 x^3+492 x^2-369 x+82\right].
\nonumber 
\end{align}

We now turn to the discussion of $I_{3,5}$. These cannot be written
in terms of MPLs since the corresponding integrands $G_{3,5}$
involve the second square root $\sqrt{1-c_{12}^2 \beta^2}$. To deal with
this case, we find it convenient to introduce another variable
transformation, namely
\begin{equation}
  \beta = \frac{2\tilde z}{\left(1+s_{12}\right)+\left(1-s_{12}\right)\tilde{z}^2},~~~~~~\tilde
  z\in (0,1).
\end{equation}
With this transformation, the second square root rationalises
\begin{equation}
  \sqrt{1-c_{12}^2\beta^2} = \frac{\left(1+s_{12}\right)-\left(1-s_{12}\right)\tilde z^2}{\left(1+s_{12}\right)+\left(1-s_{12}\right)\tilde z^2}\,,
\end{equation}
while the quadratic term under the first one turns into a quartic polynomial
in $\tilde z$
\begin{equation}
  \sqrt{1-\beta^2} = \frac{1-s_{12}}{\left(1+s_{12}\right)+\left(1-s_{12}\right)\tilde z^2} \times\sqrt{(\tilde z-q_1)(\tilde z-q_2)(\tilde z-q_3)(\tilde z-q_4)}.
\end{equation}
The branch points read
\begin{equation}
q_1=-\frac{1+s_{12}}{1-s_{12}},\;\;\;\;\;\; q_2=-1,\;\;\;\;\;\; q_3=1,\;\;\;\;\;\; q_4=\frac{1+s_{12}}{1-s_{12}}.
\label{eq:branchpointsCparam}
\end{equation}
Integrals containing a square root of a degree four polynomial can be
written in terms of elliptic multiple polylogarithms
(eMPLs)~\cite{Broedel:2017kkb,Broedel:2017siw}, see
appendix~\ref{sec:mpls_empls} for a quick review. We obtain\footnote{We point out that the combinations of the elliptic functions in eq.~\eqref{eq:i3e} in fact evaluates to polylogarithmic functions only. More precisely, the term in
the curly bracket evaluates to $\pi^2/2$. This non-trivial simplification can be derived by going to the iterated Eisenstein series representation.}
\begin{align}
  \label{eq:i3e}
I_3=&\frac{1}{\beta_{\rm min}^5}\frac{7 (x-2)}{16 x}-\frac{1}{\beta_{\rm min}^3}\frac{5 \left(31 x^3 y^2-31 x^3 y-74 x^2 y^2+74 x^2 y+23 x^2-81 x+58\right)}{288 (x (x y-1) (x y-x+1))}
\nonumber \\
&+\frac{1}{\beta_{\rm min}}\frac{1}{1152 x (x (y-1)+1)^2 (x y-1)^2}\left[81 x^5 (y-1)^2 y^2+2 x^4 y \left(-213 y^3+426 y^2\right.\right.
\nonumber \\
&\left.\left.+404 y-617\right)+x^3 \left(-4134 y^2+4134 y+1057\right)+4 x^2 \left(725 y^2-725 y-1099\right)\right.
\nonumber \\
&\left.+5621 x-2282\right]+\frac{\omega_{\rm max}}{\lambda}\frac{4}{3 x} \left[\log{s_{12}}-\log{\frac{\lambda}{2\omega_{\rm max}}}-1\right]
\\
&+\frac{(x-1)^2}{512 s_{12} x (x y-1)^3 (x y-x+1)^3}\left(160 x^4 y^4-320 x^4 y^3+160 x^4 y^2+192 x^3 y^2\right.
\nonumber \\
&\left.-192 x^3 y-68 x^2 y^2+68 x^2 y+37 x^2+45 x-82\right)\left\{\frac{\pi^2}{4}+\Gmpl{\left(0,\frac{s_{12}+1}{1-s_{12}},1\right)}\right.
\nonumber \\
&-\Gmpl{\left(0,\frac{s_{12}+1}{s_{12}-1},1\right)}+\Empl_4{\left(\begin{smallmatrix} -1 & 1 \\ \infty & \frac{s_{12}+1}{1-s_{12}} \end{smallmatrix};1,\vec{q}\right)}-\Empl_4{\left(\begin{smallmatrix} -1 & 1 \\ \infty & \frac{s_{12}+1}{s_{12}-1} \end{smallmatrix};1,\vec{q}\right)}
\nonumber \\
&+\Empl_4{\left(\begin{smallmatrix} -1 & 1 \\ 0 & \frac{s_{12}+1}{1-s_{12}}\end{smallmatrix};1,\vec{q}\right)}-\Empl_4{\left(\begin{smallmatrix}-1 & 1 \\ 0 & \frac{s_{12}+1}{s_{12}-1}\end{smallmatrix};1,\vec{q}\right)}
+\Empl_4{\left(\begin{smallmatrix} -1 & 1 \\ \infty & -1 \end{smallmatrix};1,\vec{q}\right)}
\nonumber \\
&\left.-\Empl_4{\left(\begin{smallmatrix} -1 & 1 \\ \infty & 1 \end{smallmatrix};1,\vec{q}\right)}+\Empl_4{\left(\begin{smallmatrix} -1 & 1 \\ 0 & -1 \end{smallmatrix}; 1,\vec{q}\right)}-\Empl_4{\left(\begin{smallmatrix} -1 & 1 \\ 0 & 1 \end{smallmatrix}; 1,\vec{q}\right)}\right\},
\nonumber 
\end{align}
and\footnote{We note that eq.~\eqref{eq:i5e} contains terms which are individually ill-defined, e.g. $\Empl_4{\left(\begin{smallmatrix} 2 & 1 \\1 & 1 \end{smallmatrix};1,\vec{q}\right)}$. Upon introducing regularised functions and extracting the divergent pieces, it is straightforward to show that all divergent terms cancel and that the final expression is finite.}
\begin{align}
  \label{eq:i5e}
I_5=&-\frac{1}{\beta_{\rm min}^5}\frac{7 (x-2)}{16 x}+\frac{1}{\beta_{\rm min}^3}\frac{5 \left(31 x^3 (y-1) y+x^2 \left(-74 y^2+74 y+23\right)-81 x+58\right)}{288 x (x (y-1)+1) (x y-1)}
\nonumber \\
&+\frac{1}{\beta_{\rm min}}\frac{1}{1152 x (x (y-1)+1)^2 (x y-1)^2}\left[-81 x^5 (y-1)^2 y^2+2 x^4 y \left(213 y^3-426 y^2\right.\right.
\nonumber \\
&\left.\left.-404 y+617\right)+x^3 \left(4134 y^2-4134 y-1057\right)+x^2 \left(-2900 y^2+2900 y+4396\right)\right.
\nonumber\\
&\left.-5621 x+2282\right]-\frac{\pi^2 (x-1)^2}{1024 s_{12} x (x (y-1)+1)^3 (x y-1)^3}\times
\nonumber \\
&\times\left(160 x^4 (y-1)^2 y^2+192 x^3 (y-1) y+x^2 \left(-68 y^2+68 y+37\right)+45 x-82\right)
\nonumber \\
&+\frac{1}{1024 x (x y-1)^2 (x y-x+1)^2}\left[3 x^5 y^4-6 x^5 y^3+3 x^5 y^2-2 x^4 y^4+4 x^4 y^3-196 x^4 y^2\right.
\nonumber \\
&\left.+194 x^4 y+430 x^3 y^2-430 x^3 y-37 x^3-236 x^2 y^2+236 x^2 y-8 x^2+127 x-82\right]\times
\nonumber \\
&\times \left[-4 \Empl_4{\left(\begin{smallmatrix} -1 & 1 \\ 0 & \frac{1+s_{12}}{1-s_{12}}\end{smallmatrix};1,\vec{q}\right)}+4 \Empl_4{\left(\begin{smallmatrix} -1 & 1 \\ 0 & \frac{1+s_{12}}{-1+s_{12}}\end{smallmatrix};1,\vec{q}\right)}-4 \Gmpl{\left(0,\frac{s_{12}+1}{1-s_{12}};1\right)}\right.
\nonumber \\
&\left.+4 \Gmpl{\left(0,\frac{s_{12}+1}{s_{12}-1};1\right)}+\pi^2\right]+\Empl_4{\left(\begin{smallmatrix} 0 \\ 0 \end{smallmatrix};1,\vec{q}\right)} \left[\frac{5 \pi ^2 (x-1)^2 \left(x^2 (y-1) y+1\right)}{8 s_{12} x (x (y-1)+1)^2 (x y-1)^2}\frac{\eta_1}{\omega_1}\right.
\nonumber \\
&\left.+\frac{5 \pi ^2 (x-1)^2 \left(2 x^4 (y-1)^2 y^2+7 x^3 (y-1) y+x^2 \left(-8 y^2+8 y+3\right)-5 x+2\right)}{96 s_{12} x (x (y-1)+1)^3 (x y-1)^3}\right]
\nonumber \\
&+\left(\frac{5 (x-1)^2 \left(x^2 (y-1) y+1\right)}{8 s_{12} x (x (y-1)+1)^2 (x y-1)^2}+\frac{5 (x-1)^2 \left(x^2 (y-1) y+1\right)}{8 x (x (y-1)+1)^2 (x y-1)^2}\right) \Empl_4{\left(\begin{smallmatrix}-1 & 1 \\ \infty & \frac{s_{12}+1}{1-s_{12}} \end{smallmatrix};1,\vec{q}\right)}
\nonumber \\
&+\left(-\frac{5 (x-1)^2 \left(x^2 (y-1) y+1\right)}{8 s_{12} x (x (y-1)+1)^2 (x y-1)^2}-\frac{5 (x-1)^2 \left(x^2 (y-1) y+1\right)}{8 x (x (y-1)+1)^2 (x y-1)^2}\right)\times
\nonumber \\
&\times \Empl_4{\left(\begin{smallmatrix} -1 & 1 \\ \infty & \frac{s_{12}+1}{s_{12}-1} \end{smallmatrix};1,\vec{q}\right)}+\left(\frac{5 (x-1)^2 \left(x^2 (y-1) y+1\right)}{8 s_{12} x (x (y-1)+1)^2 (x y-1)^2}\right.
\nonumber \\
&\left.+\frac{-3 x^3 (y-1) y+x^2 \left(2 y^2-2 y+37\right)-115 x+78}{256 x (x (y-1)+1) (x y-1)}\right) \Empl_4{\left(\begin{smallmatrix} -1 & 1 \\ \infty & -1 \end{smallmatrix};1,\vec{q}\right)}
\nonumber \\
&+\Empl_4{\left(\begin{smallmatrix} -1 & 1 \\ \infty & 1 \end{smallmatrix};1,\vec{q}\right)}\left(\frac{3 x^3 (y-1) y+x^2 \left(-2 y^2+2 y-37\right)+115 x-78}{256 x (x (y-1)+1) (x y-1)}\right.
\\
&\left.-\frac{5 (x-1)^2 \left(x^2 (y-1) y+1\right)}{8 s_{12} x (x (y-1)+1)^2 (x y-1)^2}\right)+\frac{5 (x-1)^2 \left(x^2 y^2-x^2 y+1\right)}{16 s_{12} x (x y-1)^2 (x y-x+1)^2}\times
\nonumber \\
&\times\left[-\Empl_4{\left(\begin{smallmatrix} 2 & 1 \\ \frac{s_{12}+1}{1-s_{12}} & \frac{s_{12}+1}{1-s_{12}} \end{smallmatrix};1,\vec{q}\right)}+ \Empl_4{\left(\begin{smallmatrix} 2 & 1 \\ \frac{s_{12}+1}{1-s_{12}} & \frac{s_{12}+1}{s_{12}-1} \end{smallmatrix};1,\vec{q}\right)}+ \Empl_4{\left( \begin{smallmatrix} 2 & 1 \\ \frac{s_{12}+1}{s_{12}-1} & \frac{s_{12}+1}{1-s_{12}} \end{smallmatrix};1,\vec{q}\right)}\right.
\nonumber \\
&- \Empl_4{\left(\begin{smallmatrix} 2 & 1 \\ \frac{s_{12}+1}{s_{12}-1} & \frac{s_{12}+1}{s_{12}-1} \end{smallmatrix};1,\vec{q}\right)}+ \Empl_4{\left(\begin{smallmatrix} 2 & 1 \\ -1 & -1 \end{smallmatrix};1,\vec{q}\right)}- \Empl_4{\left(\begin{smallmatrix} 2 & 1 \\ -1 & 1 \end{smallmatrix};1,\vec{q}\right)}- \Empl_4{\left(\begin{smallmatrix} 2 & 1 \\ 1 & -1 \end{smallmatrix};1,\vec{q}\right)}
\nonumber \\
&+ \Empl_4{\left(\begin{smallmatrix} 2 & 1 \\ 1 & 1 \end{smallmatrix};1,\vec{q}\right)}+ Z_4{\left(1\right)}\left( \Gmpl{\left(\frac{s_{12}+1}{s_{12}-1},\frac{s_{12}+1}{s_{12}-1};1\right)}+ \Gmpl{\left(\frac{s_{12}+1}{1-s_{12}},\frac{s_{12}+1}{1-s_{12}};1\right)}-\frac{\pi^2}{12}\right.
\nonumber \\
&- \Gmpl{\left(\frac{s_{12}+1}{1-s_{12}},\frac{s_{12}+1}{s_{12}-1},1\right)} - \Gmpl{\left(\frac{s_{12}+1}{s_{12}-1},\frac{s_{12}+1}{1-s_{12}};1\right)}+\Empl_4{\left(\begin{smallmatrix} 1 & 1 \\ 1 & -1 \end{smallmatrix};1,\vec{q}\right)}
\nonumber \\
&\left.\left.-\Empl_4{\left(\begin{smallmatrix} 1 & 1 \\1 & 1 \end{smallmatrix};1,\vec{q}\right)}\right)\right],
\nonumber 
\end{align}
where $Z_4$, $\omega_1$ and $\eta_1$ are defined for a generic elliptic
curve in eqs.~(\ref{eq:definZ4}, \ref{eq:period1}, \ref{eq:quasi1}).
In our case they read
\begin{equation}
  Z_4(1) = \frac{2\pi i}{\omega_1},
  ~~~~~
    \omega_1 =  2 K(c_{12}^2),
  ~~~~~
  \eta_1 = E(c_{12}^2)-\frac{1+s_{12}^2}{3}K(c_{12}^2),
\end{equation}
where $K$ and $E$ are the complete elliptic integrals of the first
and second kind, see eq.~\eqref{eq:EK}. 

Combining the five contributions eqs.~(\ref{eq:i1e}, \ref{eq:i2e},
\ref{eq:i4e}, \ref{eq:i3e}, \ref{eq:i5e}), we observe that terms that
are singular in the limit $\beta_{\rm min} \to 0$ cancel out so that
this limit can be taken safely.  Unfortunately, apart from this, no
significant simplifications are observed and we are left with a
complicated linear combination of elliptic multiple polylogarithms
$\Empl_4$.

\subsection{Result in terms of Eisenstein integrals}
\label{sec:eisen}
To proceed further, we rewrite the result of the previous
section in terms of the so-called $\tilde\Gamma$ iterated integrals
\begin{equation}
  \tilde{\Gamma}{\left(\begin{smallmatrix} n_1 & ... & n_m \\ z_1 &
      ... & z_m \end{smallmatrix};z,\tau\right)}=\int \limits_0^z dz'
  g^{\left(n_1\right)}{\left(z'-z_1,\tau\right)}\tilde{\Gamma}{\left(\begin{smallmatrix}
      n_2 & ... & n_m \\ z_2 & ...  &
      z_m \end{smallmatrix};z',\tau\right)},
\end{equation}
see
refs.~\cite{Broedel:2017kkb,Broedel:2017siw,Broedel:2018iwv,Broedel:2018qkq,Broedel:2019hyg,Weinzierl:2022eaz}
for details and appendix~\ref{sec:eisen_rev} for a quick overview. Here,
$\tau$ is the ratio of the \emph{periods} of the torus associated with the elliptic
curve -- see eq.~\eqref{eq:taudef} -- and the $g^{(n)}$ functions are
integration kernels whose exact form is irrelevant for our
discussion. What is important is that in our case the kernels only
have to be evaluated at specific, \emph{rational} points. Indeed, upon
inspection we find that in our calculation we only need to consider
the origin and three half periods
\begin{equation}
z_{q_1}=0,\hspace{1cm} z_{q_2}=\frac{\tau}{2},\hspace{1cm} z_{q_3}=\frac{\tau}{2}+\frac{1}{2},\hspace{1cm}
z_{q_4}=\frac{1}{2},
\label{eq:e20}
\end{equation}
as well as three additional points 
\begin{equation}
  z_0=\frac{1}{4}+\frac{\tau}{2},\hspace{1cm}
  z_{\infty}=\frac{1}{4}, \hspace{1cm} z_{-\infty}=\frac{1}{4}.
  \label{eq:Abelmapval2}
\end{equation}
Because of this, major simplifications happen. Indeed, the full result
can be written as an integral of modular forms and can be expressed
in terms of the so-called iterated Eisenstein integrals
\begin{equation}
  I{\left(f_1,...,f_n;\tau\right)}
  =\int \limits_{i\infty}^{\tau} \frac{d\tau'}{2\pi i}f_1{\left(\tau'\right)}I{\left(f_2,...,f_n;\tau'\right)},
\label{eq:iterEisenIntDef0}
\end{equation}
see
refs.~\cite{Adams:2017ejb,Broedel:2018iwv,Abreu:2019fgk,Duhr:2019rrs,Weinzierl:2022eaz,Abreu:2022vei}
for details and appendix~\ref{sec:eisen_rev} for a brief review. In
terms of these functions, we can write our result for the linear
$\lambda$ term of $I_C$ in eq.~\eqref{eq:3.3} as
\begin{equation}
  \mathcal T_\lambda\left[I_C\right] =
  -\frac{3}{4\pi^3}\frac{\lambda}{q}(t_1+t_2+t_3), 
\end{equation}
with
\begin{align}
t_1=&\frac{5 (x-1)^2 \left(x^2 (y-1) y+1\right)}{4 \pi ^2 \omega_1 s_{12} x (x (y-1)+1)^2 (x y-1)^2}\times
\nonumber \\
&\times\left[I{\left(1,h^{(3)}_{4,1,2},\tau\right)} \left(-2 \omega_1 (s_{12}+1) g^{\left(1\right)}{\left(\frac{1}{4},\tau\right)}-g^{\left(2\right)}{\left(\frac{3}{4},\tau\right)}\right.\right.
\nonumber \\
&\left.\hspace{0.6cm}+g^{\left(2\right)}{\left(-\frac{1}{4},\tau\right)}+\omega_1^2 (s_{12}+1)^2\right)
\label{eqf4}  \\
&\hspace{0.6cm}+i \pi  I{\left(h^{(3)}_{4,1,0},\tau\right)} \left(-g^{\left(1\right)}{\left(\frac{1}{4},\tau\right)}+g^{\left(1\right)}{\left(\frac{3}{4},\tau\right)}+\omega_1 s_{12}+\omega_1\right)
\nonumber \\
&\hspace{0.6cm}\left.+i \pi  I{\left(h^{(3)}_{4,1,2},\tau\right)} \left(g^{\left(1\right)}{\left(\frac{\tau}{2}+\frac{1}{4},\tau\right)}-g^{\left(1\right)}{\left(\frac{\tau}{2}+\frac{3}{4},\tau\right)}+\omega_1 (s_{12}-1)\right)\right],
\nonumber \\
\nonumber \\
t_2=&\frac{1}{384 \pi  s_{12} x \left(x^2 (y-1) y+x-1\right)^3}\times
\nonumber \\
&\times\left[-\frac{6 \tilde{C} (x (y-1)+1) (x y-1)}{\omega_1} \left(\omega_1 g^{\left(1\right)}{\left(\frac{1}{4},\tau\right)} \left(s_{12} (x (y-1)+1) (x y-1) \times \right.\right.\right.
\nonumber \\
&\left.\times (x (x ((3 x-2) (y-1) y-37)+115)-78)-160 (x-1)^2 \left(x^2 (y-1) y+1\right)\right)
\nonumber \\
&\left.+80 (x-1)^2 \left(x^2 (y-1) y+1\right) \left(g^{\left(2\right)}{\left(\frac{\tau}{2}-\frac{1}{4},\tau\right)}-g^{\left(2\right)}{\left(\frac{\tau}{2}+\frac{3}{4},\tau\right)}\right)\right)
\label{eqf5} \\
&+3 \tilde{C} \omega_1 (s_{12}+1) (x (y-1)+1) (x y-1) \left(s_{12} (x (y-1)+1) (x y-1) \times \right.
\nonumber \\
&\left.\times(x (x ((3 x-2) (y-1) y-37)+115)-78)-160 (x-1)^2 \left(x^2 (y-1) y+1\right)\right)
\nonumber \\
&+5 \pi ^3 (x-1)^2 \left(12 \eta_1 (x (y-1)+1) (x y-1) \left(x^2 (y-1) y+1\right)\right.
\nonumber \\
&\left.\left.+\omega_1 (x (x ((y-1) y (x (2 x (y-1) y+7)-8)+3)-5)+2)\right)\right],
\nonumber \\
\nonumber \\
t_3=&-\frac{1}{128 \pi ^2 \omega_1 s_{12} x \left(x^2 (y-1) y+x-1\right)^2} I{\left(1,h^{(3)}_{4,1,0},\tau\right)} \times 
\nonumber \\
&\times \left[2 \omega_1 g^{\left(1\right)}{\left(\frac{1}{4},\tau\right)} \left(s_{12} (x (y-1)+1) (x y-1) (x (x ((3 x-2) (y-1) y-37)\right.\right.
\nonumber \\
&\left.+115)-78)\right)-160 (x-1)^2 \left(x^2 (y-1) y+1\right)\times
\label{eqf6}
\\
&\times\left(2 \omega_1 g^{\left(1\right)}{\left(\frac{1}{4},\tau\right)}+g^{\left(2\right)}{\left(\frac{\tau}{2}+\frac{3}{4},\tau\right)}-g^{\left(2\right)}{\left(\frac{\tau}{2}-\frac{1}{4},\tau\right)}\right)
\nonumber \\
&+\omega_1^2 (s_{12}+1) \left(160 (x-1)^2 \left(x^2 (y-1) y+1\right)\right.
\nonumber \\
&\left.\left.-s_{12} (x (y-1)+1) (x y-1) (x (x ((3 x-2) (y-1) y-37)+115)-78)\right)\right].
\nonumber 
\end{align}
In these equations, $h^{(n)}_{N,r,s}$ are specific kernels of the iterated
integrals (labeled generically $f_1$ in
eq.~\eqref{eq:iterEisenIntDef0}, see appendix~\ref{sec:eisen_rev} for
their definition) and $\tilde{C}\approx 0.915966$ stands for the
Catalan constant.

Expressing our result in terms of iterated Eisenstein integrals serves
two purposes: on the one hand, it allows us to write $\Empl_4$
integrals in terms of integrals for which rapidly-convergent series
representations exist and, on the other hand, it may allow us to find
non-trivial relations between $\Empl_4$ integrals that may lead to
significant simplifications in the above expressions. This is because
the kernel $g^{(n)}$ evaluated at rational points are related
to the periods of the elliptic curve. In our case, we find the following
relations
\begin{align}
g^{\left(1\right)}{\left(\frac{3}{4},\tau\right)}=&-\frac{1}{2} \omega_1(s_{12}+1),
\\
g^{\left(1\right)}{\left(\frac{\tau}{2}+\frac{1}{4},\tau\right)}
=& g^{\left(1\right)}{\left(\frac{\tau}{2}+\frac{3}{4},\tau\right)}-\omega_1 (s_{12}-1),
\\
g^{\left(2\right)}{\left(\frac{\tau}{2}-\frac{1}{4},\tau\right)}=& g^{\left(2\right)}{\left(\frac{\tau}{2}+\frac{3}{4},\tau\right)},
\\
g^{\left(2\right)}{\left(-\frac{1}{4},\tau\right)}=& g^{\left(2\right)}{\left(\frac{3}{4},\tau\right)},
\\
g^{\left(1\right)}{\left(\frac{1}{4},\tau\right)}=& -g^{\left(1\right)}{\left(\frac{3}{4},\tau\right)}.
\end{align}
Upon using these relations all iterated Eisenstein integrals cancel
out and we obtain a remarkably simple expression shown in
eq.~(\ref{eq:tlIc}).

\section{Brief introduction to elliptic multiple polylogarithms}
\label{sec:brief}
In this appendix, we give a very brief overview of the techniques that
we have used for the calculation in appendix~\ref{sec:ellint}. First, in
appendix~\ref{sec:mpls_empls} we recall the definition of multiple
polylogarithms and their elliptic generalisation, and discuss some
properties of the latter. Then, in appendix~\ref{sec:eisen_rev} we discuss
the so-called $\tilde\Gamma$ representation and the simplifications
that occur if one only needs to evaluate $\tilde\Gamma$ iterated
integrals at special points, as is the case in our calculation.

\subsection{Multiple polylogarithms and elliptic multiple polylogarithms}
\label{sec:mpls_empls}
Multiple polylogarithms (MPLs) are by-now standards in the high-energy
physics community. They can be defined recursively as iterated integrals
of the form~\cite{Goncharov:1998kja,Goncharov:2001iea}
\begin{equation}
\Gmpl{\left(a_1,...,a_n;z\right)}=\int \limits_{0}^{z} dt \frac{1}{t-a_1}\Gmpl{\left(a_2,...,a_n;t\right)}.
\label{eqc1}
\end{equation}
To terminate the recursion, we require  $\Gmpl{\left(;t'\right)}=1$.
In the special case when all indices $a_j$, $j \le n$ are equal to zero, we define 
\begin{equation}
\Gmpl{\left(\vec{0}_n;t\right)}=\frac{1}{n!}\log^n{t}, 
\end{equation}
where $\vec{0}_n$ is a set of $n$ zeroes. Public tools for efficiently
dealing with MPLs are
available~\cite{Panzer:2014caa,Duhr:2019tlz}. For our calculations, we
used the \texttt{PolyLogTools} package~\cite{Duhr:2019tlz}.

Multiple polylogarithms, as defined above, can be used to compute
integrals $I_1$, $I_2$ and $I_4$. To this end, we need to write $\ln[
  (1+z)/(1-z)]$ and $\ln^2[(1+z)/(1-z)]$ in terms of MPLs, do a
partial fractioning with respect to $z$ and then integrate the
resulting expressions using eq.~\eqref{eqc1}. If the required
integrals are not in the canonical form, we integrate by parts several
times until the canonical form is reached.  Applying this procedure to
the calculation of functions $I_{1,2,4}$, we obtain a result written
in terms of MPLs which depends upon $\lambda/\omega_{\rm max}$ and
$\beta_{\rm min}$.  However, since both of these quantities only
appear in the final result because of the $z$-integration boundaries,
it is relatively straightforward to derive an expansion of the
integrals $I_{1,2,4}$ in these parameters.

As we have already mentioned, if an additional square root is present
as is the case for $G_3$ and $G_5$, it is not possible to write the
result of the integration in terms of multiple polylogarithms and the
so-called elliptic multiple polylogarithms (eMPLs) are needed.
Although elliptic functions are well-known in the mathematical
community, they are relatively new to particle theorists and their
application to particle physics problems was started being explored
only recently. For further reading and additional references, we refer
the reader to
refs.~\cite{Broedel:2017kkb,Broedel:2017siw,Broedel:2018iwv,Broedel:2018qkq,Broedel:2019hyg,Weinzierl:2022eaz}.
Below we give a brief review on elliptic multiple polylogarithms.

Elliptic curves are characterised by either a cubic or a quartic
polynomial under the square root. In the calculation of Feynman
integrals, the quartic case is much more common and for this reason we
will focus on it in what follows.
Similarly to MPLs, eMPLs \cite{Broedel:2017kkb,Broedel:2017siw} can be
defined recursively using the following equation
\begin{equation}
  \Empl_4{\left(\begin{smallmatrix} n_1 & ... & n_m \\ c_1 & ... & c_m \end{smallmatrix}; z, \vec{q}\right)}=\int \limits_0^z dt\, \psi_{n_1}{\left(c_1,t,\vec{q}\right)} \Empl_4{\left(\begin{smallmatrix} n_2 & ... & n_m \\ c_2 & ... & c_m \end{smallmatrix}; t, \vec{q}\right)},
  \label{eqiter}
\end{equation}
where  the recursion terminates with $\Empl_4{\left(; t, \vec{q}\right)}=1$.
We discuss the kernels  $\psi_n{\left(c_1,t,\vec{q}\right)}$ below. It is clear that these kernels should contain a square root of a quartic polynomial that we denote as $P_4(x)$. To simplify the
notation, we define an ``elliptic curve''  
\begin{equation}
y^2=P_4{\left(x\right)}=\left(x-q_1\right)\left(x-q_2\right)\left(x-q_3\right)\left(x-q_4\right).
\end{equation}
For our case, cf. eq.~\eqref{eq:branchpointsCparam}, all roots are
real. We enumerate them in such a way that
\begin{equation}
q_1<q_2<q_3<q_4,
\label{eq:c6}
\end{equation}
and define a vector $\vec q = (q_1,q_2,q_3,q_4)$. 
We note that the ordering eq.~(\ref{eq:c6}) applies if $s_{12} > 0$, which
is always the case in our situation.

In spite of the fact that we are interested in integration over the
real axis, it is important to consider the integration in definition
eq.~\eqref{eqiter} in the complex plane. To define
an analytic function in the complex plane, we require three cuts and
we choose them as cuts along the real axis at the intervals $[q_4, +
  \infty]$, $[q_2,q_3]$ and $[-\infty,q_1]$. The integration contour
is then defined through the following values of the elliptic curve on
these intervals \cite{Broedel:2019hyg}
\begin{equation}
y=\sqrt{P_4{\left(x\right)}}=\sqrt{\left\vert P_4{\left(x\right)}\right\vert}\times\begin{cases}-1 & x\leq q_1\lor x> q_4, \\
-i & q_1 < x\leq q_2, \\
1 & q_2 < x\leq q_3, \\
i & q_3 < x\leq q_4.
\end{cases}
\end{equation}

The choice of kernels $\psi$ is dictated by the fact that we should
parameterise all relevant independent integrands that we can face in
our computation, modulo integration by parts.  It turns out that for
our purposes the following kernels are sufficient
\begin{equation}
\begin{split}
& \psi_0{\left(0,x,\vec{q}\right)}=\frac{c_4}{y}, \qquad  \qquad \qquad \;\;\;\;\;\;\;\;\psi_1{\left(c,x,\vec{q}\right)}=\frac{1}{x-c},
\\
& \psi_{-1}{\left(c,x,\vec{q}\right)}=\frac{y_c}{y\left(x-c\right)}-\frac{\delta_{0c}}{x},
\qquad 
\psi_1{\left(\infty,x,\vec{q}\right)}=\frac{c_4}{y}Z_4{\left(x\right)},
\\
& \psi_{-1}{\left(\infty,x,\vec{q}\right)}= \frac{x}{y},
\qquad \quad
\psi_{2}{\left(c,x,\vec{q}\right)}=\frac{1}{x-c}Z_4{\left(x\right)}
-\Phi_4{\left(x\right)}-\frac{\delta_{0c}}{x}Z_4{\left(0\right)},
\end{split}
\label{eq:kernellist}
\end{equation}
where we defined  the following quantities
\begin{align}
& y_c=\sqrt{P_4{\left(c\right)}},\;\;\;\;\;\;\;\;\;\;\;\;\;  c_4=\frac{1}{2}\sqrt{q_{13}q_{24}},\;\;\;\;\;\;\;\;\;\;\;\;\; q_{ij}=q_i-q_j,
    \\
& Z_4{\left(x\right)}=\int \limits_{q_1}^x dx' \Phi_4{\left(x'\right)},
\;\;\;\;\;\;\;\;\;\;\; \Phi_4{\left(x\right)}=\tilde{\Phi}_4{\left(x\right)}+4c_4 \frac{\eta_1}{\omega_1}\frac{1}{y},
\label{eq:definZ4}
\\
& \tilde{\Phi}_4{\left(x\right)}=\frac{1}{c_4 y}\left(x^2
-\frac{x}{2} \sum \limits_{i=1}^{4} q_i +\frac{1}{6}\sum_{i=1}^4 q_i \sum_{j>i}^4 q_j\right).
\end{align}
We also note that the kernel $\psi_1{\left(c,x,\vec{q}\right)}$ can be
identified as the usual polylogarithmic kernel and that MPLs that we
defined in the preceding section form a subset of eMPLs. Explicitly,
we have
\begin{equation}
\Empl_4{\left(\begin{smallmatrix} \vec{1}_m \\ \vec{c}_m \end{smallmatrix}; z, \vec{q}\right)}=\Gmpl{\left(\vec{c}_m; z\right)}.
\end{equation}

Finally, we note that integrals of the function $y(x)$ between its various  roots are called periods and
quasi-periods. They are defined as 
\begin{align}
\omega_1=&2c_4\int \limits_{q_2}^{q_3} \frac{ dx }{y}=2 K{\left(\rho\right)},
\label{eq:period1}
\\
\omega_2=&2c_4\int \limits_{q_1}^{q_2} \frac{ dx}{y}=2i K{\left(1-\rho\right)},
\label{eq:period2}
\\
\eta_1=&-\frac{1}{2}\int \limits_{q_2}^{q_3} dx  \tilde{\Phi}_4{\left(x\right)}=E{\left(\rho\right)}-\frac{2-\rho}{3}K{\left(\rho\right)}
\label{eq:quasi1}
\\
\eta_2=&-\frac{1}{2}\int \limits_{q_1}^{q_2}dx \tilde{\Phi}_4{\left(x\right)}=-i E{\left(1-\rho\right)}+i\frac{1+\rho}{3}K{\left(1-\rho\right)},
\label{eq:quasi2}
\end{align}
where 
\begin{equation}
\rho=\frac{q_{14}q_{23}}{q_{13}q_{24}}, 
\end{equation}
and the complete elliptic integrals of the first and second kind read  
\begin{align}
K{\left(\rho\right)}=&\int \limits_{0}^{1}  \frac{dx}{\sqrt{\left(1-x^2\right)\left(1-\rho x^2\right)}},\;\;\;\;\;
E{\left(\rho\right)}=\int \limits_{0}^{1} dx  \frac{\sqrt{1-\rho x^2}}{\sqrt{1-x^2}}.
\end{align}
A relation which we will use later is the so-called Legendre equation
which relates the elliptic periods to the elliptic quasi-periods:
\begin{equation}
\omega_1 \eta_2 -\omega_2 \eta_1=-i\pi.
\label{eq:legendreeq}
\end{equation}

We note that with this formalism, it is straightforward to integrate
$G_3$ and $G_5$ over $\beta$ (or $z$) in a fully-algorithmic way.  The
integration amounts to an identification of the iterative structures
as in eq.~\eqref{eqiter} and writing the resulting integrals in the
canonical form. This way, the problem of integration is mapped into a
much simpler algebraic one. There are currently no public tools to deal with eMPL functions. We employ our own codes to perform the integrations involving elliptic functions. These build on an extended version of the \texttt{PolyLogTools} package~\cite{Duhr:2019tlz}.\footnote{We are grateful to the authors of \cite{Duhr:2019tlz} for making it available.}

\subsection{From eMPLs to iterated Eisenstein integrals}
\label{sec:eisen_rev}
There is a profound connection between elliptic multiple
polylogarithms and the geometry of the torus, and there exists a
particular representation of these functions, called $\tilde{\Gamma}$
representation, that is related to this geometry. In the following we
summarise key properties of the $\tilde{\Gamma}$ representation and
explain why it is useful for simplifying elliptic integrals $\Empl_4$
discussed in the preceding section. We refer to the literature for
additional details
\cite{Broedel:2017kkb,Broedel:2017siw,Broedel:2018iwv,Broedel:2018qkq,Broedel:2019hyg,Weinzierl:2022eaz}.

The connection between the torus geometry and elliptic curves is most clearly illustrated by the Weierstrass
function defined on a torus. To introduce this function, we define a torus as a two-dimensional lattice
\begin{equation}
\Lambda_{\tau}=\mathbb{Z}+\mathbb{Z}\tau=\{m+n\tau\vert m,n\in \mathbb{Z}\},
\end{equation}
where $\tau$ is a complex number given by the ratio of  two elliptic periods
$\omega_{1,2}$, cf. eqs.~(\ref{eq:period1}, \ref{eq:period2})
\begin{equation}
  \tau=\frac{\omega_2}{\omega_1}.
  \label{eq:taudef}
\end{equation}
A torus is double-periodic with the periods one and $\tau$. One can
think about points on a torus as points on a complex plane subject to
double-periodicity conditions.

The Weierstrass function $\wp$ is defined on the torus. It reads
\begin{equation}
\wp{\left(z\right)}=\frac{1}{\omega_1^2}\left[\frac{1}{z^2}+\sum_{\left(m,n\right)\neq
    \left(0,0\right)}\left(\frac{1}{\left(z+m+n
    \tau\right)^2}-\frac{1}{\left(m+n \tau\right)^2}\right)\right].
\label{eq:definweierstrass}
\end{equation}
This function is double-periodic with respect to shifts $z\rightarrow
z+m+n\tau$ for $n,m\in \mathbb{Z}$. It is an even function
$\wp{\left(-z\right)}=\wp{\left(z\right)}$ and has a double pole at
each lattice point in $\Lambda_{\tau}$. In contrast to this, its
derivative $\wp'$ is an odd function and has a triple pole at each
lattice point. For convenience, we introduce a re-scaled derivative of
the Weierstrass function $\tilde{\wp}'{\left(z\right)}$ defined as
\begin{equation}
\tilde{\wp}'{\left(z\right)}=\frac{\wp'{\left(z\right)}}{\omega_1}=-\frac{2}{\omega_1^3}\left[\frac{1}{z^3}+\sum_{\left(m,n\right)\neq
    \left(0,0\right)}\frac{1}{\left(z+m+n \tau\right)^3}\right].
\end{equation}
The important role of the Weierstrass function and its derivative is
emphasised by the so-called Weierstrass equation\footnote{
Specifically, $\wp_{1} = \wp(1/2)$, $\wp_{2} = \wp(\tau/2)$ and
$\wp_3 = \wp\big((1+\tau)/2\big)$.}
\begin{equation}
\tilde{\wp}'^2{\left(z\right)}=4\left(\wp{\left(z\right)}-\wp_1\right)\left(\wp{\left(z\right)}-\wp_2\right)\left(\wp{\left(z\right)}-\wp_3\right), \hspace{0.5cm}\text{
  with }\hspace{0.5cm} \wp_1+\wp_2+\wp_3=0,
\end{equation}
which is indeed very similar to an elliptic curve described by a cubic
polynomial
\begin{equation}
  y^2=4\left(x-\tilde{q}_1\right)\left(x-\tilde{q}_2\right)\left(x-\tilde{q}_3\right), \hspace{0.5cm}\text{ with }\hspace{0.5cm} \tilde{q}_1+\tilde{q}_2+\tilde{q}_3=0.
  \label{ecurve3}
\end{equation}

To make this connection explicit, we identify $x$ with
$\wp{\left(z\right)}$ and $y$ with $\tilde{\wp}'{\left(z\right)}$,
which defines a map from a point on the torus $z$ to a point on an
elliptic curve
\begin{equation}
\left(x,y\right)=\left(\wp{\left(z\right)},\tilde{\wp}'{\left(z\right)}\right), 
\label{eq:mappingxyw}
\end{equation}
with  $\wp_{i} = \tilde q_i$, $i=1,2,3$. 
This transformation is convenient, as it allows us to express elliptic
integrals involving $dx/y$ directly as a simple integral between two
points on the torus
\begin{equation}
\int \frac{dx}{y}=\omega_1 \int dz.
\end{equation}
As outlined
in the previous section,  we need several kernels to construct a basis for elliptic integrals, so it is important to understand whether all of these kernels can be simplified by introducing the mapping as in eq.~(\ref{eq:mappingxyw}).
To show that this is indeed possible, 
we consider  the cubic elliptic curve as in  eq.~(\ref{ecurve3}) 
and  discuss the following integral
\begin{equation}
I_e=\int \frac{dx}{y}\frac{y_c}{x-c}.
\label{eq:intellkernel}
\end{equation}
Using the mapping in eq.~\eqref{eq:mappingxyw}, we re-write it as
\begin{equation}
I_e=\int dz \mu(z), \;\;\;\;\; \mu(z) = \frac{\wp'{\left(z_c\right)}}{\wp{\left(z\right)}-\wp{\left(z_c\right)}}.
\label{eq:phikernelint}
\end{equation}
The point $z_c$ is defined implicitly by the equation $c=\wp{\left(z_c\right)}$.

We now discuss properties of the integrand in
eq.~\eqref{eq:phikernelint}.  It is clear that $\mu{\left(z\right)}$
is a double-periodic function of $z$ with a simple pole at $z_c$. In
fact, since the function $\wp{\left(z\right)}$ is an even function of
$z$, there is also a pole at $z=-z_c$. By expanding around $z=\pm z_c$
and using the fact that the derivative is an odd function of $z$, one
can show that the residue at $z=\pm z_c$ is $\pm 1$.  Therefore, we
can write $\mu(z)$ as
\begin{equation}
\mu{\left(z\right)}=\frac{r{\left(z,z_c,\tau\right)}}{\left(z-z_c\right)\left(z+z_c\right)},
\end{equation}
where the function $r{\left(z,z_c,\tau\right)}$ is regular in $z$ and
has a double zero at $z=0$.\footnote{To see this, we note that
$\wp{\left(z\right)}$ scales as $1/z^2$ in
eq.~\eqref{eq:definweierstrass}. Therefore, $r\left(z,z_c,\tau\right)$
scales as $\propto z^2 z_c^2$.}
We now need to look for a suitable function on the torus that allows
us to parameterise $\mu{\left(z\right)}$. For this, we introduce the
periodic function $E{\left(z\right)}$ that has a simple pole at $z =
0$ with residue $+1$.  It is defined as\footnote{Note that one has
to choose a particular order of the sums as they do not commute. If
the summation over $m$ is performed first, $E{\left(z\right)}$ is
periodic with respects to shifts $z\rightarrow z+m$ but not with
respect to shifts $z\rightarrow z+n\tau$.  In the latter case, $E(z)$
acquires an additional term $-2\pi n i$.}
\begin{equation}
E{\left(z\right)}=\sum_{\left(m,n\right)}\frac{1}{\left(z+m+n \tau\right)}.
\end{equation}
Taking into account the two poles at $z=\pm z_c$ and the corresponding residues, we can write
 $\mu{\left(z\right)}$ as
\begin{equation}
  \mu{\left(z\right)}=E{\left(z-z_c\right)}-E{\left(z+z_c\right)}+d{\left(z\right)},
  \label{eq:e15}
\end{equation}
where $d{\left(z\right)}$ is a regular function. It is straightforward to verify that, thanks
to the relative
sign between two  $E{\left(z\right)}$ functions, the first two terms on the right hand side
of eq.~(\ref{eq:e15}),  when taken together, 
are indeed double-periodic
under shifts of $z$. What remains to be done, is to determine the function $d{\left(z\right)}$.
Since all functions except $d(z)$ in eq.~(\ref{eq:e15}) are
double-periodic, $d(z)$ must be double-periodic as well.  The function
$d(z)$ has no poles. Since this is an elliptic function, it must have
the same number of poles as the number of zeroes. Since $d(z)$ has no
poles and no zeroes, it must be a constant and its value can be
determined from the boundary condition $\mu{\left(0\right)}=0$. Hence,
we obtain
\begin{equation}
\mu{\left(z\right)}=E{\left(z-z_c\right)}-E{\left(z+z_c\right)}+2E{\left(z_c\right)},
\end{equation}
where we used the fact that $E{\left(-z_c\right)}=-E{\left(z_c\right)}$.
It follows that for the integral $I_e$ in eq.~(\ref{eq:intellkernel}), the function $E(z)$ defines a suitable
kernel on the torus. 

One can generalise this procedure to other elliptic kernels by introducing functions
on the torus that exhibit higher-order poles.
 The corresponding functions  are given by
the  $g^{\left(n\right)}$ kernels which are the coefficients of
the so-called Kronecker-Eisenstein series $F{\left(z,\alpha,\tau\right)}$.
They  read 
\begin{equation}
F{\left(z,\alpha,\tau\right)}=\frac{1}{\alpha}\sum_{n\geq 0} g^{\left(n\right)}{\left(z,\tau\right)}\frac{\alpha^n}{\omega_1^{n}}=\frac{1}{\alpha}\exp{\left[-\sum_{j\geq 1}\frac{\left(-\alpha\right)^j}{\omega_1^j}\frac{\big(E_j{\left(z\right)}-G_j{\left(z\right)}\big)}{j}\right]},
\end{equation}
where
\begin{align}
E_j{\left(z\right)}=\sum_{\left(m,n\right)}\frac{1}{\left(z+m+n \tau\right)^j}, \quad G_j{\left(z\right)}=\sum_{\left(m,n\right) \neq \left(0,0\right)}\frac{1}{\left(m+n \tau\right)^j}.
\end{align}
  Similar to our earlier discussion, sums that define function
  $E_{1,2}$ and $G_{1,2}$ do not commute.  In these cases, one first
  performs the sum over $m$ and then the sum over $n$.\footnote{ For
  completeness, we give below the generic structure of the $g$-kernels
  \cite{Broedel:2014vla,Broedel:2015hia,Broedel:2017jdo,Broedel:2018qkq}
\begin{align}
g^{\left(0\right)}{\left(z,\tau\right)}=&1,
\nonumber
\\
g^{\left(1\right)}{\left(z,\tau\right)}=&\pi \cot{\pi z}+4\pi \sum_{r=1}^{\infty} \sin{\left(2\pi r z\right)}\sum_{p=1}^{\infty}q^{rp},
\nonumber
\\
g^{\left(n\right)}{\left(z,\tau\right)}=&\begin{cases} \text{even } n & -2\left[\zeta_n+\frac{\left(2 \pi i\right)^n}{\left(n-1\right)!} \sum_{r=1}^{\infty}\cos{\left(2\pi r z\right)}\sum_{p=1}^{\infty}p^{n-1}q^{rp}\right],\\ \text{odd } n & -2i\frac{\left(2 \pi i\right)^n}{\left(n-1\right)!}\sum_{r=1}^{\infty}\sin{\left(2\pi r z\right)}\sum_{p=1}^{\infty}p^{n-1}q^{rp},\end{cases}
\nonumber
\end{align}
with $q=e^{2\pi i \tau}$.
}
Making use of these kernels, we can rewrite
the integral in eq.~\eqref{eq:intellkernel} that we discussed earlier as 
\begin{equation}
\int \frac{dx}{y}\frac{y_c}{x-c}=\int dz \left(g^{\left(1\right)}{\left(z-z_c,\tau\right)}-g^{\left(1\right)}{\left(z+z_c,\tau\right)}+2 g^{\left(1\right)}{\left(z_c,\tau\right)}\right).
\end{equation}
Finally,  we can use the $g$-kernels to define the so-called $\tilde{\Gamma}$ iterated integrals. 
They read 
\begin{equation}
  \tilde{\Gamma}{\left(\begin{smallmatrix} n_1 & ... & n_m \\ z_1 & ... & z_m \end{smallmatrix};z,\tau\right)}=\int \limits_0^z dz' g^{\left(n_1\right)}{\left(z'-z_1,\tau\right)}\tilde{\Gamma}{\left(\begin{smallmatrix} n_2 & ... & n_m \\ z_2 & ...
      & z_m \end{smallmatrix};z',\tau\right)}.
\label{eq:gammatildedef}
\end{equation}
These integrals are identical to elliptic $\Empl_4$-integrals
introduced earlier except that our discussion so far was limited to
elliptic curves defined by cubic polynomials of a particular type,
cf. eq.~(\ref{ecurve3}).

To generalise this discussion to quartic elliptic curves, we follow
refs.~\cite{Broedel:2017kkb,Broedel:2018qkq} and map each point on an
elliptic curve onto a point on a torus via the Abel map
\begin{equation}
z_{x}=\frac{c_4}{\omega_1}\int \limits_{q_1}^x \frac{dx'}{y}, \mod \Lambda_{\tau}.
\label{eq:Abelmap}
\end{equation}
Then, we can use a mapping that is similar but more complex than the one shown in eq.~(\ref{eq:mappingxyw}), 
to connect $y(x)$ and $x$ with functions of $z$~\cite{Broedel:2017kkb,Broedel:2018qkq} and in this
way relate $\Empl_4$ integrals with $\tilde \Gamma$ integrals also for quartic elliptic curves.  

Although, in general, working with $\tilde \Gamma$ integrals does not
offer particular advantages as compared to elliptic
$\Empl_4$-integrals,\footnote{ We note, however, that in contrast to
$\Empl_4$-integrals, $\tilde \Gamma$ integrals can be represented by
convergent series which simplifies their numerical evaluation.}  we
deal with a special case since, upon inspection, we find that to
compute ${\cal O}(\lambda)$ corrections to the $C$-parameter, we only
need to know $\tilde \Gamma$ functions at \emph{rational} points on
the torus, see eqs.~(\ref{eq:e20}, \ref{eq:Abelmapval2}).
  Although it is not obvious from eq.~(\ref{eq:gammatildedef}) that to
  compute $\tilde \Gamma$ at rational points we only need to know
  kernels $g^{(n)}$ at rational points, it is actually true.  To see
  this, we note that the total derivative of the $\tilde{\Gamma}$
  integrals with respect to $\tau$ can be schematically represented as
  follows~\cite{Broedel:2018iwv}
\begin{equation}
  \frac{ {\rm d} \tilde{\Gamma}{\left(\begin{smallmatrix} \vec{n}
        \\ \vec{z} \end{smallmatrix};z_0,\tau\right)}}{{\rm d} \tau}
  \sim \sum \;g^{\left(n_i\right)}(\tilde{z},\tau) \times
  \tilde{\Gamma}{\left(\begin{smallmatrix} \vec{n}'
      \\ \vec{z}' \end{smallmatrix};z_0,\tau\right)},
\label{eqe25}
\end{equation}
where ${\vec n}'$ and $\vec{z}'$ describe original vectors ${\vec n}$
and ${\vec z}$ with one entry removed, and $\tilde z$ stands for
various points shown in eqs.~(\ref{eq:e20}, \ref{eq:Abelmapval2}) and
their combinations.  Eq.~(\ref{eqe25}) confirms that the $g$-kernels
required to compute ${\rm d} \tilde \Gamma/{\rm d} \tau$ are evaluated
at rational points of the form $\tilde{z}=r/N+s\tau /N$ with $0 \leq
r,s<N$.
 A straightforward integration of eq.~(\ref{eqe25}) over $\tau$ gives
\begin{equation}
  \tilde{\Gamma}{\left(\begin{smallmatrix} \vec{n}
      \\ \vec{z} \end{smallmatrix};z_0,\tau\right)}=\text{Cusp}{\left(\tilde{\Gamma}{\left(\begin{smallmatrix}
        \vec{n}
        \\ \vec{z} \end{smallmatrix};z_0,\tau\right)}\right)}+\int
  \limits_{i\infty}^{\tau} \frac{
    d\tilde{\Gamma}{\left(\begin{smallmatrix} \vec{n}
        \\ \vec{z}(\tau')
      \end{smallmatrix};z_0(\tau'),\tau'\right)}}{{\rm d} \tau'} \; {\rm d \tau'},
  \label{eq:e26}
\end{equation}
where the additional term on the right-hand side is called the cusp
which accounts for differences between $z$ and $\tau$ integration
boundaries.  It follows from eq.~(\ref{eq:e26}) that the cusp is given
by $ \lim_{\tau \to i\infty} \tilde \Gamma{\left(\begin{smallmatrix}
    \vec{n} \\ \vec{z} \end{smallmatrix};z_0,\tau\right)}$. To
evaluate it, we can go back to the $z$-representation for $\tilde
\Gamma$ integrals, c.f. eq.~(\ref{eq:gammatildedef}), and make use of
the fact that, in the $\tau \to i \infty$ limit, all the $g$-kernels
simplify making calculation of the cusp a relatively straightforward
endeavor.

The benefit of the representation eq.~(\ref{eq:e26}) can be seen from
the fact that $g$-kernels at rational points can be written as an
expansion
 \begin{equation}
g^{\left(n\right)}{\left(\frac{r}{N}+\frac{s}{N}\tau,\tau\right)}=\sum_{k=0}^n \frac{\left(-2\pi i s\right)^k}{k! N^k} h_{N,r,s}^{\left(n-k\right)}{\left(\tau\right)},
\label{eq:grelEisensteinh}
\end{equation}
where $h_{N,r,s}^{\left(n\right)}(\tau)$ are the so-called Eisenstein
series functions. They are defined as follows
\begin{equation}
h_{N,r,s}^{\left(n\right)}{\left(\tau\right)}=-\sum_{\substack{\left(a,b\right)\in \mathbb{Z}^2 \\ \left(a,b\right)\neq \left(0,0\right)}} \frac{e^{2\pi i\frac{\left(bs-ar\right)}{N}}}{\left(a\tau+b\right)^n}.
\end{equation}
  It should be clear from eqs.~(\ref{eqe25}, \ref{eq:e26},
  \ref{eq:grelEisensteinh}) that the integration over $\tau$ has an
  iterative structure and that kernels of iterated integrals can be
  associated with the Eisenstein series functions.  To accommodate
  this iterative structure, we define the iterated Eisenstein
  integrals. They read
\begin{equation}
  I{\left(f_1,...,f_n;\tau\right)}
  =\int \limits_{i\infty}^{\tau} \frac{d\tau'}{2\pi i}f_1{\left(\tau'\right)}I{\left(f_2,...,f_n;\tau'\right)},
\label{eq:iterEisenIntDef}
\end{equation}
where the recursion starts with $I{\left(;\tau\right)}=1$ and the
kernels $f_i$ stand for the $h_{N,r,s}^{\left(n\right)}$
functions. These integrals belong to the special class of iterated
integrals of modular forms. For further details and applications, we
refer to the literature
\cite{Adams:2017ejb,Broedel:2018iwv,Abreu:2019fgk,Duhr:2019rrs,Weinzierl:2022eaz,Abreu:2022vei}.
The important point for us is that the above construction allows us to
rewrite all $\Empl_4$ integrals in terms of iterated Eisenstein
integrals in a systematic manner. 

\section{Power corrections to the $C$-parameter for $N$-jet final states}
\label{sec:cn}
In this appendix, we provide further detail on how we obtain the
result of eq.~\eqref{eq:Wcij} for the linear power corrections to the
$C$-parameter in the generic $N$-jet region. Our starting point is
the quantity $W_C^{ij,(k)}$ in eq.~\eqref{eq:5.36}, with momenta
parametrised as in eq.~\eqref{eq:genSudakov}. Using the following
expressions for scalar products
\begin{equation}
  \begin{split}
    \big(p_k\ltone\,\big) =& \,p_{k,t}
    \big(\cosh{(\eta-\eta_k)}-\cos{(\varphi-\varphi_k)}\big), \\ \big(
    \ltone q\big)=& \,
    \frac{q}{s_{ij}}\big(\cosh{(\eta-\eta_q)}-c_{ij}\cos{(\varphi-\varphi_q)}
    \big),
  \end{split}
  \label{eq:momenrel2pN1}
\end{equation}
with $s_{ij}$ and $c_{ij}$ defined in eq.~\eqref{eq:5.37}, we can write
$W_C^{ij,(k)}$ as
\begin{equation}
  W_C^{ij,(k)}=\frac{p^{2}_{k,t} s_{ij}}{2 q}\int \mathd\eta\,
  \mathd\varphi\,\frac{\big(\cosh{(\eta-\eta_k)}
    -\cos{(\varphi-\varphi_k)}\big)^2}
  {\cosh{(\eta-\eta_q)}-c_{ij}\cos{(\varphi-\varphi_q)}}
  e^{-\epsilon|\eta-\eta_q|}.
  \label{eq5.9}
\end{equation}
We note that $p_{k,t}$ implicitly depends on $ij$ through the decomposition
eq.~\eqref{eq:genSudakov}. 

To proceed further, it is convenient to shift the integration variables 
\begin{equation}
\eta\rightarrow \tilde{\eta}+\eta_q,
\;\;\;\;\;\varphi\rightarrow \tilde{\varphi}+\varphi_q, 
\end{equation}
and  define the following quantities 
\begin{equation}
\eta_{qk}=\eta_q-\eta_k, \;\;\;\;\;
\varphi_{qk}=\varphi_q-\varphi_k,
\end{equation}
to simplify the  integrand in eq.~(\ref{eq5.9}). We find 
\begin{equation}
  W_C^{ij,(k)}=\frac{p^{2}_{k,t} s_{ij}}{2 q}\int
  \limits_{-\infty}^{\infty} \mathd\tilde{\eta} \int \limits_0^{2\pi}
  \mathd\tilde{\varphi}\,\frac{\big(\cosh{(\tilde{\eta}+\eta_{qk})}
    -\cos{(\tilde{\varphi}+\varphi_{qk})}\big)^2}{\left(\cosh{\tilde{\eta}}-c_{ij}\cos{\tilde{\varphi}}\right)}e^{-\epsilon|\tilde{\eta}|}.
\label{eq:d5}
\end{equation}
We now integrate eq.~\eqref{eq:d5} over $\tilde\ourphi$. Using
\begin{equation}
  \int\limits_0^{2\pi} \frac{\mathd \tilde\ourphi}{a-b\cos\tilde\ourphi}
  = \frac{2\pi}{\sqrt{a^2-b^2}},
\end{equation}
and neglecting all odd terms that vanish
upon $\tilde \eta$ integration, we arrive at
\begin{equation}
\begin{split}
  W_C^{ij,(k)}=&\frac{p^2_{k,t} s_{ij}}{ q}\frac{\pi}{c_{ij}^2} \int
  \limits_{-\infty}^{\infty} \mathd\tilde{\eta}\; \Bigg [
    -\cosh{\tilde{\eta}} (\cos{2 \varphi_{qk}}-2 c_{ij}
    \cos{\varphi_{qk}} \cosh{\eta_{qk}}) \\ &+\frac{1}{2
      \sqrt{\cosh^2{\tilde{\eta}}-c_{ij}^2}}\Bigg (-c_{ij}^2 \cos{2
      \varphi_{qk}}+2 c_{ij}^2 \cosh^2{\tilde{\eta}}
    \sinh^2{\eta_{qk}}-2 c_{ij}^2 \sinh^2{\eta_{qk}} \\ &+2 c_{ij}^2
    \cosh^2{\tilde{\eta}} \cosh^2{\eta_{qk}}+c_{ij}^2-4 c_{ij}
    \cos{\varphi_{qk}} \cosh^2{\tilde{\eta}} \cosh{\eta_{qk}} \\ & +2
    \cos{2 \varphi_{qk}} \cosh^2{\tilde{\eta}}\Bigg )\Bigg
  ]e^{-\epsilon|\tilde{\eta}|}.
\end{split}
\label{eq5.15}
\end{equation}

It is straightforward to see that the first term in the integrand in
eq.~(\ref{eq5.15}) vanishes upon integration over $\tilde{\eta}$
thanks to the analytic regulator. For the other terms, we use the
symmetry of the integrand to restrict the integral to the positive semi-axis,
and write
\begin{equation}
\begin{split}
  W_C^{ij,(k)}=&\frac{p^{2}_{k,t} \pi s_{ij}}{c_{ij}^2 q} \int
  \limits_{0}^{\infty} \mathd\tilde{\eta}\;\frac{e^{-\epsilon
      \tilde{\eta}}}{\sqrt{\cosh^2{\tilde{\eta}}-c_{ij}^2}} \Bigg
  (-c_{ij}^2 \cos{2 \varphi_{qk}}+2 c_{ij}^2 \cosh^2{\tilde{\eta}}
  \sinh^2{\eta_{qk}} \\ &-2 c_{ij}^2 \sinh^2{\eta_{qk}} +2 c_{ij}^2
  \cosh^2{\tilde{\eta}} \cosh^2{\eta_{qk}}+c_{ij}^2 \\ &-4 c_{ij}
  \cos{\varphi_{qk}} \cosh^2{\tilde{\eta}} \cosh{\eta_{qk}} +2 \cos{2
    \varphi_{qk}} \cosh^2{\tilde{\eta}} \Bigg ).
\end{split}
\end{equation}
To proceed further, we change variable
\begin{align}
\tilde{\eta}=&\ln{\frac{1+\sqrt{1-\xi^2}}{\xi}}
\end{align}
and find an expression that is very similar to the three-jet case. We
write it as
\begin{equation}
\begin{split}
  W_C^{ij,(k)}=&\frac{p^{2}_{k,t} \pi s_{ij}}{c_{ij}^2 q} \int
  \limits_{0}^{1}
  \mathd\xi\;\frac{\xi^{\epsilon-2}\left(1+\sqrt{1-\xi^2}\right)^{-\epsilon}}{\sqrt{1-\xi^2}\sqrt{1-c_{ij}^2
      \xi^2}}\times\left(A+\xi^2 B\right),
\end{split}
\label{eq:d10}
\end{equation}
where the two $\xi$-independent quantities  $A$ and $B$ read
\begin{align}
  A=&
  2 c_{ij}^2 \sinh^2{\eta_{qk}}+2 c_{ij}^2 \cosh^2{\eta_{qk}}-4 c_{ij} \cos{\varphi_{qk}} \cosh{\eta_{qk}}+2 \cos{2 \varphi_{qk}},
\\
B=&-2 c_{ij}^2 \sinh^2{\eta_{qk}}-c_{ij}^2 \cos{2 \varphi_{qk}}+c_{ij}^2.
\end{align}
Using eqs.~(\ref{eq:splitX1X2int2}, \ref{eq:splitX1X2int3}), we can further
write eq.~\eqref{eq:d10} as
\begin{equation}
  W_C^{ij,(k)}=\frac{p^{2}_{k,t} \pi s_{ij}}{c_{ij}^2 q} \left(
  \frac{A}{4} \; Z_1^a + \frac{A+B}{2} \; Z_1^b\right) =
  \frac{p^{2}_{k,t} \pi s_{ij}}{c_{ij}^2 q} \bigg[
  (A+B)K(c_{12}^2) - A E(c_{12}^2)\bigg].
\end{equation}
Putting everything together, we obtain the following result for
$W^{ij,(k)}_C$
\begin{equation}
\begin{split}
W_C^{ij,(k)}&=\frac{p^2_{k,t} \pi s_{ij}}{c_{ij}^2 q}
\Bigg [
  K{\left(c_{ij}^2\right)}\Bigg (2 c_{ij}^2 \cosh^2{\eta_{qk}}-4 c_{ij} \cos{\varphi_{qk}} \cosh{\eta_{qk}}
  \\
& +\left(2-c_{ij}^2\right) \cos{2 \varphi_{qk}} +c_{ij}^2 \Bigg )
  +E{\left (c_{ij}^2\right)} \Bigg (-2 c_{ij}^2 \cosh{2\eta_{qk}}
  \\
& +4 c_{ij} \cos{\varphi_{qk}} \cosh{\eta_{qk}} -2 \cos{2 \varphi_{qk}}\Bigg ) \Bigg ].
\end{split}
\label{eq:doubleWCkpre}
\end{equation}

It turns out that the expression for $W_C^{ij,(k)}$ in
eq.~(\ref{eq:doubleWCkpre}) can be simplified if written through sines
and cosines of the angles between the directions of the hard partons
in the rest frame of $q$.  To introduce them, we take the four-momenta of
two partons, $k$ and $m$, and write
\begin{equation}
(p_k p_m) = E_k E_m (1-\cos \theta_{km}) = 2 E_k E_m
  \sin^2(\theta_{km}/2) = 2 E_k E_m s_{km}^2,
\end{equation}
where energies and angles are defined in the $q$ rest frame.  Since
$E_{k,m} = (p_k q)/\sqrt{q^2}$, we find
\begin{equation}
(p_k p_m) = \frac{ 2 (p_k q) (p_m q)}{q^2} s_{km}^2.
\end{equation}
Similar to the three-jet case, we parameterise energy fractions of
the various partons as follows
\begin{equation}
(p_k q) = \frac{q^2}{2} (1-x_k), ~~~~ \sum_{k=1}^{N}(1-x_k) = 2,
\label{eq:pkqdotproddef}
\end{equation}
to obtain
\begin{equation}
(p_k p_m) = \frac{q^2}{2} (1-x_k) (1-x_m)  s_{km}^2.
\label{eq5.29}
\end{equation}
We can employ the relative half-angles $s_{ij}$ to parameterise the various
quantities that appear in the expression for the function
$W_C^{ij,(k)}$ eq.~\eqref{eq:doubleWCkpre}. We obtain
\begin{equation}
\begin{split} 
  p^{2}_{k,t}=& q^2 \left(1-x_k\right)^2\frac{s_{ik}^2
    s_{jk}^2}{s_{ij}^2}, \;\;\;\; \eta_{qk} =
  \frac{1}{2}\log{\left(\frac{s_{ik}^2}{s_{jk}^2}\right)}, \;\;\;\;
  \cos{\varphi_{qk}}= \frac{s_{ik}^2+s_{jk}^2-s_{ij}^2}{2c_{ij} s_{ik}
    s_{jk}}.
\end{split}
\end{equation}
We use these expressions in eq.~\eqref{eq:doubleWCkpre} and finally
find
\begin{equation}
W_C^{ij,(k)}=\frac{q \pi s_{ij} \left(1-x_k\right)^2}{2 c_{ij}^4} \tilde{W}_C^{ij, (k)},
\end{equation}
where 
\begin{equation}
\begin{split} 
  \tilde{W}_C^{ij, (k)}=& K{\left(c_{ij}^2\right)}\Bigg [
    s_{ij}^4+s_{ij}^2\left(1+ s_{ik}^4+s_{jk}^4 -4 \left (s_{ik}^2+s_{jk}^2\right)+6 s_{ik}^2 s_{jk}^2\right)
\\
&+\left(s_{ik}^2-s_{jk}^2\right)^2\Bigg ] -2 E{\left(c_{ij}^2\right)}\Bigg [ s_{ij}^2\left(1- s_{ik}^2 - s_{jk}^2 + s_{ik}^4+s_{jk}^4 \right)
\\
&-s_{ik}^4-s_{jk}^4 - s_{ik}^2 - s_{jk}^2 + 4 s_{ik}^2 s_{jk}^2 +\frac{\big(s_{ik}^2-s_{jk}^2\big)^2}{s_{ij}^2}\Bigg ].
\end{split}
\label{eq5.31}
\end{equation}

We are now in position to evaluate $W_C^{ij}$, cf.
eq.~(\ref{eq:5.36}). It can be written as
\begin{equation}
  W_C^{ij} = -\frac{3}{8\pi^2 q}\frac{s_{ij}}{c_{ij}^4}
  \sum_{k=1}^{N}(1-x_k)\tilde W^{ij,(k)}_{C}.
  \label{eq:d22}
\end{equation}
Inspecting eq.~(\ref{eq:d22}), we observe that we need to evaluate
the following sums
\begin{equation}
  \begin{gathered}
    \sum_{k=1}^N \left(1-x_k\right)=2,
    ~~~~~~~~~~~
    \sum_{k=1}^N \left(1-x_k\right) s_{(i,j)k}^2=1,
    \\
    \sum_{k=1}^N \left(1-x_k\right) s_{(i,j)k}^4=n_3^{i(j)},
    ~~~~~~~~~~~
    n_4^{ij}=\sum_{k=1}^N \left(1-x_k\right) s_{ik}^2 s_{jk}^2=n_4^{ij},
  \end{gathered}
\end{equation}
where $n_3^{i(j)}$ and $n_4^{ij}$ are defined in eq.~\eqref{eq:5.38}.
Using these relations, it is straightforward to obtain our final result
eq.~\eqref{eq:Wcij}.

\bibliographystyle{JHEP}
\bibliography{renII}

\providecommand{\href}[2]{#2}\begingroup\raggedright\begin{thebibliography}{10}

\bibitem{ParticleDataGroup:2020ssz}
{\scshape Particle Data Group} collaboration, P.~A. Zyla et~al., \emph{{Review
  of Particle Physics}},
  \href{http://dx.doi.org/10.1093/ptep/ptaa104}{\emph{PTEP} {\bf 2020} (2020)
  083C01}.

\bibitem{Bethke:2009ehn}
{\scshape JADE} collaboration, S.~Bethke, S.~Kluth, C.~Pahl and J.~Schieck,
  \emph{{Determination of the Strong Coupling alpha(s) from hadronic Event
  Shapes with O(alpha**3(s)) and resummed QCD predictions using JADE Data}},
  \href{http://dx.doi.org/10.1140/epjc/s10052-009-1149-1}{\emph{Eur. Phys. J.
  C} {\bf 64} (2009) 351--360}, [\href{http://arxiv.org/abs/0810.1389}{{\tt
  0810.1389}}].

\bibitem{Dissertori:2009ik}
G.~Dissertori, A.~Gehrmann-De~Ridder, T.~Gehrmann, E.~W.~N. Glover,
  G.~Heinrich, G.~Luisoni et~al., \emph{{Determination of the strong coupling
  constant using matched NNLO+NLLA predictions for hadronic event shapes in
  e+e- annihilations}},
  \href{http://dx.doi.org/10.1088/1126-6708/2009/08/036}{\emph{JHEP} {\bf 08}
  (2009) 036}, [\href{http://arxiv.org/abs/0906.3436}{{\tt 0906.3436}}].

\bibitem{Kardos:2018kqj}
A.~Kardos, S.~Kluth, G.~Somogyi, Z.~Tulip\'ant and A.~Verbytskyi,
  \emph{{Precise determination of $\alpha _{S}(M_Z)$ from a global fit of
  energy\textendash{}energy correlation to NNLO+NNLL predictions}},
  \href{http://dx.doi.org/10.1140/epjc/s10052-018-5963-1}{\emph{Eur. Phys. J.
  C} {\bf 78} (2018) 498}, [\href{http://arxiv.org/abs/1804.09146}{{\tt
  1804.09146}}].

\bibitem{Akhoury:1995fb}
R.~Akhoury and V.~I. Zakharov, \emph{{Leading power corrections in QCD: From
  renormalons to phenomenology}},
  \href{http://dx.doi.org/10.1016/0550-3213(96)00056-9}{\emph{Nucl. Phys. B}
  {\bf 465} (1996) 295--314}, [\href{http://arxiv.org/abs/hep-ph/9507253}{{\tt
  hep-ph/9507253}}].

\bibitem{Salam:2001bd}
G.~P. Salam and D.~Wicke, \emph{{Hadron masses and power corrections to event
  shapes}}, \href{http://dx.doi.org/10.1088/1126-6708/2001/05/061}{\emph{JHEP}
  {\bf 05} (2001) 061}, [\href{http://arxiv.org/abs/hep-ph/0102343}{{\tt
  hep-ph/0102343}}].

\bibitem{Hoang:2007vb}
A.~H. Hoang and I.~W. Stewart, \emph{{Designing gapped soft functions for jet
  production}},
  \href{http://dx.doi.org/10.1016/j.physletb.2008.01.040}{\emph{Phys. Lett. B}
  {\bf 660} (2008) 483--493}, [\href{http://arxiv.org/abs/0709.3519}{{\tt
  0709.3519}}].

\bibitem{Abbate:2010xh}
R.~Abbate, M.~Fickinger, A.~H. Hoang, V.~Mateu and I.~W. Stewart, \emph{{Thrust
  at $N^{3}LL$ with Power Corrections and a Precision Global Fit for
  $\alpha_{s}(mZ)$}},
  \href{http://dx.doi.org/10.1103/PhysRevD.83.074021}{\emph{Phys. Rev. D} {\bf
  83} (2011) 074021}, [\href{http://arxiv.org/abs/1006.3080}{{\tt 1006.3080}}].

\bibitem{Gehrmann:2012sc}
T.~Gehrmann, G.~Luisoni and P.~F. Monni, \emph{{Power corrections in the
  dispersive model for a determination of the strong coupling constant from the
  thrust distribution}},
  \href{http://dx.doi.org/10.1140/epjc/s10052-012-2265-x}{\emph{Eur. Phys. J.
  C} {\bf 73} (2013) 2265}, [\href{http://arxiv.org/abs/1210.6945}{{\tt
  1210.6945}}].

\bibitem{Mateu:2012nk}
V.~Mateu, I.~W. Stewart and J.~Thaler, \emph{{Power Corrections to Event Shapes
  with Mass-Dependent Operators}},
  \href{http://dx.doi.org/10.1103/PhysRevD.87.014025}{\emph{Phys. Rev. D} {\bf
  87} (2013) 014025}, [\href{http://arxiv.org/abs/1209.3781}{{\tt 1209.3781}}].

\bibitem{Hoang:2015hka}
A.~H. Hoang, D.~W. Kolodrubetz, V.~Mateu and I.~W. Stewart, \emph{{Precise
  determination of $\alpha_s$ from the $C$-parameter distribution}},
  \href{http://dx.doi.org/10.1103/PhysRevD.91.094018}{\emph{Phys. Rev. D} {\bf
  91} (2015) 094018}, [\href{http://arxiv.org/abs/1501.04111}{{\tt
  1501.04111}}].

\bibitem{Gracia:2021nut}
N.~G. Gracia and V.~Mateu, \emph{{Toward massless and massive event shapes in
  the large-\ensuremath{\beta}$_{0}$ limit}},
  \href{http://dx.doi.org/10.1007/JHEP07(2021)229}{\emph{JHEP} {\bf 07} (2021)
  229}, [\href{http://arxiv.org/abs/2104.13942}{{\tt 2104.13942}}].

\bibitem{Manohar:1994kq}
A.~V. Manohar and M.~B. Wise, \emph{{Power suppressed corrections to hadronic
  event shapes}},
  \href{http://dx.doi.org/10.1016/0370-2693(94)01504-6}{\emph{Phys. Lett. B}
  {\bf 344} (1995) 407--412}, [\href{http://arxiv.org/abs/hep-ph/9406392}{{\tt
  hep-ph/9406392}}].

\bibitem{Webber:1994cp}
B.~R. Webber, \emph{{Estimation of power corrections to hadronic event
  shapes}}, \href{http://dx.doi.org/10.1016/0370-2693(94)91147-9}{\emph{Phys.
  Lett. B} {\bf 339} (1994) 148--150},
  [\href{http://arxiv.org/abs/hep-ph/9408222}{{\tt hep-ph/9408222}}].

\bibitem{Dokshitzer:1995qm}
Y.~L. Dokshitzer, G.~Marchesini and B.~R. Webber, \emph{{Dispersive approach to
  power behaved contributions in QCD hard processes}},
  \href{http://dx.doi.org/10.1016/0550-3213(96)00155-1}{\emph{Nucl. Phys. B}
  {\bf 469} (1996) 93--142}, [\href{http://arxiv.org/abs/hep-ph/9512336}{{\tt
  hep-ph/9512336}}].

\bibitem{Nason:1995np}
P.~Nason and M.~H. Seymour, \emph{{Infrared renormalons and power suppressed
  effects in e+ e- jet events}},
  \href{http://dx.doi.org/10.1016/0550-3213(95)00461-Z}{\emph{Nucl. Phys. B}
  {\bf 454} (1995) 291--312}, [\href{http://arxiv.org/abs/hep-ph/9506317}{{\tt
  hep-ph/9506317}}].

\bibitem{Dasgupta:1996ki}
M.~Dasgupta and B.~R. Webber, \emph{{Power corrections and renormalons in
  $e^{+} e^{-}$ fragmentation functions}},
  \href{http://dx.doi.org/10.1016/S0550-3213(96)00622-0}{\emph{Nucl. Phys. B}
  {\bf 484} (1997) 247--264}, [\href{http://arxiv.org/abs/hep-ph/9608394}{{\tt
  hep-ph/9608394}}].

\bibitem{Nason:1996pk}
P.~Nason and B.~R. Webber, \emph{{Nonperturbative corrections to heavy quark
  fragmentation in e+ e- annihilation}},
  \href{http://dx.doi.org/10.1016/S0370-2693(97)00129-9}{\emph{Phys. Lett. B}
  {\bf 395} (1997) 355--363}, [\href{http://arxiv.org/abs/hep-ph/9612353}{{\tt
  hep-ph/9612353}}].

\bibitem{Beneke:1997sr}
M.~Beneke, V.~M. Braun and L.~Magnea, \emph{{Phenomenology of power corrections
  in fragmentation processes in e+ e- annihilation}},
  \href{http://dx.doi.org/10.1016/S0550-3213(97)00251-4}{\emph{Nucl. Phys. B}
  {\bf 497} (1997) 297--333}, [\href{http://arxiv.org/abs/hep-ph/9701309}{{\tt
  hep-ph/9701309}}].

\bibitem{Dokshitzer:1997ew}
Y.~L. Dokshitzer and B.~R. Webber, \emph{{Power corrections to event shape
  distributions}},
  \href{http://dx.doi.org/10.1016/S0370-2693(97)00573-X}{\emph{Phys. Lett. B}
  {\bf 404} (1997) 321--327}, [\href{http://arxiv.org/abs/hep-ph/9704298}{{\tt
  hep-ph/9704298}}].

\bibitem{Dokshitzer:1997iz}
Y.~L. Dokshitzer, A.~Lucenti, G.~Marchesini and G.~P. Salam,
  \emph{{Universality of 1/Q corrections to jet-shape observables rescued}},
  \href{http://dx.doi.org/10.1016/S0550-3213(97)00650-0}{\emph{Nucl. Phys. B}
  {\bf 511} (1998) 396--418}, [\href{http://arxiv.org/abs/hep-ph/9707532}{{\tt
  hep-ph/9707532}}].

\bibitem{Dokshitzer:1998pt}
Y.~L. Dokshitzer, A.~Lucenti, G.~Marchesini and G.~P. Salam, \emph{{On the
  universality of the Milan factor for 1 / Q power corrections to jet shapes}},
  \href{http://dx.doi.org/10.1088/1126-6708/1998/05/003}{\emph{JHEP} {\bf 05}
  (1998) 003}, [\href{http://arxiv.org/abs/hep-ph/9802381}{{\tt
  hep-ph/9802381}}].

\bibitem{Korchemsky:1999kt}
G.~P. Korchemsky and G.~F. Sterman, \emph{{Power corrections to event shapes
  and factorization}},
  \href{http://dx.doi.org/10.1016/S0550-3213(99)00308-9}{\emph{Nucl. Phys. B}
  {\bf 555} (1999) 335--351}, [\href{http://arxiv.org/abs/hep-ph/9902341}{{\tt
  hep-ph/9902341}}].

\bibitem{Korchemsky:2000kp}
G.~P. Korchemsky and S.~Tafat, \emph{{On power corrections to the event shape
  distributions in QCD}},
  \href{http://dx.doi.org/10.1088/1126-6708/2000/10/010}{\emph{JHEP} {\bf 10}
  (2000) 010}, [\href{http://arxiv.org/abs/hep-ph/0007005}{{\tt
  hep-ph/0007005}}].

\bibitem{Gardi:2001ny}
E.~Gardi and J.~Rathsman, \emph{{Renormalon resummation and exponentiation of
  soft and collinear gluon radiation in the thrust distribution}},
  \href{http://dx.doi.org/10.1016/S0550-3213(01)00284-X}{\emph{Nucl. Phys. B}
  {\bf 609} (2001) 123--182}, [\href{http://arxiv.org/abs/hep-ph/0103217}{{\tt
  hep-ph/0103217}}].

\bibitem{Gardi:2003iv}
E.~Gardi and L.~Magnea, \emph{{The C parameter distribution in e+ e-
  annihilation}},
  \href{http://dx.doi.org/10.1088/1126-6708/2003/08/030}{\emph{JHEP} {\bf 08}
  (2003) 030}, [\href{http://arxiv.org/abs/hep-ph/0306094}{{\tt
  hep-ph/0306094}}].

\bibitem{Bauer:2003di}
C.~W. Bauer, C.~Lee, A.~V. Manohar and M.~B. Wise, \emph{{Enhanced
  nonperturbative effects in Z decays to hadrons}},
  \href{http://dx.doi.org/10.1103/PhysRevD.70.034014}{\emph{Phys. Rev. D} {\bf
  70} (2004) 034014}, [\href{http://arxiv.org/abs/hep-ph/0309278}{{\tt
  hep-ph/0309278}}].

\bibitem{Lee:2006nr}
C.~Lee and G.~F. Sterman, \emph{{Momentum Flow Correlations from Event Shapes:
  Factorized Soft Gluons and Soft-Collinear Effective Theory}},
  \href{http://dx.doi.org/10.1103/PhysRevD.75.014022}{\emph{Phys. Rev. D} {\bf
  75} (2007) 014022}, [\href{http://arxiv.org/abs/hep-ph/0611061}{{\tt
  hep-ph/0611061}}].

\bibitem{Luisoni:2020efy}
G.~Luisoni, P.~F. Monni and G.~P. Salam, \emph{{$C$-parameter hadronisation in
  the symmetric 3-jet limit and impact on $\alpha_s$ fits}},
  \href{http://dx.doi.org/10.1140/epjc/s10052-021-08941-z}{\emph{Eur. Phys. J.
  C} {\bf 81} (2021) 158}, [\href{http://arxiv.org/abs/2012.00622}{{\tt
  2012.00622}}].

\bibitem{Caola:2021kzt}
F.~Caola, S.~Ferrario~Ravasio, G.~Limatola, K.~Melnikov and P.~Nason, \emph{{On
  linear power corrections in certain collider observables}},
  \href{http://dx.doi.org/10.1007/JHEP01(2022)093}{\emph{JHEP} {\bf 01} (2022)
  093}, [\href{http://arxiv.org/abs/2108.08897}{{\tt 2108.08897}}].

\bibitem{Kluth:2006bw}
S.~Kluth, \emph{{Tests of Quantum Chromo Dynamics at e+ e- Colliders}},
  \href{http://dx.doi.org/10.1088/0034-4885/69/6/R04}{\emph{Rept. Prog. Phys.}
  {\bf 69} (2006) 1771--1846}, [\href{http://arxiv.org/abs/hep-ex/0603011}{{\tt
  hep-ex/0603011}}].

\bibitem{OPAL:2005vad}
{\scshape OPAL} collaboration, G.~Abbiendi et~al., \emph{{Measurement of the
  Strong Coupling alpha(s) from four-jet observables in e+ e- annihilation}},
  \href{http://dx.doi.org/10.1140/epjc/s2006-02581-y}{\emph{Eur. Phys. J. C}
  {\bf 47} (2006) 295--307}, [\href{http://arxiv.org/abs/hep-ex/0601048}{{\tt
  hep-ex/0601048}}].

\bibitem{Schieck:2006tc}
{\scshape JADE} collaboration, J.~Schieck, S.~Bethke, O.~Biebel, S.~Kluth,
  P.~A. Movilla~Fernandez and C.~Pahl, \emph{{Measurement of the strong
  coupling alpha(s) from the four-jet rate in e+ e- annihilation using JADE
  data}}, \href{http://dx.doi.org/10.1140/epjc/s2006-02625-4}{\emph{Eur. Phys.
  J. C} {\bf 48} (2006) 3--13}, [\href{http://arxiv.org/abs/0707.0392}{{\tt
  0707.0392}}].

\bibitem{Beneke:1998ui}
M.~Beneke, \emph{{Renormalons}},
  \href{http://dx.doi.org/10.1016/S0370-1573(98)00130-6}{\emph{Phys. Rept.}
  {\bf 317} (1999) 1--142}, [\href{http://arxiv.org/abs/hep-ph/9807443}{{\tt
  hep-ph/9807443}}].

\bibitem{Broedel:2017kkb}
J.~Broedel, C.~Duhr, F.~Dulat and L.~Tancredi, \emph{{Elliptic polylogarithms
  and iterated integrals on elliptic curves. Part I: general formalism}},
  \href{http://dx.doi.org/10.1007/JHEP05(2018)093}{\emph{JHEP} {\bf 05} (2018)
  093}, [\href{http://arxiv.org/abs/1712.07089}{{\tt 1712.07089}}].

\bibitem{Broedel:2017siw}
J.~Broedel, C.~Duhr, F.~Dulat and L.~Tancredi, \emph{{Elliptic polylogarithms
  and iterated integrals on elliptic curves II: an application to the sunrise
  integral}}, \href{http://dx.doi.org/10.1103/PhysRevD.97.116009}{\emph{Phys.
  Rev. D} {\bf 97} (2018) 116009}, [\href{http://arxiv.org/abs/1712.07095}{{\tt
  1712.07095}}].

\bibitem{Broedel:2018iwv}
J.~Broedel, C.~Duhr, F.~Dulat, B.~Penante and L.~Tancredi, \emph{{Elliptic
  symbol calculus: from elliptic polylogarithms to iterated integrals of
  Eisenstein series}},
  \href{http://dx.doi.org/10.1007/JHEP08(2018)014}{\emph{JHEP} {\bf 08} (2018)
  014}, [\href{http://arxiv.org/abs/1803.10256}{{\tt 1803.10256}}].

\bibitem{Broedel:2018qkq}
J.~Broedel, C.~Duhr, F.~Dulat, B.~Penante and L.~Tancredi, \emph{{Elliptic
  Feynman integrals and pure functions}},
  \href{http://dx.doi.org/10.1007/JHEP01(2019)023}{\emph{JHEP} {\bf 01} (2019)
  023}, [\href{http://arxiv.org/abs/1809.10698}{{\tt 1809.10698}}].

\bibitem{Broedel:2019hyg}
J.~Broedel, C.~Duhr, F.~Dulat, B.~Penante and L.~Tancredi, \emph{{Elliptic
  polylogarithms and Feynman parameter integrals}},
  \href{http://dx.doi.org/10.1007/JHEP05(2019)120}{\emph{JHEP} {\bf 05} (2019)
  120}, [\href{http://arxiv.org/abs/1902.09971}{{\tt 1902.09971}}].

\bibitem{Weinzierl:2022eaz}
S.~Weinzierl, \emph{{Feynman Integrals}},
  \href{http://arxiv.org/abs/2201.03593}{{\tt 2201.03593}}.

\bibitem{FerrarioRavasio:2018ubr}
S.~Ferrario~Ravasio, P.~Nason and C.~Oleari, \emph{{All-orders behaviour and
  renormalons in top-mass observables}},
  \href{http://dx.doi.org/10.1007/JHEP01(2019)203}{\emph{JHEP} {\bf 01} (2019)
  203}, [\href{http://arxiv.org/abs/1810.10931}{{\tt 1810.10931}}].

\bibitem{Dasgupta:1999mb}
M.~Dasgupta, L.~Magnea and G.~Smye, \emph{{Universality of 1/Q corrections
  revisited}},
  \href{http://dx.doi.org/10.1088/1126-6708/1999/11/025}{\emph{JHEP} {\bf 11}
  (1999) 025}, [\href{http://arxiv.org/abs/hep-ph/9911316}{{\tt
  hep-ph/9911316}}].

\bibitem{Smye:2001gq}
G.~E. Smye, \emph{{On the 1/Q correction to the C - parameter at two loops}},
  \href{http://dx.doi.org/10.1088/1126-6708/2001/05/005}{\emph{JHEP} {\bf 05}
  (2001) 005}, [\href{http://arxiv.org/abs/hep-ph/0101323}{{\tt
  hep-ph/0101323}}].

\bibitem{Dasgupta:2009tm}
M.~Dasgupta and Y.~Delenda, \emph{{On the universality of hadronisation
  corrections to QCD jets}},
  \href{http://dx.doi.org/10.1088/1126-6708/2009/07/004}{\emph{JHEP} {\bf 07}
  (2009) 004}, [\href{http://arxiv.org/abs/0903.2187}{{\tt 0903.2187}}].

\bibitem{Catani:1996vz}
S.~Catani and M.~H. Seymour, \emph{{A General algorithm for calculating jet
  cross-sections in NLO QCD}},
  \href{http://dx.doi.org/10.1016/S0550-3213(96)00589-5}{\emph{Nucl. Phys. B}
  {\bf 485} (1997) 291--419}, [\href{http://arxiv.org/abs/hep-ph/9605323}{{\tt
  hep-ph/9605323}}].

\bibitem{Dokshitzer:1995zt}
Y.~L. Dokshitzer and B.~R. Webber, \emph{{Calculation of power corrections to
  hadronic event shapes}},
  \href{http://dx.doi.org/10.1016/0370-2693(95)00548-Y}{\emph{Phys. Lett. B}
  {\bf 352} (1995) 451--455}, [\href{http://arxiv.org/abs/hep-ph/9504219}{{\tt
  hep-ph/9504219}}].

\bibitem{Akhoury:1995sp}
R.~Akhoury and V.~I. Zakharov, \emph{{On the universality of the leading, 1/Q
  power corrections in QCD}},
  \href{http://dx.doi.org/10.1016/0370-2693(95)00866-J}{\emph{Phys. Lett. B}
  {\bf 357} (1995) 646--652}, [\href{http://arxiv.org/abs/hep-ph/9504248}{{\tt
  hep-ph/9504248}}].

\bibitem{Dasgupta:2020fwr}
M.~Dasgupta, F.~A. Dreyer, K.~Hamilton, P.~F. Monni, G.~P. Salam and G.~Soyez,
  \emph{{Parton showers beyond leading logarithmic accuracy}},
  \href{http://dx.doi.org/10.1103/PhysRevLett.125.052002}{\emph{Phys. Rev.
  Lett.} {\bf 125} (2020) 052002}, [\href{http://arxiv.org/abs/2002.11114}{{\tt
  2002.11114}}].

\bibitem{Forshaw:2020wrq}
J.~R. Forshaw, J.~Holguin and S.~Pl\"atzer, \emph{{Building a consistent parton
  shower}}, \href{http://dx.doi.org/10.1007/JHEP09(2020)014}{\emph{JHEP} {\bf
  09} (2020) 014}, [\href{http://arxiv.org/abs/2003.06400}{{\tt 2003.06400}}].

\bibitem{FerrarioRavasio:2020guj}
S.~Ferrario~Ravasio, G.~Limatola and P.~Nason, \emph{{Infrared renormalons in
  kinematic distributions for hadron collider processes}},
  \href{http://dx.doi.org/10.1007/JHEP06(2021)018}{\emph{JHEP} {\bf 06} (2021)
  018}, [\href{http://arxiv.org/abs/2011.14114}{{\tt 2011.14114}}].

\bibitem{Besier:2018jen}
M.~Besier, D.~Van~Straten and S.~Weinzierl, \emph{{Rationalizing roots: an
  algorithmic approach}},
  \href{http://dx.doi.org/10.4310/CNTP.2019.v13.n2.a1}{\emph{Commun. Num.
  Theor. Phys.} {\bf 13} (2019) 253--297},
  [\href{http://arxiv.org/abs/1809.10983}{{\tt 1809.10983}}].

\bibitem{Besier:2019kco}
M.~Besier, P.~Wasser and S.~Weinzierl, \emph{{RationalizeRoots: Software
  Package for the Rationalization of Square Roots}},
  \href{http://dx.doi.org/10.1016/j.cpc.2020.107197}{\emph{Comput. Phys.
  Commun.} {\bf 253} (2020) 107197},
  [\href{http://arxiv.org/abs/1910.13251}{{\tt 1910.13251}}].

\bibitem{Goncharov:1998kja}
A.~B. Goncharov, \emph{{Multiple polylogarithms, cyclotomy and modular
  complexes}}, \href{http://dx.doi.org/10.4310/MRL.1998.v5.n4.a7}{\emph{Math.
  Res. Lett.} {\bf 5} (1998) 497--516},
  [\href{http://arxiv.org/abs/1105.2076}{{\tt 1105.2076}}].

\bibitem{Goncharov:2001iea}
A.~B. Goncharov, \emph{{Multiple polylogarithms and mixed Tate motives}},
  \href{http://arxiv.org/abs/math/0103059}{{\tt math/0103059}}.

\bibitem{Adams:2017ejb}
L.~Adams and S.~Weinzierl, \emph{{Feynman integrals and iterated integrals of
  modular forms}},
  \href{http://dx.doi.org/10.4310/CNTP.2018.v12.n2.a1}{\emph{Commun. Num.
  Theor. Phys.} {\bf 12} (2018) 193--251},
  [\href{http://arxiv.org/abs/1704.08895}{{\tt 1704.08895}}].

\bibitem{Abreu:2019fgk}
S.~Abreu, M.~Becchetti, C.~Duhr and R.~Marzucca, \emph{{Three-loop
  contributions to the $\rho$ parameter and iterated integrals of modular
  forms}}, \href{http://dx.doi.org/10.1007/JHEP02(2020)050}{\emph{JHEP} {\bf
  02} (2020) 050}, [\href{http://arxiv.org/abs/1912.02747}{{\tt 1912.02747}}].

\bibitem{Duhr:2019rrs}
C.~Duhr and L.~Tancredi, \emph{{Algorithms and tools for iterated Eisenstein
  integrals}}, \href{http://dx.doi.org/10.1007/JHEP02(2020)105}{\emph{JHEP}
  {\bf 02} (2020) 105}, [\href{http://arxiv.org/abs/1912.00077}{{\tt
  1912.00077}}].

\bibitem{Abreu:2022vei}
S.~Abreu, M.~Becchetti, C.~Duhr and M.~A. Ozcelik, \emph{{Two-loop master
  integrals for pseudo-scalar quarkonium and leptonium production and decay}},
  \href{http://dx.doi.org/10.1007/JHEP09(2022)194}{\emph{JHEP} {\bf 09} (2022)
  194}, [\href{http://arxiv.org/abs/2206.03848}{{\tt 2206.03848}}].

\bibitem{Panzer:2014caa}
E.~Panzer, \emph{{Algorithms for the symbolic integration of hyperlogarithms
  with applications to Feynman integrals}},
  \href{http://dx.doi.org/10.1016/j.cpc.2014.10.019}{\emph{Comput. Phys.
  Commun.} {\bf 188} (2015) 148--166},
  [\href{http://arxiv.org/abs/1403.3385}{{\tt 1403.3385}}].

\bibitem{Duhr:2019tlz}
C.~Duhr and F.~Dulat, \emph{{PolyLogTools \textemdash{} polylogs for the
  masses}}, \href{http://dx.doi.org/10.1007/JHEP08(2019)135}{\emph{JHEP} {\bf
  08} (2019) 135}, [\href{http://arxiv.org/abs/1904.07279}{{\tt 1904.07279}}].

\bibitem{Broedel:2014vla}
J.~Broedel, C.~R. Mafra, N.~Matthes and O.~Schlotterer, \emph{{Elliptic
  multiple zeta values and one-loop superstring amplitudes}},
  \href{http://dx.doi.org/10.1007/JHEP07(2015)112}{\emph{JHEP} {\bf 07} (2015)
  112}, [\href{http://arxiv.org/abs/1412.5535}{{\tt 1412.5535}}].

\bibitem{Broedel:2015hia}
J.~Broedel, N.~Matthes and O.~Schlotterer, \emph{{Relations between elliptic
  multiple zeta values and a special derivation algebra}},
  \href{http://dx.doi.org/10.1088/1751-8113/49/15/155203}{\emph{J. Phys. A}
  {\bf 49} (2016) 155203}, [\href{http://arxiv.org/abs/1507.02254}{{\tt
  1507.02254}}].

\bibitem{Broedel:2017jdo}
J.~Broedel, N.~Matthes, G.~Richter and O.~Schlotterer, \emph{{Twisted elliptic
  multiple zeta values and non-planar one-loop open-string amplitudes}},
  \href{http://dx.doi.org/10.1088/1751-8121/aac601}{\emph{J. Phys. A} {\bf 51}
  (2018) 285401}, [\href{http://arxiv.org/abs/1704.03449}{{\tt 1704.03449}}].

\end{thebibliography}\endgroup

\end{document}